\documentclass[aps,prb,twocolumn,superscriptaddress]{revtex4-1}
\usepackage[utf8]{inputenc}
\usepackage{amsfonts,amsmath,amssymb,amsthm} % Math packages
\usepackage{wasysym} % to load more math symbols
\usepackage{bbold}
\usepackage[usenames,dvipsnames]{xcolor}
\usepackage[pdftex]{graphicx}
\usepackage{tikz} % for the tree graph
\usetikzlibrary{arrows.meta}
\usetikzlibrary{shapes}
\usepackage{bm}
\usepackage[pdftex]{hyperref}
{\end{pmatrix}\end{medsize}} % Use accordingly to nccmath 
\usepackage{nccmath} % To resize the matrices (creating new environments).
\usepackage{natbib} % To get all the nice features/freedom when citing
\usepackage{physics} % To access Dirac notation
\usepackage{ulem}
\usepackage{diagbox} % For use of backslashbox

\hypersetup{colorlinks=true,linkcolor=red, citecolor=blue, urlcolor=blue,colorlinks=true,bookmarks=true,breaklinks=true}

\usepackage{array}
\newcolumntype{L}[1]{>{\raggedright\let\newline\\\arraybackslash\hspace{0pt}}m{#1}}
\newcolumntype{C}[1]{>{\centering\let\newline\\\arraybackslash\hspace{0pt}}m{#1}}
\newcolumntype{R}[1]{>{\raggedleft\let\newline\\\arraybackslash\hspace{0pt}}m{#1}}

\newcommand{\htcal}[1]{\hat{\mathcal{#1}}}
\newcommand{\tc}{\mathcal{T}_{\mathcal{C}}}

\newcommand{\cc}{\hat{c}}
\newcommand{\gp}{\mathcal{G}^{\phi}}

\begin{document}
% Title layout
%\title{Non-equilibrium spectral characteristics of the $\pi$-ton}
\title{Dynamical Mean Field Theory extension to the nonequilibrium Two-Particle Self-Consistent approach}
\author{Olivier \surname{Simard}}
\affiliation{Department of Physics, University of Fribourg, 1700 Fribourg, Switzerland}
\author{Philipp \surname{Werner}}
\affiliation{Department of Physics, University of Fribourg, 1700 Fribourg, Switzerland}
\date{\today}

%  Abstract
\begin{abstract}
Nonlocal correlations play an essential role in correlated electron systems, especially in the vicinity of phase transitions and crossovers, where two-particle correlation functions display a distinct momentum dependence. In nonequilibrium settings, the effect of nonlocal correlations on dynamical phase transitions, prethermalization phenomena and trapping in metastable states is not well understood. In this paper, we introduce a dynamical mean field theory (DMFT) extension to the nonequilibrium Two-particle Self-Consistent (TPSC) approach, which allows to perform nonequilibrium simulations capturing short- and long-ranged nonlocal correlations in the weak-, intermediate- and strong-correlation regime. The method self-consistently computes local spin and charge vertices, from which a momentum-dependent self-energy is constructed. Replacing the local part of the self-energy by the DMFT result within this self-consistent scheme provides an improved description of local correlation effects. We explain the details of the formalism and the implementation, and demonstrate the versatility of DMFT+TPSC with lattice hopping quenches and dimensional crossovers in the Hubbard model.
\end{abstract}
\maketitle

%%%%%%%%%%%%%%%%%%%%%%%%%%%%%%%%%%%%%%%%%%%%%%%% Introduction %%%%%%%%%%%%%%%%%%%%%%%%%%%%%%%%%%%%%%%%%%%%%%%%%%%%%%%%%%%%%%%%%%
%%

\section{Introduction}
\label{sec:Introduction}

Correlated electron materials are often characterized by competing or correlated degrees of freedom whose interplay can give rise to remarkable physical properties and symmetry-broken states. This competition involves spin, orbital, charge and lattice degrees of freedom which may be active at comparable energy scales.~\cite{PhysRevLett.120.166401} One way to disentangle competing or cooperative effects is via fine-tuned laser pulse excitation of correlated systems, which allow to reveal characteristic timescales, coupling constants and collective modes,\cite{Smallwood2016} and in some cases hidden nonthermal states.~\cite{PhysRevLett.127.227401} 
%Moreover, dimensionality is an important aspect when it comes to describing the phases of matter. In low-dimensional correlated systems, close to phase instabilities and crossovers, nonlocal correlations become essential for capturing the underlying physics.~\cite{Rohringer2018,PhysRevLett.123.097601,kauch_pitons_2019,https://doi.org/10.48550/arxiv.2112.15323,PhysRevB.104.245127_non_eq_piton_simard} 
Up to hundreds of femtoseconds after an impulsive excitation, the order parameter involved in a dynamical phase transition exhibits distinctly nonthermal scaling relations and fluctuations~\cite{Tsuji2013,Maklar2021,delaTorre2022} and the electronic band structure can be strongly modified.~\cite{PhysRevX.12.011013,PhysRevB.102.035109,doi:10.1126/sciadv.abd9275} 
These effects are expected to be particularly prominent in low-dimensional systems, where nonlocal correlations govern the physics close to phase instabilities and crossovers.~\cite{Rohringer2018,PhysRevLett.123.097601,kauch_pitons_2019,https://doi.org/10.48550/arxiv.2112.15323,PhysRevB.104.245127_non_eq_piton_simard} 
To capture the effect of nonlocal correlations, single- and two-particle correlation functions need to be calculated consistently, and this is challenging for several reasons. There is a lack of out-of-equilibrium methods that incorporate both local and nonlocal correlations and which allow to access the strongly-correlated regime. Dynamical Mean Field theory (DMFT) only captures local correlations,~\cite{RevModPhys.86.779_non_eq_review} the nonlocal components of GW+DMFT only charge fluctuations,\cite{Biermann2003,Ayral2013,PhysRevB.100.235117} the phenomenological time-dependent Ginzburg-Landau (tdGL) only considers low-order microscopic electronic fluctuations~\cite{PhysRevB.8.3423} and time-dependent Density Functional Theory (tdDFT) cannot properly describe strong correlation effects and does not capture the scattering processes which are relevant for thermalization at long times.~\cite{PhysRevLett.52.997}

The development of reliable, yet computationally efficient numerical methods, is crucial if we want to simulate nonthermal phenomena, including symmetry-broken states, up to experimentally relevant times of the order of picoseconds. Such methods would allow one to study accurately the destruction of thermal states and formation of nonthermal phases triggered by impulsive excitations,\cite{Strand2015,Bauer2015,Stahl2021} and possibly shed light on the mechanisms which underlie photoinduced metastable states, such as the superconducting-like states observed in  K$_3$C$_{60}$~\cite{Budden2021} and $\kappa$-organic compounds.~\cite{Buzzi2020} They would also allow to address fundamental questions such as the effect of long- and short-range correlations in the formation and (de)stabilization of prethermal and hidden states,\cite{Berges2004,Moeckel2008,Eckstein2009,Werner2020} and they would allow to study the role of order parameter fluctuations in nonthermal phase transitions beyond tdGL.

The challenge is to devise nonequilibrium numerical many-body methods for treating nonlocal correlations that are, on the one hand, computationally tractable and, on the other hand, accurate enough to capture the relevant physics. A promising method, which has recently been extended to the nonequilibrium domain,~\cite{https://doi.org/10.48550/arxiv.2205.13813} is the so-called Two-Particle Self-Consistent approach (TPSC).~\cite{Vilk_1997} TPSC correctly reproduces the pseudogap in models for cuprates~\cite{Vilk_1996} and the growth of antiferromagnetic (AFM) correlations as the renormalized classical regime -- where the AFM correlation length exceeds the de Broglie wave length -- is approached.~\cite{Vilk_1997} %
It can also deal with superconducting phases,~\cite{PhysRevB.68.174502_tpsc_superconductivity} two-particle vertex corrections~\cite{Bergeron_2011_optical_cond} and multi-orbital systems.\cite{https://doi.org/10.1002/andp.202000399} TPSC has been used in conjunction with Density Functional theory (DFT) in equilibrium to calculate the renormalization of the bands of iron pnictides and chalcogenides.~\cite{PhysRevB.102.035109} The main drawback of TPSC is that is does not fully capture strong local correlations, so that the method does not give access to the renormalized classical regime or Mott physics. To better account for strong local correlations while at the same time keeping track of the nonlocal correlations, a combination of DMFT and nonequilibrium TPSC is proposed in this paper, which resembles in spirit the recently developed equilibrium approaches of Refs.~\onlinecite{https://doi.org/10.48550/arxiv.2211.01919} and \onlinecite{https://doi.org/10.48550/arxiv.2211.01400}, and is applied to the single-band Hubbard model in the context of hopping and interaction quenches. We will in particular study the time-dependent spin and charge correlation functions and the pseudogap phase in the weak-to-intermediate coupling regime. 

The paper is structured as follows: In Sec.~\ref{subsec:Hubbard_model}, we present the Hamiltonian of the model and the methods used to solve it. More specifically, the nonequilibrium DMFT, nonequilibrium TPSC and nonequilibrium DMFT+TPSC are presented in Secs.~\ref{subsec:DMFT}, \ref{subsec:TPSC_variants} and \ref{ch:dmft_tpsc}, respectively. The results are shown and discussed in Sec.~\ref{sec:results}. We give our conclusions in Sec.~\ref{sec:conclusion}.
%%%%%%%%%%%%%%%%%%%%%%%%%%%%%%%%%%%%%%%%%%%%%%%% Model and Method %%%%%%%%%%%%%%%%%%%%%%%%%%%%%%%%%%%%%%%%%%%%%%%%%%%%%%%%%%%%%%
%%

\section{Model and methods}
\label{sec:Models_and_methods}

\subsection{Hubbard model}
\label{subsec:Hubbard_model}

We consider a single-band Hubbard model with time-dependent hopping parameters 
\begin{align}
\label{eq:Hubbard_model_intro}
\hat{\mathcal{H}}(t)=&-\sum_{ij,\sigma}t^{\text{hop}}_{ij}(t)\left(\hat{c}_{i,\sigma}^{\dagger}\hat{c}_{j,\sigma}+\text{H.c.}\right)+U\sum_i\hat{n}_{i,\uparrow}\hat{n}_{i,\downarrow}\nonumber\\
&-\mu \sum_i (\hat{n}_{i,\uparrow} + \hat{n}_{i,\downarrow}).
\end{align} 
Here, $t^{\text{hop}}_{ij}$ denotes the hopping amplitudes between sites $j$ and $i$, $\sigma \in \{\uparrow,\downarrow\}$ the spin, $\hat{c}^{(\dagger)}_{i,\sigma}$ are annihilation (creation) operators for site $i$, while $\hat{n}_{i\sigma}=\hat{c}^{\dagger}_{i,\sigma}\hat{c}_{i,\sigma}$ is the number operator, $U$ the local Hubbard repulsion and $\mu$ the chemical potential. We will consider ramps from a 2D square lattice to a 3D cubic lattice (and vice-versa) with in-plane nearest-neighbor hoppings $t^{\text{hop}}$, and use $t^{\text{hop}}$ as the unit of energy ($\hbar/t^{\text{hop}}$ as the unit of time). The ramps are implemented for the $z$-axis hopping, so that the corresponding time-dependent bare electronic dispersion reads 

\begin{align}
\label{eq:dispersion_relation}
\epsilon_{\mathbf{k}}(t) = -2t^{\text{hop}}\left(\cos{k_x}+\cos{k_y}\right) - 2t_{z}^{\text{hop}}(t)\cos{k_z},
\end{align}
where $-\pi \le k_x,k_y,k_z\le \pi$ defines the Brillouin zone. This implies that the bare bandwidth $W$ of the Hubbard model~\eqref{eq:Hubbard_model_intro} changes from $8t^{\text{hop}}$ (2D) to $12t^{\text{hop}}$ (3D) and vice-versa. Note that we have set the fundamental constants like $\hbar$, $k_B$, the electric charge $e$ and the lattice spacings $a$ to unity.

\subsection{Nonequilibrium DMFT}
\label{subsec:DMFT}

\subsubsection{General formalism}

Nonequilibrium DMFT is an implementation of the DMFT equations on the Kadanoff-Baym contour $\mathcal{C}$.~\cite{RevModPhys.86.779_non_eq_review,PhysRevLett.97.266408_freericks_non_eq} In DMFT, the lattice model is self-consistently mapped onto a single-site Anderson impurity model, where upon convergence the time-dependent hybridization function captures the effects of the lattice environment.~\cite{Georges_1996} The action of the nonequilibrium Anderson impurity problem is
\begin{align}
\label{eq:DMFT:DMFT_action}
\mathcal{S}[\Delta] =& -\int_{\mathcal{C}}\mathrm{d}z \ \hat{\mathcal{H}}_{\text{loc}}(z) \nonumber\\
& - \int_{\mathcal{C}}\mathrm{d}z \int_{\mathcal{C}} \mathrm{d}z^{\prime} \sum_\sigma \hat{c}^{\dagger}_\sigma(z)\Delta_\sigma(z,z^{\prime})\hat{c}_\sigma(z^{\prime}),
\end{align} 
where $ \hat{\mathcal{H}}_{\text{loc}}$ is the same local term as in the lattice model, $\hat{c}^{(\dagger)}_\sigma$ annihilates (creates) an electron with spin $\sigma$ on the impurity and $z\in \mathcal{C}$. The hybridization function is denoted by $\Delta^\sigma(z,z^{\prime})$, and the integrals span over the entire Kadanoff-Baym contour $\mathcal{C}$.

%\begin{figure}[t]
%  \centering
%    \includegraphics[width=1.0\linewidth]{kadanoff_baym_contour-crop.pdf}
%      \caption{Illustration of the Kadanoff-Baym contour. This contour starts off at time $t_0$ and goes to some time $t_{\text{max}}$, returns to $t_0$, and then extends along the imaginary-time axis to $t_0-i\beta$, where $\beta$ is the inverse temperature of the initial equilibrium state. The arrows indicate the contour ordering.
%      }
%  \label{fig:kadanoff_baym_contour}
%\end{figure}
%
With the nonequilibrium action \eqref{eq:DMFT:DMFT_action}, one can define the nonequilibrium impurity Green's function

\begin{align}
\label{eq:DMFT:Greens_function}
\mathcal{G}_{\text{imp}}^{\sigma}(z,z^{\prime}) = -i\text{Tr}\left[\mathcal{T}_{\mathcal{C}}e^{i\mathcal{S}[\Delta]}\hat{c}_{\sigma}(t)\hat{c}^{\dagger}_\sigma(t^{\prime})\right]/\mathcal{Z}[\Delta],
\end{align}
where $\mathcal{T}_{\mathcal{C}}$ is the time-ordering operator defined on the Kadanoff-Baym contour and $\mathcal{Z}[\Delta] = \text{Tr}\left[\mathcal{T}_{\mathcal{C}}e^{i\mathcal{S}[\Delta]}\right]$ is the partition function. The operator $\mathcal{T}_{\mathcal{C}}$ orders strings of operators according to the contour $\mathcal{C}$, which includes the forward branch $\mathcal{C}_1$, the backward branch $\mathcal{C}_2$ and the imaginary time branch ($\mathcal{C}$: $\mathcal{C}_1 \prec \mathcal{C}_2 \prec \mathcal{C}_3$). The impurity Green's function $\mathcal{G}^{\sigma}_{\text{imp}}$ will be computed using the third-order iterated perturbation theory (IPT) method, adapted to the nonequilibrium formalism (see Sec.~\ref{subsubsec:IPT}). When compared to second-order IPT, the additional third-order diagrams to the impurity solver allow to push $U/W$ to larger values and to dope away from half-filling.~\citep{tsuji_nonequilibrium_2013}

\subsubsection{Paramagnetic self-consistency}
\label{subsubsec:PM}

In nonequilibrium DMFT, the lattice self-energy is assumed to be local and identified with the impurity self-energy, $\Sigma^\sigma_{ij}(z,z^{\prime})=\Sigma_{\text{imp}}^\sigma(z,z^{\prime})\delta_{ij}$, which is an approximation in systems with finite coordination number.~\cite{Mueller_1989,Georges_1996} Moreover, to attain the self-consistency condition, the impurity Green's function 
$\mathcal{G}^{\sigma}_{\text{imp}}(z,z^{\prime})$ must be identical to the local lattice Green's function $\mathcal{G}^{\sigma}_\text{loc}(z,z^{\prime})$. 
This self-consistency condition determines the hybridization function $\Delta^\sigma(z,z')$ appearing in the impurity action~\eqref{eq:DMFT:DMFT_action}, which plays the role of a dynamical mean field. 

In impurity solvers based on weak-coupling perturbation theory, it is more convenient to work with the so-called Weiss Green's function $\mathcal{G}^{\sigma}_{0}$, which is related to the hybridization function via the Dyson equation
\begin{align}
\label{eq:PM:Weiss_Green_hyb}
\left[i\partial_z+\mu\right]\mathcal{G}^{\sigma}_0(z,z^{\prime}) - \Delta^{\sigma}(z,\bar{z})\mathcal{G}^0_{\sigma}(\bar{z},z^{\prime})=\delta^{\mathcal{C}}(z,z^{\prime}),
\end{align} 
and which contains the same information. 
Here, $\delta^{\mathcal{C}}(z,z')$ represents the delta function on the Kadanoff-Baym contour. The convolution along the contour $\mathcal{C}$ will sometimes be denoted by the operator ``$\ast$''. Contour-time arguments $z$ featuring an over-bar are integrated over $\mathcal{C}$.

The impurity Dyson equation for the interacting problem connects the impurity Green's function $\mathcal{G}^{\sigma}_{\text{imp}}$, the impurity self-energy $\Sigma_{\text{imp}}^{\sigma}$ and the Weiss Green's function $\mathcal{G}_{\sigma}^0$ as follows: 
\begin{align}
\label{eq:PM:Dyson_eq}
&\mathcal{G}_{\text{imp}}^{\sigma}(z,z^{\prime}) = \mathcal{G}_{0}^{\sigma}(z,z^{\prime}) + \mathcal{G}_{0}^{\sigma}(z,\bar{z})\Sigma_{\text{imp}}^{\sigma}(\bar{z},\bar{z}^{\prime})\mathcal{G}_{\text{imp}}^{\sigma}(\bar{z}^{\prime},z^{\prime}).
\end{align}
As pointed out for example in Ref.~\onlinecite{PhysRevB.104.245127_non_eq_piton_simard}, the formulation of the impurity solver in terms of the Weiss Green's functions, $\Sigma_{\text{imp}}=\Sigma_{\text{imp}}[\mathcal{G}_0]$, violates the energy conservation principle (in the absence of an external field) because the self-energy is not expressed in terms of the interacting Green's functions. However, it turns out that combining DMFT with TPSC improves the energy conservation such that one can perform meaningful simulations up to longer times. Hence, it is not necessary to resort to an impurity solver which expresses the self-energy in terms of the interacting impurity Green's function, $\Sigma_{\text{imp}}=\Sigma_{\text{imp}}[\mathcal{G}_{\text{imp}}]$, which can lead to poor results already for the short time dynamics \cite{Eckstein2010ipt} and which does not correctly reproduce the energy scale $\omega\sim W$ of the onset of the asymptotic behavior (high-frequency and atomic limits) of the Hubbard model self-energy~\cite{Vilk_1997} (see Sec.~\ref{subsubsec:IPT}).

The lattice Green's function $\mathcal{G}_{\mathbf{k}}^{\sigma}$ is related to the impurity self-energy via the lattice Dyson equation
\begin{align}
\label{eq:PM:projected_green_function_impurity}
&\left[i\partial_z+\mu-\epsilon(\mathbf{k})-\Sigma^{\delta,\sigma}_{\text{imp}}(z)\right]\mathcal{G}_{\mathbf{k}}^{\sigma}(z,z^{\prime})\notag\\
&\hspace{1.0cm}- \Sigma_{\text{imp}}^{\sigma}(z,\bar{z})\mathcal{G}_{\mathbf{k}}^{\sigma}(\bar{z},z^{\prime})=\delta^{\mathcal{C}}(z,z^{\prime}),
\end{align}
where $\epsilon(\mathbf{k})$ is the bare electronic dispersion written in Eq.~\eqref{eq:dispersion_relation}, $\mu$ is the impurity chemical potential and $\Sigma^{\delta}_{\text{imp}}$ represents the time-local impurity self-energy diagrams, denoted by $\Sigma_{H}$ in Sec.~\ref{subsubsec:IPT}. 

Owing to the DMFT self-consistency condition, the impurity Dyson equation \eqref{eq:PM:Dyson_eq} can be rewritten as a Volterra integral equation where the impurity Green's function $\mathcal{G}_{\text{imp}}^{\sigma}$ is replaced by the $\mathbf{k}$-averaged lattice Green's function $\mathcal{G}^{\sigma}_\text{loc}$:

\begin{align}
\label{eq:PM:impurity_self_energy}
\mathcal{G}_0^{\sigma}(z,\bar{z})\left[\delta^{\mathcal{C}}(\bar{z},z^{\prime}) + F^{\sigma}(\bar{z},z^{\prime})\right] = \mathcal{G}^{\sigma}_\text{loc}(z,z^{\prime}),
\end{align} where $F^{\sigma}(z,z^{\prime})\equiv \Sigma_{\text{imp}}^{\sigma}(z,\bar{z}^{\prime})\mathcal{G}^{\sigma}_\text{loc}(\bar{z}^{\prime},z^{\prime})$. Equations \eqref{eq:PM:projected_green_function_impurity} and \eqref{eq:PM:impurity_self_energy}, along with the diagrammatic expression for the impurity self-energy, form a closed set of equations determining $\mathcal{G}^{\sigma}_0$.~\cite{Eckstein2010ipt,tsuji_nonequilibrium_2013} The weak-coupling impurity self-energy $\Sigma_{\text{imp}}^{\sigma}$ enters Eq.~\eqref{eq:PM:projected_green_function_impurity} and the impurity Dyson equation~\eqref{eq:PM:Dyson_eq}, and the DMFT equations are iterated until $\mathcal{G}^{\sigma}_0$ has converged. To solve the Dyson equations \eqref{eq:PM:Dyson_eq}, \eqref{eq:PM:projected_green_function_impurity} and the Volterra integral equation \eqref{eq:PM:impurity_self_energy}, we use the NESSi package.~\cite{Nessi} For the paramagnetic solutions considered in this study, all quantities are independent of the spin projection, \textit{i.e.} we have that $\Sigma_{\text{imp}}^{\sigma}=\Sigma_{\text{imp}}^{-\sigma}$ and the same holds for $\mathcal{G}_{\text{imp}}^{\sigma}$ and $\Delta^{\sigma}$.

\subsubsection{Impurity solver}
\label{subsubsec:IPT} 

Since we work in the weak coupling regime ($U \lesssim W/2$), we use a weak-coupling impurity solver based on an expansion of the self-energy up to $3^{\text{rd}}$ order in the interaction $U$.~\cite{tsuji_nonequilibrium_2013} This approach is a generalization of the second-order iterated perturbation theory (IPT) for the Anderson impurity model.~\cite{kajueter_new_1996,arsenault_benchmark_2012} In the ``bare IPT" formalism, the self-energy $\Sigma_{\text{imp}}[\mathcal{G}_0]$ is approximated as a functional of the Weiss Green's function defined in Eq.~\eqref{eq:PM:Weiss_Green_hyb}. Alternatively, one can define a ``bold IPT," where $\mathcal{G}_0^{\sigma}$ in the self-energy diagrams is replaced by the dressed impurity Green's function $\mathcal{G}^{\sigma}_{\text{imp}}$ obtained from Eq.~\eqref{eq:PM:Dyson_eq}. This replacement has a detrimental effect on the short-time dynamics, but it yields a conserving approximation, which means that the total energy after a perturbation is conserved under the time evolution.~\citep{Eckstein2010ipt} In this paper, we will use the ``bare IPT" formalism within the nonequilibrium DMFT+TPSC scheme introduced in Sec.~\ref{ch:dmft_tpsc}, since it turns out that this scheme conserves the energy to a very good approximation in the considered parameter range.

By making use of Hedin's equations,~\cite{Hedin1965} one can generate systematically, order by order, the Feynman diagrams that characterize single- and two-particle correlation functions. This, however, %cannot be done for arbitrary 
becomes impractical for 
high expansion orders in the interaction $U$, since one would have to deal with a large set of diagrams. We thus only consider diagrams up to the third order. In the case of the Hubbard model, the Fock interaction term vanishes and this leads (in addition to the first-order Hartree diagram) to two self-energy diagrams of order $\mathcal{O}(U^2)$ and eight diagrams of order $\mathcal{O}(U^3)$. These leading diagrams are derived in detail in Appendix~\ref{appendice:ch:weak_coupling_self_expansion}. In this section, we present the formulas for the different contributions and their diagrammatic representations. Note that at half-filling, we choose $\mu=U/2$, so that the Hartree terms vanish in the paramagnetic state. However, the Hartree diagrams and those containing Hartree insertions do not vanish if the system is doped away from half-filling.\cite{tsuji_nonequilibrium_2013}

%%%%%%%%%%%%%%%%%%%%%%%%%%%%%%%%%%%%%%%%%%%%%%%% 2nd order IPT %%%%%%%%%%%%%%%%%%%%%%%%%%%%%%%%%%

\paragraph{$2^{\text{nd}}$-order IPT}
\label{ch:2nd_order_IPT}

\begin{figure}[t!]
  \centering
    \includegraphics[scale=1.2]{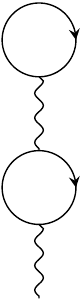}
      \caption{$2^{\text{nd}}$-order Hartree self-energy diagram. The fermionic propagators represent the Weiss Green's functions $\mathcal{G}^0$.}
  \label{fig:2nd_Hartree_diagram}
\end{figure}

To second order, the Hartree contribution $\Sigma_H^{(2)}$ reads

\begin{align}
\label{eq:nonequilibrium_quantum_many_body_physics:IPT:Hartree_2nd_order}
&\Sigma_{H,\sigma}^{(2)}(z,z^{\prime}) = (-i)^2U(z)\mathcal{G}_0^{-\sigma}(z,\bar{z})U(\bar{z})\mathcal{G}_0^{\sigma}(\bar{z},\bar{z}^{+})\notag\\
&\times\mathcal{G}_{0}^{-\sigma}(\bar{z},z^+)\delta^{\mathcal{C}}(z,z^{\prime}).
\end{align}
The diagram representing Eq.~\eqref{eq:nonequilibrium_quantum_many_body_physics:IPT:Hartree_2nd_order} is shown in Fig.~\ref{fig:2nd_Hartree_diagram} and is a combination of two Hartree diagrams.
\begin{figure}[t!]
  \centering
    \includegraphics[scale=1.2]{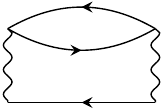}
      \caption{$2^{\text{nd}}$-order self-energy diagram.}
  \label{fig:2nd_order_self}
\end{figure}
The term $\Sigma_{H,\sigma}^{(2)}$ is necessary to spontaneously break the SU(2) spin symmetry within DMFT, since it confers different effective chemical potentials to the different spin projections.~\cite{PhysRevB.103.104415_Simard_pi_ton} 

The remaining second-order diagram comprises one particle-hole bubble diagram, as depicted in Fig.~\ref{fig:2nd_order_self}, and reads
\begin{align}
\label{eq:nonequilibrium_quantum_many_body_physics:IPT:second_order_bubble}
\Sigma^{(2)}_{\sigma}(z,z^{\prime}) = U(z)\mathcal{G}_0^{\sigma}(z,z^{\prime})U(z^{\prime})\mathcal{G}_0^{-\sigma}(z^{\prime},z^+)\mathcal{G}^{-\sigma}_0(z,{z^{\prime}}^+).
\end{align}
The self-energy~\eqref{eq:nonequilibrium_quantum_many_body_physics:IPT:second_order_bubble}, expressed as a functional of the Weiss Green's function, $\Sigma^{(2)}[\mathcal{G}_0]$, captures the Mott transition and crossover because Eq.~\eqref{eq:nonequilibrium_quantum_many_body_physics:IPT:second_order_bubble} correctly reproduces the large-$U$ limit of the Hubbard model~\eqref{eq:Hubbard_model_intro},\footnote{See Appendix~\ref{appendice:ch:weak_coupling_self_expansion} and the discussion below for more details.} which coincides with the atomic limit, at half-filling. On the other hand, the self-energy expressed in terms of the boldified Green's function, $\Sigma^{(2)}[\mathcal{G}_{\text{imp}}]$, does not allow to describe the Mott transition. This is due to the fact that, even though the perturbation theory expressed in terms of the interacting Green's functions leads to the correct asymptotics at half-filling, it does not set in at $\omega\sim W$, but rather at $\omega\gg U$, which is too high and contradicts the Pauli exclusion principle.~\cite{Vilk_1997} In order to carry out the perturbation theory using the dressed Green's functions, one would need to consider the frequency dependent G-skeletonic two-particle vertex corrections to get physically sound results. In the weak-coupling regime $U\lesssim W/2$, both schemes however lead to similar results for short times.~\cite{PhysRevB.104.245127_non_eq_piton_simard}

%%%%%%%%%%%%%%%%%%%%%%%%%%%%%%%%%%%%%%%% 3rd order IPT %%%%%%%%%%%%%%%%%%%%%%%%%%%%%%%%%%

\paragraph{$3^{\text{rd}}$-order solver}
\label{ch:3rd_order_IPT}

\begin{figure}[t!]
  \begin{minipage}{.5\columnwidth}
  \centering
  \includegraphics[scale=1.0]{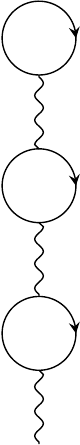}
  \end{minipage}\hfill
  \begin{minipage}{.5\columnwidth}
  \centering
  \includegraphics[scale=1.0]{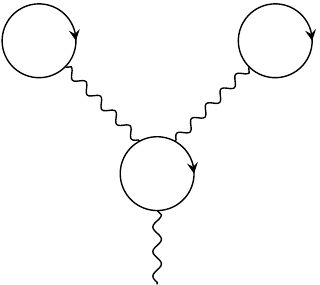}%
  \end{minipage}%
  \par
  \begin{minipage}{\columnwidth}
  \centering
  \includegraphics[scale=1.0]{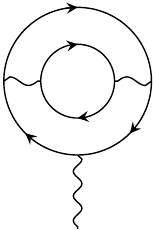}%
  \end{minipage}%
  \caption{$3^{\text{rd}}$-order diagrams $\Sigma_H^{3a}$ (top left corner), $\Sigma_H^{3b}$ (top right corner) and $\Sigma_H^{3c}$ (bottom).}
  \label{fig:H_3_diagram}
\end{figure}

We next describe the $3^{\text{rd}}$-order self-energy diagrams. There are three diagrams contributing to the time-local component of the self-energy. The first one is obtained by attaching a Hartree diagram to the top propagator of the $2^{\text{nd}}$-order diagram \eqref{eq:nonequilibrium_quantum_many_body_physics:IPT:Hartree_2nd_order}. This produces the diagram shown in the top left corner of Fig.~\ref{fig:H_3_diagram}, which corresponds to the expression
\begin{align}
\label{eq:nonequilibrium_quantum_many_body_physics:IPT:Hartree_3a}
&\Sigma^{3a}_{H,\sigma}(z,z^{\prime}) = (-i)^3U(z)\mathcal{G}_0^{-\sigma}(z,\bar{z})U(\bar{z})\mathcal{G}_0^{\sigma}(\bar{z},\bar{z}^{\prime})U(\bar{z}^{\prime})\notag\\
&\times\mathcal{G}_{0}^{-\sigma}(\bar{z}^{\prime},\bar{z}^{\prime +})\mathcal{G}_0^{\sigma}(\bar{z}^{\prime},\bar{z}^{+})\mathcal{G}_0^{-\sigma}(\bar{z},z^+)\delta^{\mathcal{C}}(z,z^{\prime}).
\end{align}
The second time-local $3^{\text{rd}}$-order self-energy diagram stems from two Hartree self-energy corrections to the first-order Hartree term. This gives the diagram shown in the top right corner of Fig.~\ref{fig:H_3_diagram}, namely
\begin{align}
\label{eq:nonequilibrium_quantum_many_body_physics:IPT:Hartree_3b}
&\Sigma^{3b}_{H,\sigma}(z,z^{\prime}) = (-i)^3U(z)\mathcal{G}_0^{-\sigma}(z,\bar{z})U(\bar{z})\mathcal{G}_0^{\sigma}(\bar{z},\bar{z}^{+})\mathcal{G}_{0}^{-\sigma}(\bar{z},\bar{z}^{\prime})\notag\\
&\times U(\bar{z}^{\prime})\mathcal{G}_0^{\sigma}(\bar{z}^{\prime},\bar{z}^{\prime +})\mathcal{G}_0^{-\sigma}(\bar{z}^{\prime},z^+)\delta^{\mathcal{C}}(z,z^{\prime}).
\end{align}
The third diagram comes from the insertion of the bare $2^{\text{nd}}$-order self-energy diagram \eqref{eq:nonequilibrium_quantum_many_body_physics:IPT:second_order_bubble} into the first-order Hartree propagator, giving the bottom diagram of Fig.~\ref{fig:H_3_diagram},
\begin{align}
\label{eq:nonequilibrium_quantum_many_body_physics:IPT:Hartree_3c}
&\Sigma^{3c}_{H,\sigma}(z,z^{\prime}) = -iU(z)\mathcal{G}_0^{-\sigma}(z,\bar{z})U(\bar{z})\mathcal{G}_0^{\sigma}(\bar{z},\bar{z}^{\prime})U(\bar{z}^{\prime})\notag\\
&\times\mathcal{G}_{0}^{-\sigma}(\bar{z}^{\prime},\bar{z}^{+})\mathcal{G}_0^{-\sigma}(\bar{z},\bar{z}^{\prime})\mathcal{G}_0^{\sigma}(\bar{z}^{\prime},z^+)\delta^{\mathcal{C}}(z,z^{\prime}).
\end{align}
The set of diagrams corresponding to Eqs.~\eqref{eq:nonequilibrium_quantum_many_body_physics:IPT:Hartree_3a}, \eqref{eq:nonequilibrium_quantum_many_body_physics:IPT:Hartree_3b} and \eqref{eq:nonequilibrium_quantum_many_body_physics:IPT:Hartree_3c} represent a $3^{\text{rd}}$-order shift of the chemical potential.

\begin{figure}[t!]
    \centering
    \begin{minipage}{0.5\columnwidth}
        \centering
        \includegraphics[scale=1.2]{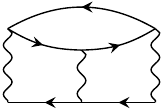}
    \end{minipage}%
    \begin{minipage}{0.5\columnwidth}
        \centering
        \includegraphics[scale=1.2]{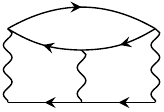}
    \end{minipage}
    \caption{$3^{\text{rd}}$-order diagrams $\Sigma^{3a}$ (left) and $\Sigma^{3b}$ (right).}
    \label{fig:3_diagram}
\end{figure}

Another category of diagrams originates from the consideration of the second-order self-energy diagram~\eqref{eq:nonequilibrium_quantum_many_body_physics:IPT:second_order_bubble} in the vertex function $\Gamma\equiv-\frac{\delta\Sigma}{\delta\mathcal{G}}$ discussed in details in Sec.~\ref{subsec:TPSC_variants}. This gives three distinct vertex terms out of which two lead to a nonzero contribution.\footnote{To obtain those diagrams, the lowest-order diagram (particle-hole bubble in $\mathcal{G}_0^{\sigma}$) in the Bethe-Salpeter equation~\eqref{eq:bethe_Salpeter_equation} is used in Eq.~\eqref{eq:self_longitudinal_derivation}.} The first of those diagrams reads
\begin{align}
\label{eq:nonequilibrium_quantum_many_body_physics:IPT:3a_diagram}
&\Sigma_{\sigma}^{3a}(z,z^{\prime}) = iU(z)U(z^{\prime})\mathcal{G}_0^{-\sigma}(z,z^{\prime})\mathcal{G}_0^{\sigma}(z,\bar{z})\mathcal{G}_0^{\sigma}(\bar{z},z^{\prime})U(\bar{z})\notag\\
&\times\mathcal{G}_0^{-\sigma}(z^{\prime},\bar{z})\mathcal{G}_0^{-\sigma}(\bar{z},z^{+}),
\end{align}
and the second diagram of this category reads
\begin{align}
\label{eq:nonequilibrium_quantum_many_body_physics:IPT:3b_diagram}
&\Sigma_{\sigma}^{3b}(z,z^{\prime}) = iU(z)U(z^{\prime})\mathcal{G}_0^{-\sigma}(z^{\prime},z^+)\mathcal{G}_0^{-\sigma}(z,\bar{z}^{+})\mathcal{G}_0^{\sigma}(z,\bar{z})\notag\\
&\times U(\bar{z})\mathcal{G}_0^{\sigma}(\bar{z},z^{\prime})\mathcal{G}_0^{-\sigma}(\bar{z},{z^{\prime}}^{+}).
\end{align}
The diagram representing Eq.~\eqref{eq:nonequilibrium_quantum_many_body_physics:IPT:3a_diagram} is shown on the left of Fig.~\ref{fig:3_diagram} and the one representing Eq.~\eqref{eq:nonequilibrium_quantum_many_body_physics:IPT:3b_diagram} is shown on the right of Fig.~\ref{fig:3_diagram}.

\begin{figure}[h!]
  \begin{minipage}{.5\columnwidth}
  \includegraphics[scale=1.2]{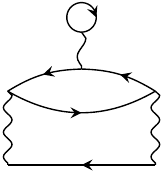}
  \end{minipage}\hfill
  \begin{minipage}{.5\columnwidth}
  \includegraphics[scale=1.2]{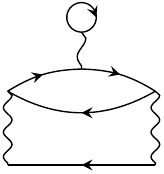}%
  \end{minipage}%
  \par
  \begin{minipage}{\columnwidth}
  \centering
  \includegraphics[scale=1.2]{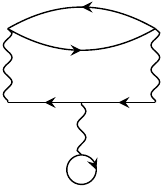}%
  \end{minipage}%
  \caption{$3^{\text{rd}}$-order diagrams $\Sigma^{3d}$ (top left corner), $\Sigma^{3c}$ (top right corner) and $\Sigma^{3e}$ (bottom).}
  \label{fig:3_diagram_2}
\end{figure}

The next (and last) series of $3^{\text{rd}}$-order Feynman diagrams come from the insertion of Hartree-type self-energy corrections into the Green's functions of the second-order self-energy \eqref{eq:nonequilibrium_quantum_many_body_physics:IPT:second_order_bubble}. The first such diagram (top left of Fig.~\ref{fig:3_diagram_2}) reads
\begin{align}
\label{eq:nonequilibrium_quantum_many_body_physics:IPT:3c_diagram}
&\Sigma_{\sigma}^{3c}(z,z^{\prime}) = -iU(z)U(z^{\prime})\mathcal{G}_0^{\sigma}(z,z^{\prime})\mathcal{G}_0^{-\sigma}(z^{\prime},z^{+})\mathcal{G}_0^{-\sigma}(z,\bar{z})\notag\\
&\times U(\bar{z})\mathcal{G}_0^{\sigma}(\bar{z},\bar{z}^{+})\mathcal{G}_0^{-\sigma}(\bar{z},{z^{\prime}}).
\end{align}
As second diagram (top right of Fig.~\ref{fig:3_diagram_2}) we obtain
\begin{align}
\label{eq:nonequilibrium_quantum_many_body_physics:IPT:3d_diagram}
&\Sigma_{\sigma}^{3d}(z,z^{\prime}) = -iU(z)U(z^{\prime})\mathcal{G}_0^{\sigma}(z,z^{\prime})\mathcal{G}_0^{-\sigma}(z^{\prime},\bar{z})U(\bar{z})\mathcal{G}_0^{\sigma}(\bar{z},\bar{z}^{+})\notag\\
&\times\mathcal{G}_0^{-\sigma}(\bar{z},z^{+})\mathcal{G}_0^{-\sigma}(z,{z^{\prime}})
\end{align}
and the third diagram (bottom of Fig.~\ref{fig:3_diagram_2}) is
\begin{align}
\label{eq:nonequilibrium_quantum_many_body_physics:IPT:3e_diagram}
&\Sigma_{\sigma}^{3e}(z,z^{\prime}) = -iU(z)U(z^{\prime})\mathcal{G}_0^{\sigma}(z,\bar{z})U(\bar{z})\mathcal{G}_0^{-\sigma}(\bar{z},\bar{z}^{+})\mathcal{G}_0^{\sigma}(\bar{z},z^{\prime})\notag\\
&\times\mathcal{G}_0^{-\sigma}(z^{\prime},z^{+})\mathcal{G}_0^{-\sigma}(z,{z^{\prime}}^+).
\end{align}

As shown in Ref.~\onlinecite{tsuji_nonequilibrium_2013}, the addition of the third-order self-energy diagrams allows one to access higher values of $U/W$ (compared to second-order IPT) and to dope the systems with electrons or holes away from half-filling. The inclusion of these extra self-energy diagrams however does not improve the IPT impurity solver in the strong-coupling regime ($U>W$), which is why we restrict the current study to the weak-to-intermediate correlation regime. Moreover, the fact that the TPSC self-energy introduced below and the second-order IPT self-energy \eqref{eq:nonequilibrium_quantum_many_body_physics:IPT:second_order_bubble} share the same asymptotics when $U \rightarrow 0$ makes it natural to combine these two diagrammatic approaches.~\cite{doi:10.1143/JPSJ.69.3912}

\subsection{Nonequilibrium TPSC and TPSC+GG}
\label{subsec:TPSC_variants}

\subsubsection{General formalism}
\label{ch:general_formalism}

In this section, we derive in detail the nonequilibrium Two-Particle Self-Consistent approach and a variant proposed in Ref.~\onlinecite{https://doi.org/10.48550/arxiv.2205.13813}, namely TPSC+GG. The formalism and the steps in the derivation follow Refs.~\onlinecite{https://doi.org/10.48550/arxiv.2205.13813} and \onlinecite{Vilk_1997}. We first briefly introduce the nonequilibrium generating functional formalism.~\cite{PhysRev.115.1342} The nonequilibrium Green's function can be used to express arbitrary order correlation functions between particles on the Kadanoff-Baym (KB) contour and these can be generated by the functional 

\begin{align}
\label{eq:generating_corr_func_out_equilibrium}
&\mathcal{Z}[\phi] = \text{Tr}\biggl[\tc e^{-i\int_{\mathcal{C}} \mathrm{d}z \ \htcal{H}(z)}\underbrace{e^{-i \cc^{\dagger}_{\bar{\alpha}}(\bar{z}_1)\phi_{\bar{\alpha}\bar{\beta}}(\bar{z}_1,\bar{z}_2)\cc_{\bar{\beta}}(\bar{z}_2)}}_{\equiv S[\phi]} \biggr],
\end{align} where $\mathcal{C}$ stands for the KB contour and $\htcal{H}$ is the Hubbard Hamiltonian~\eqref{eq:Hubbard_model_intro} whose equations of motion we want to derive. $\tc$ is the time-ordering operator on $\mathcal{C}$ and $\phi$ is a source field defined on the contour. The Greek indices represent arbitrary degrees of freedom, such as lattice site or spin, and $S[\phi]$ is a functional of the source field $\phi$. Just like for the contour-time arguments, the bar over the indices means that they are summed over. The trace in Eq.~\eqref{eq:generating_corr_func_out_equilibrium} spans over the eigenstates in Fock space. According to Eq.~\eqref{eq:generating_corr_func_out_equilibrium}, the contour Green's function reads

\begin{align}
\label{eq:contour_G_phi}
\gp_{\epsilon\zeta}(z_1,z_2) = -\frac{\delta\ln{\mathcal{Z}[\phi]}}{\delta\phi_{\zeta\epsilon}(z_2,z_1)} = -i\langle\tc \cc_{\epsilon}(z_1)\cc^{\dagger}_{\zeta}(z_2)\rangle_{\phi}.
\end{align}
In Eq.~\eqref{eq:contour_G_phi}, the grand-canonical ensemble average is

\begin{align}
\label{eq:def_ensemble_average}
\langle \cdots \rangle_{\phi} = \frac{1}{\mathcal{Z}[\phi]}\sum_{i}\bra{\Psi_i}e^{-i\int_{\mathcal{C}}\mathrm{d}\bar{z}\htcal{H}(\bar{z})} S[\phi]\cdots\ket{\Psi_i},
\end{align} with the $\{\ket{\Psi_i}\}$ a set of eigenstates of the Fock space. Using Eq.~\eqref{eq:contour_G_phi}, we perform a second functional derivative
\begin{align}
\label{eq:second_der_phi_partition_function_1}
&\frac{\delta \gp_{\epsilon\zeta}(z_1,z_2)}{\delta\phi_{\gamma\delta}(z_4,z_3)} = \gp_{\delta\gamma}(z_3,z_4)\gp_{\epsilon\zeta}(z_1,z_2)\notag\\
&\hspace{1.0cm} - \langle\cc_{\gamma}^{\dagger}(z_4)\cc_{\delta}(z_3)\cc_{\epsilon}(z_1)\cc^{\dagger}_{\zeta}(z_2) \rangle_{\phi},
\end{align} which, defining the two-particle correlation function $\chi\equiv -i\frac{\delta\mathcal{G}}{\delta\phi}$ (\textit{cf.} Eq.~(12.18) in Ref.~\onlinecite{stefanucci_van_leeuwen_2013}), leads to
\begin{align}
\label{eq:second_der_phi_partition_function}
&\chi^{\phi}_{\epsilon\zeta,\gamma\delta}(z_1,z_2;z_4,z_3) =  i\langle\tc\cc^{\dagger}_{\gamma}(z_4)\cc_{\delta}(z_3)\cc_{\epsilon}(z_1)\cc^{\dagger}_{\zeta}(z_2) \rangle_{\phi}\notag\\
&\hspace{1.0cm} - i\gp_{\delta\gamma}(z_3,z_4)\gp_{\epsilon\zeta}(z_1,z_2).
\end{align}\noindent
Note that Eq.~\eqref{eq:second_der_phi_partition_function_1} corresponds to Eq.~(15.11) in Ref.~\onlinecite{stefanucci_van_leeuwen_2013}. Another important result originates from the ``closure relation''
\begin{align}
\label{eq:closure_relation}
\frac{\delta \left(\gp_{\epsilon\bar{\alpha}}(z_1,\bar{z}_3)\gp_{\bar{\alpha}\eta}(\bar{z}_3,z_2)^{-1}\right) }{\delta \phi_{\gamma\delta}(z_4,z_3)} = 0.
\end{align} 
%allowing us to skip the summation and integral symbols. 
Equation~\eqref{eq:closure_relation} gives
\begin{align}
\label{eq:closure_relation_expanded}
\frac{\delta\gp_{\epsilon\zeta}(z_1,z_2)}{\delta\phi_{\gamma\delta}(z_4,z_3)} = -\gp_{\epsilon\bar{\alpha}}(z_1,\bar{z}_3)\frac{\delta\gp_{\bar{\alpha}\bar{\eta}}(\bar{z}_3,\bar{z}_5)^{-1}}{\delta\phi_{\gamma\delta}(z_4,z_3)}\gp_{\bar{\eta}\zeta}(\bar{z}_5,z_2),
\end{align} 
and the modified Dyson equation with the source field reads
\begin{align}
\label{eq:modified_Dyson_equation}
&\gp_{\alpha\eta}(z_3,z_5)^{-1}\notag\\
&= {\mathcal{G}^0_{\alpha\eta}(z_3,z_5)}^{-1} - \phi_{\alpha\eta}(z_3,z_5) - \Sigma^{\phi}_{\alpha\eta}(z_3,z_5).
\end{align} 
Equation~\eqref{eq:modified_Dyson_equation} appears naturally when deriving the equations of motion of Eq.~\eqref{eq:contour_G_phi}, as will be shown later on. 
%In this section, the noninteracting Green's function $\mathcal{G}_0^{\sigma}$ is not a functional of the hybridization function. 
In this section, $\mathcal{G}_0^{\sigma}$ denotes the noninteracting lattice Green's function.  
Note that all the two-time objects introduced hitherto can be expressed in a $3\times 3$ matrix form, as described in Ref.~\onlinecite{RevModPhys.86.779_non_eq_review}. Inserting Eq.~\eqref{eq:modified_Dyson_equation} into Eq.~\eqref{eq:closure_relation_expanded}, we get
\begin{align}
\label{eq:closure_relation_expanded_2}
&-i\frac{\delta\gp_{\epsilon\zeta}(z_1,z_2)}{\delta\phi_{\gamma\delta}(z_4,z_3)} = -i\gp_{\epsilon\gamma}(z_1,z_4)\gp_{\delta\zeta}(z_3,z_2)\notag\\
& -i \gp_{\epsilon\bar{\alpha}}(z_1,\bar{z}_3)\frac{\delta\Sigma^{\phi}_{\bar{\alpha}\bar{\eta}}(\bar{z}_3,\bar{z}_5)}{\delta\gp_{\bar{\theta}\bar{\omega}}(\bar{z}_6,\bar{z}_7)}\frac{\delta\gp_{\bar{\theta}\bar{\omega}}(\bar{z}_6,\bar{z}_7)}{\delta\phi_{\gamma\delta}(z_4,z_3)}\gp_{\bar{\eta}\zeta}(\bar{z}_5,z_2),
\end{align} 
where we used the chain rule for the self-energy $\Sigma[\mathcal{G}]$. Defining the two-particle irreducible G-skeletonic vertex function $\Gamma \equiv -\frac{\delta\Sigma}{\delta\mathcal{G}}$ (\textit{cf.} Eq.~(12.34) in Ref.~\onlinecite{stefanucci_van_leeuwen_2013}), we get the Bethe-Salpeter equation (\textit{cf.} Eq.~(12.17) in Ref.~\onlinecite{stefanucci_van_leeuwen_2013})
\begin{align}
\label{eq:bethe_Salpeter_equation}
&\chi^{\phi}_{\epsilon\zeta,\gamma\delta}(z_1,z_2; z_4,z_3) = -i\gp_{\epsilon\gamma}(z_1,z_4)\gp_{\delta\zeta}(z_3,z_2) \notag\\
&\quad -\gp_{\epsilon\bar{\alpha}}(z_1,\bar{z}_3)\Gamma^{\phi}_{\bar{\alpha}\bar{\eta},\bar{\theta}\bar{\omega}}(\bar{z}_3,\bar{z}_5;\bar{z}_6,\bar{z}_7)\chi^{\phi}_{\bar{\theta}\bar{\omega},\gamma\delta}(\bar{z}_6,\bar{z}_7;z_4,z_3)\notag\\
&\quad \times\gp_{\bar{\eta}\zeta}(\bar{z}_5,z_2).
\end{align}
We then finally note that Eqs.~\eqref{eq:second_der_phi_partition_function} and \eqref{eq:bethe_Salpeter_equation} can be combined to give
\begin{align}
\label{eq:combination_bethe_Salpeter_two_particle_corr}
&i\langle\tc\cc^{\dagger}_{\gamma}(z_4)\cc_{\delta}(z_3)\cc_{\epsilon}(z_1)\cc^{\dagger}_{\zeta}(z_2) \rangle_{\phi}\notag\\
& = i\gp_{\delta\gamma}(z_3,z_4)\gp_{\epsilon\zeta}(z_1,z_2)-i\gp_{\epsilon\gamma}(z_1,z_4)\gp_{\delta\zeta}(z_3,z_2) \notag\\
&\quad -\gp_{\epsilon\bar{\alpha}}(z_1,\bar{z}_3)\Gamma^{\phi}_{\bar{\alpha}\bar{\eta},\bar{\theta}\bar{\omega}}(\bar{z}_3,\bar{z}_5;\bar{z}_6,\bar{z}_7)\chi^{\phi}_{\bar{\theta}\bar{\omega},\gamma\delta}(\bar{z}_6,\bar{z}_7;z_4,z_3)\notag\\
&\quad \times\gp_{\bar{\eta}\zeta}(\bar{z}_5,z_2).
\end{align}
Equation~\eqref{eq:combination_bethe_Salpeter_two_particle_corr} allows us to determine the equations of motion of the Hubbard model~\eqref{eq:Hubbard_model_intro} and to calculate the TPSC self-energy.

%%%%%%%%%%%%%%%%%%%%%%%%%%%%%%%%%%%% Equations of motion %%%%%%%%%%%%%%%%%%%%%%%%%%%%%%%%%%%%%%%%%%

\subsubsection{Equations of motion}
\label{ch:eq_of_motion}
To properly account for the different degrees of freedom defining the Hubbard model, the Greek indices in Eq.~\eqref{eq:contour_G_phi} will be replaced by tuples of lattice sites represented by Latin letters and spin represented by $\sigma$. %From now on, the different degrees of freedom are split off by a comma. 
To obtain the equations of motion, we differentiate the contour one-body Green's function \eqref{eq:contour_G_phi}:

\begin{align}
\label{eq:differentiation_greens_function_for_equation_motion}
&i\partial_{z_1}\mathcal{G}^{\phi}_{lm,\sigma}(z_1,z_2) = \partial_{z_1}\langle\tc \cc_{l,\sigma}(z_1)\cc_{m,\sigma}^{\dagger}(z_2)\rangle_{\phi}\notag\\
&= \delta^{\mathcal{C}}(z_1,z_2)\langle\{\cc_{l,\sigma},\cc_{m,\sigma}^{\dagger}\} \rangle_{\phi} + \left<\tc\partial_{z_1}S[\phi]\cc_{l,\sigma}(z_1)\cc_{m,\sigma}^{\dagger}(z_2)\right>_{\phi}\notag\\
&\hspace{4mm}+i\big\langle \tc [\htcal{H},\cc_{l,\sigma}](z_1)\cc_{m,\sigma}^{\dagger}(z_2)\big\rangle_{\phi},
\end{align} where the chemical potential term is absorbed into the Hamiltonian $\htcal{H}\to \htcal{H}-\mu\hat{N}$ since we work in the grand-canonical ensemble.~\footnote{$\hat{N}$ is the total number operator and $\mu$ is the chemical potential.} The first term of the development \eqref{eq:differentiation_greens_function_for_equation_motion} yields the identity matrix. The second term has to be dealt with carefully because the differentiation involves the source field $S[\phi]$:

\begin{align}
\label{eq:eq_motion_source_field_differentiation}
&\left<\tc\partial_{z_1}S[\phi]\cc_{l,\sigma}(z_1)\cc^{\dagger}_{m,\sigma}(z_2)\right>_{\phi}=\notag\\
&=i\phi_{\bar{a}\bar{b},\bar{\sigma}^{\prime}\bar{\sigma}^{\prime\prime}}(z_1,\bar{z}_4)\notag\\
&\hspace{0.4cm}\times\left<\tc\left[\cc^{\dagger}_{\bar{a},\bar{\sigma}^{\prime}}(z_1)\cc_{\bar{b},\bar{\sigma}^{\prime\prime}}(\bar{z}_4),\cc_{l,\sigma}(z_1)\right]\cc_{m,\sigma}^{\dagger}(z_2)\right>_{\phi}\notag\\
&=\phi_{l\bar{b},\sigma\bar{\sigma}^{\prime\prime}}(z_1,\bar{z}_4)\gp_{\bar{b}m,\bar{\sigma}^{\prime\prime}\sigma}(\bar{z}_4,z_2).
\end{align} Here, we used the fact that, in the exponential representing the time-evolution operators, $\partial_x\int^{x^{\prime}}_{x}\mathrm{d}x^{\prime\prime}f^{\prime}(x^{\prime\prime}) = -f^{\prime}(x)$, and also the relation $[AB,C]=A\{B,C\}-\{A,C\}B=A[B,C]+[A,C]B$. The annihilation operator in the exponential anticommutes with $\cc_{l,\sigma}(z_1)$ which is taken care of by the contour-ordering operator. There is no global sign associated with shifting around $S[\phi]$ within the thermal average, since its arguments consist of an even number of annihilation and creation operators.

Finally, after evaluating the commutator in Eq.~\eqref{eq:differentiation_greens_function_for_equation_motion} (last term) using the Hamiltonian~\eqref{eq:Hubbard_model_intro}, the equations of motion become
\begin{align}
\label{eq:eq_motion_including_cummutator_developed}
&i\partial_{z_1}\gp_{lm,\sigma}(z_1,z_2) + t_{l\bar{b}}^{\text{hop}}(z_1)\gp_{\bar{b}m,\sigma}(z_1,z_2)\notag\\
&-\phi_{l\bar{b},\sigma\bar{\sigma}^{\prime\prime}}(z_1,\bar{z}_4)\gp_{\bar{b}m,\bar{\sigma}^{\prime\prime}\sigma}(\bar{z}_4,z_2) = \delta^{\mathcal{C}}(z_1,z_2)\delta_{lm} \notag\\
&- iU(z_1)\left<\tc\hat{n}_{l,-\sigma}(z_1)\cc_{l,\sigma}(z_1)\cc^{\dagger}_{m,\sigma}(z_2)\right>_{\phi}.
\end{align}
Note that the adjoint can be obtained in a similar fashion by acting from the right with the complex conjugate operator $-i\overleftarrow{\partial}_{z_2}$ on the single-particle Green's function. In Eq.~\eqref{eq:eq_motion_including_cummutator_developed} one can recognize the modified Dyson's equation \eqref{eq:modified_Dyson_equation}. Indeed, we have

\begin{align*}
%\label{eq:deducing_Dyson_from_eqs_motion}
&\left[{\mathcal{G}^0_{l\bar{b},\sigma\sigma^{\prime\prime}}(z_1,\bar{z}_2)}^{-1}-\phi_{l\bar{b},\sigma\bar{\sigma}^{\prime\prime}}(z_1,\bar{z}_2)\right]\gp_{\bar{b}m,\bar{\sigma}^{\prime\prime}\sigma}(\bar{z}_2,z_2)\notag\\
&=\delta^{\mathcal{C}}(z_1,z_2)\delta_{lm}+\Sigma^{\phi}_{l\bar{b},\sigma\bar{\sigma}^{\prime\prime}}(z_1,\bar{z}_2)\gp_{\bar{b}m,\bar{\sigma}^{\prime\prime}\sigma}(\bar{z}_2,z_2),
\end{align*}
such that the four-point correlation function is related to the self-energy and Green's function via

\begin{align}
\label{eq:four_point_self_G}
&\Sigma^{\phi}_{l\bar{b},\sigma\bar{\sigma}^{\prime\prime}}(z_1,\bar{z}_2)\gp_{\bar{b}m,\bar{\sigma}^{\prime\prime}\sigma}(\bar{z}_2,z_2)\notag\\
&\hspace{0.7cm}=-iU(z_1)\left<\tc\hat{n}_{l,-\sigma}(z_1)\cc_{l,\sigma}(z_1)\cc^{\dagger}_{m,\sigma}(z_2)\right>_{\phi}.
\end{align}
Equation~\eqref{eq:four_point_self_G} provides an expression for the self-energy of the model Hamiltonian we are interested in. Once the desired correlation functions have been generated, the physical results are obtained by setting the source field $\phi$ to zero. We will show below that the very same four-point correlation function can be calculated in both the longitudinal and transversal channels, \textit{i.e.} by using a source field to derive Eqs.~\eqref{eq:second_der_phi_partition_function} and \eqref{eq:bethe_Salpeter_equation} which does not induce a spin-flip ($\phi_{\sigma,\sigma}$) and one inducing a spin-flip ($\phi_{\sigma,-\sigma}$), respectively. The two expressions of the self-energy will then be averaged to restore the crossing symmetry, giving the self-energy approximation of the theory $\Sigma^{\text{TPSC},(1)}$.

%%%%%%%%%%%%%%%%%%%%%%%%%%%%%%%%%%%% Longitudinal self-energy %%%%%%%%%%%%%%%%%%%%%%%%%%%%%%%%%%%%%%%%%%

\subsubsection{Longitudinal expression of the self-energy}
\label{ch:longitudinal_self}
To get the second-level longitudinal self-energy, we need to use Eq.~\eqref{eq:combination_bethe_Salpeter_two_particle_corr} and perform the following substitutions for the indices: $\gamma\to (l,-\sigma)$, $\delta\to (l,-\sigma)$, $\epsilon\to (l,\sigma)$ and $\zeta\to (m,\sigma)$. At the same time, for the contour-time variables, we have to make the following substitutions: $z_4\to z_1^{++}$, $z_3\to z_1^+$, $z_2\to z_2$ and $z_1\to z_1$. Then, inserting the resulting four-point correlation function into Eq.~\eqref{eq:four_point_self_G}, we end up with the relation 
\begin{align}
\label{eq:self_longitudinal_derivation}
&\Sigma^{\phi,\text{long.}}_{l\bar{b},\sigma\bar{\sigma}^{\prime\prime}}(z_1,\bar{z}_2)\mathcal{G}^{\phi}_{\bar{b}m,\bar{\sigma}^{\prime\prime}\sigma}(\bar{z}_2,z_2)\notag\\
&=-iU(z_1)\gp_{ll,-\sigma-\sigma}(z_1^+,z_1^{++})\gp_{lm,\sigma\sigma}(z_1,z_2)\notag\\
&+ iU(z_1)\gp_{ll,\sigma-\sigma}(z_1,z_1^{++})\gp_{lm,-\sigma\sigma}(z_1^+,z_2)\notag\\
&+U(z_1)\gp_{(l,\sigma),\bar{\alpha}}(z_1,\bar{z}_3)\Gamma^{\phi}_{\bar{\alpha}\bar{\eta}\bar{\theta}\bar{\omega}}(\bar{z}_3,\bar{z}_5;\bar{z}_6,\bar{z}_7)\notag\\
&\quad\times\chi^{\phi}_{\bar{\theta}\bar{\omega}(l,-\sigma)(l,-\sigma)}(\bar{z}_6,\bar{z}_7;z_1^{++},z_1^+)\gp_{\bar{\eta}(m,\sigma)}(\bar{z}_5,z_2),
\end{align} 
where $z^{++}$ is placed infinitesimally later than $z^{+}$ along $\mathcal{C}$. The second term of Eq.~\eqref{eq:self_longitudinal_derivation} vanishes for the Hubbard model when the source field is spin diagonal (longitudinal channel), namely $\mathcal{G}_{\sigma-\sigma}=0$. The longitudinal component to the self-energy can then be straightforwardly isolated by multiplying by $\mathcal{G}_{\sigma}^{-1}$ from the right:
\begin{align}
\label{eq:self_longitudinal_isolated_from_derivation}
&\Sigma^{\phi,\text{long.}}_{lm,\sigma}(z_1,z_2) = -iU(z_1)\gp_{l,-\sigma}(z_1^{+},z_1^{++})\delta^{\mathcal{C}}(z_1,z_2)\delta_{l,m}+\notag\\
&U(z_1)\gp_{(l,\sigma)\bar{\alpha}}(z_1,\bar{z}_3)\Gamma^{\phi}_{\bar{\alpha}(m,\sigma)\bar{\theta}\bar{\omega}}(\bar{z}_3,z_2;\bar{z}_6,\bar{z}_7)\notag\\
&\hspace{0.2cm}\times\chi^{\phi}_{\bar{\theta}\bar{\omega}(l,-\sigma)}(\bar{z}_6,\bar{z}_7;z_1).
\end{align} In Eq.~\eqref{eq:self_longitudinal_isolated_from_derivation}, for the sake of conciseness, we have used an unambiguous notation compressing tuples of repeated indices denoting the same degree of freedom, \textit{i.e.} $\chi_{jsll,\sigma\sigma-\sigma-\sigma}(z_6,z_7;z_1^{++},z_1^+)\to\chi_{jsl,\sigma-\sigma}(z_6,z_7;z_1)$. Furthermore, by expanding the implicitly summed quantities in Eq.~\eqref{eq:self_longitudinal_isolated_from_derivation}, we obtain
\begin{align}
\label{eq:self_longitudinal_isolated_from_derivation_spin_choice}
&\Sigma^{\text{long.}}_{lm,\sigma}(z_1,z_2) = -iU(z_1)\mathcal{G}_{l,-\sigma}(z_1,z_1^+)\delta^{\mathcal{C}}(z_1,z_2)\delta_{l,m}+\notag\\
&U(z_1)\mathcal{G}_{l\bar{i},\sigma}(z_1,\bar{z}_3)\biggl[\Gamma_{\bar{i}m\bar{j}\bar{s},\sigma\sigma}(\bar{z}_3,z_2;\bar{z}_6,\bar{z}_7)\chi_{\bar{j}\bar{s}l,\sigma-\sigma}(\bar{z}_6,\bar{z}_7;z_1)\notag\\
&+\Gamma_{\bar{i}m\bar{j}\bar{s},\sigma-\sigma}(\bar{z}_3,z_2;\bar{z}_6,\bar{z}_7)\chi_{\bar{j}\bar{s}l,-\sigma-\sigma}(\bar{z}_6,\bar{z}_7;z_1)\biggr].
\end{align} Let us now define two susceptibilities, namely the charge $\chi^{\text{ch}}$ and spin $\chi^{\text{sp}}$ susceptibilities. We will use two corresponding G-skeletonic irreducible vertices, \textit{i.e.} the charge $\Gamma^{\text{ch}}$ and spin $\Gamma^{\text{sp}}$ vertices. Using spin-rotational symmetry, the spin and charge susceptibilities are defined as
\begin{align}
\label{eq:charge_spin_susceptibility_TPSC_definition}
&\chi_{ij}^{\text{ch}/\text{sp}}(z_1,z_1^+;z_2^+,z_2)\notag\\
&=-2i\left(\frac{\delta\gp_{i,\uparrow}(z_1,z_1^+)}{\delta\phi_{j,\uparrow}(z_2^+,z_2)} \pm \frac{\delta\gp_{i,\uparrow}(z_1,z_1^+)}{\delta\phi_{j,\downarrow}(z_2^+,z_2)}\right)\bigg\rvert_{\phi\to 0}.
\end{align}
The factor of $2$ comes from tracing over the spin degrees of freedom and the upper (lower) sign corresponds to the charge (spin) susceptibility. We then expand the functional derivatives using Eq.~\eqref{eq:closure_relation_expanded_2}:

\begin{align}
\label{eq:charge_spin_susceptibility_TPSC_definition_developed}
&\chi_{ij}^{\text{ch}/\text{sp}}(z_1;z_2)=-2i\mathcal{G}_{ij,\uparrow}(z_1,z_2^+)\mathcal{G}_{ji,\uparrow}(z_2,z_1^+) - 2\mathcal{G}_{i\bar{l},\uparrow}(z_1,\bar{z}_3)\notag\\
&\times\bigl[\Gamma_{\bar{l}\bar{m}\bar{n}\bar{s},\uparrow\bar{\sigma}^{\prime}\bar{\sigma}^{\prime\prime}}(\bar{z}_3,\bar{z}_5;\bar{z}_6,\bar{z}_7)\chi_{\bar{n}\bar{s}j,\bar{\sigma}^{\prime}\bar{\sigma}^{\prime\prime}\uparrow}(\bar{z}_6,\bar{z}_7;z_2)\notag\\
&\pm\Gamma_{\bar{l}\bar{m}\bar{n}\bar{s},\uparrow\bar{\sigma}^{\prime}\bar{\sigma}^{\prime\prime}}(\bar{z}_3,\bar{z}_5;\bar{z}_6,\bar{z}_7)\chi_{\bar{n}\bar{s}j,\bar{\sigma}^{\prime}\bar{\sigma}^{\prime\prime}\downarrow}(\bar{z}_6,\bar{z}_7;z_2)\bigr]\notag\\
&\hspace{0.0cm}\times\mathcal{G}_{\bar{m}i,\uparrow}(\bar{z}_5,z_1^+).
\end{align}
The summed over spin indices $\sigma^{\prime}$ and $\sigma^{\prime\prime}$ must take the same value in order to lead to a nonzero result,\footnote{The Hubbard model conserves spin as well, not just particle number.} \textit{i.e.} $\chi_{\sigma^{\prime}\sigma^{\prime\prime}\sigma}=0 \ \forall \ \sigma^{\prime}\neq \sigma^{\prime\prime}$; this allows us to conveniently collapse those two spin labels into one. In Eq.~\eqref{eq:charge_spin_susceptibility_TPSC_definition}, only the functional derivative with all the same spin projections gives a non-zero Hartree term, hence we get only one bubble term in Eq.~\eqref{eq:charge_spin_susceptibility_TPSC_definition_developed}. Using that $\Gamma^{\text{ch}/\text{sp}}\equiv \Gamma_{\uparrow\downarrow}\pm\Gamma_{\uparrow\uparrow}$ and $\chi^0\equiv -2i\mathcal{G}_{\sigma}\mathcal{G}_{\sigma}$, the spin and charge susceptibilities in the paramagnetic state read

\begin{align}
\label{eq:charge_spin_susceptibility_TPSC_definition_developed_2}
&\chi_{ij}^{\text{ch}/\text{sp}}(z_1;z_2)\notag\\
&=-2i\mathcal{G}_{ij,\uparrow}(z_1,z_2^+)\mathcal{G}_{ji,\uparrow}(z_2,z_1^+) \mp 2\mathcal{G}_{i\bar{l},\uparrow}(z_1,\bar{z}_3)\notag\\
&\times\bigl[\pm\Gamma_{\bar{l}\bar{m}\bar{n}\bar{s},\uparrow\uparrow}(\bar{z}_3,\bar{z}_5;\bar{z}_6,\bar{z}_7)+\Gamma_{\bar{l}\bar{m}\bar{n}\bar{s},\uparrow\downarrow}(\bar{z}_3,\bar{z}_5;\bar{z}_6,\bar{z}_7)\bigr]\notag\\
&\times\bigl[\chi_{\bar{n}\bar{s}j,\uparrow\uparrow}(\bar{z}_6,\bar{z}_7;z_2)\pm\chi_{\bar{n}\bar{s}j,\downarrow\uparrow}(\bar{z}_6,\bar{z}_7;z_2)\bigr]\mathcal{G}_{\bar{m}i,\uparrow}(\bar{z}_5,z_1^+)\notag\\
&=\chi^0_{ij}(z_1,z_2) \mp\frac{i}{2}\chi^0_{i\bar{l}\bar{m}}(z_1;\bar{z}_3,\bar{z}_5)\Gamma_{\bar{l}\bar{m}\bar{n}\bar{s}}^{\text{ch}/\text{sp}}(\bar{z}_3,\bar{z}_5;\bar{z}_6,\bar{z}_7)\notag\\
&\hspace{0.4cm}\times\chi_{\bar{n}\bar{s}j}^{\text{ch}/\text{sp}}(\bar{z}_6,\bar{z}_7;z_2).
\end{align}
In Eq.~\eqref{eq:charge_spin_susceptibility_TPSC_definition_developed_2}, the spin rotational invariance allowed us to factorize $\chi$ and $\Gamma$ into their spin and charge components. Now, if we write out $\Gamma_{\text{ch}}\chi_{\text{ch}}+\Gamma_{\text{sp}}\chi_{\text{sp}}$, we symbolically get
\begin{align}
\label{eq:gamma_chi_combination_equivalence}
&\Gamma_{\text{ch}}\chi_{\text{ch}} + \Gamma_{\text{sp}}\chi_{\text{sp}} = 2\left[\Gamma_{\uparrow\uparrow}+\Gamma_{\uparrow\downarrow}\right]\left[\chi_{\uparrow\uparrow}+\chi_{\uparrow\downarrow}\right]\notag\\
&+2\left[\Gamma_{\uparrow\downarrow}-\Gamma_{\uparrow\uparrow}\right]\left[\chi_{\uparrow\uparrow}-\chi_{\uparrow\downarrow}\right]\notag\\
&=4\Gamma_{\uparrow\uparrow}\chi_{\uparrow\downarrow} + 4\Gamma_{\uparrow\downarrow}\chi_{\uparrow\uparrow}.
\end{align} 
The result \eqref{eq:gamma_chi_combination_equivalence} can be substituted into the longitudinal expression for the self-energy \eqref{eq:self_longitudinal_isolated_from_derivation_spin_choice}. Doing so, the physical longitudinal self-energy can be expressed as ($\phi\to 0$)
\begin{align}
\label{eq:longitudinal_self_energy_reexpressed}
&\Sigma^{\text{long.}}_{lm,\uparrow}(z_1,z_2) = U(z_1)n_{l,\downarrow}(z_1,z_1^+)\delta^{\mathcal{C}}(z_1,z_2)\delta_{l,m}\notag\\
&+\frac{U(z_1)}{4}\mathcal{G}_{l\bar{i},\uparrow}(z_1,\bar{z}_3)\biggl[\Gamma^{\text{ch}}_{\bar{i}m\bar{j}\bar{s}}(\bar{z}_3,z_2;\bar{z}_6,\bar{z}_7)\chi^{\text{ch}}_{\bar{j}\bar{s}l}(\bar{z}_6,\bar{z}_7;z_1)\notag\\
&+\Gamma^{\text{sp}}_{\bar{i}m\bar{j}\bar{s}}(\bar{z}_3,z_2;\bar{z}_6,\bar{z}_7)\chi^{\text{sp}}_{\bar{j}\bar{s}l}(\bar{z}_6,\bar{z}_7;z_1)\biggr].
\end{align} 
If we replace the irreducible vertices in Eq.~\eqref{eq:longitudinal_self_energy_reexpressed} with local ones, namely 
\begin{align}
\label{eq:irr_vertex_expression}
&\Gamma^{{\text{ch}/\text{sp}}}_{imjs}(z_3,z_2;z_6,z_7)\to \Gamma_m^{{\text{ch}/\text{sp}}}(z_2)\delta^{\mathcal{C}}(z_2,z_6)\notag\\
&\quad\times\delta^{\mathcal{C}}(z_2^+,z_7)\delta^{\mathcal{C}}(z_2,z_3)\delta_{m,j}\delta_{m,s}\delta_{m,i},
\end{align} 
we get \cite{Bergeron_2011_optical_cond,Vilk_1997}
\begin{align}
\label{eq:self_energy_approx_TPSC_locality_logitudinal}
&\Sigma^{\text{long.}}_{lm,\sigma}(z_1,z_2)\notag\\
&= U(z_1)n_{l,-\sigma}(z_1)\delta^{\mathcal{C}}(z_1,z_2)\delta_{l,m}+\frac{U(z_1)}{4}\mathcal{G}_{lm,\sigma}(z_1,z_2)\notag\\
&\quad\times\left[\Gamma_m^{\text{ch}}(z_2)\chi^{\text{ch}}_{ml}(z_2,z_1) + \Gamma_m^{\text{sp}}(z_2)\chi^{\text{sp}}_{ml}(z_2,z_1)\right].
\end{align}

%%%%%%%%%%%%%%%%%%%%%%%%%%%%%%%%%%%% Transversal self-energy %%%%%%%%%%%%%%%%%%%%%%%%%%%%%%%%%%%%%%%%%%

\subsubsection{Transversal expression of the self-energy}
\label{ch:transversal_self}
The four-point correlation function appearing in Eq.~\eqref{eq:four_point_self_G} can also be obtained by employing a transversal field.~\cite{senechal_bourbonnais_tremblay_2004} To see that, we return to Eq.~\eqref{eq:combination_bethe_Salpeter_two_particle_corr} expressing the four-point correlation function in terms of the self-energy and Green's function. We first notice that, in a transverse field, when we work out Eq.~\eqref{eq:second_der_phi_partition_function} using an off-diagonal source field $\phi_{\sigma-\sigma}$ in spin, we have 
\begin{align}
\label{eq:second_der_phi_partition_function_transversal_field}
&\chi^{\phi,\sigma-\sigma\sigma-\sigma}_{abdc}(z_1,z_2;z_4,z_3)\notag\\
&= i\langle\tc\cc^{\dagger}_{d,\sigma}(z_4)\cc_{c,-\sigma}(z_3)\cc_{a,\sigma}(z_1)\cc^{\dagger}_{b,-\sigma}(z_2) \rangle_{\phi}\notag\\
&\quad - i\gp_{cd,-\sigma\sigma}(z_3,z_4)\gp_{ab,\sigma-\sigma}(z_1,z_2),
\end{align}
where we have rendered the notation more compact by turning the spin subscripts into superscripts. Furthermore, to get Eq.~\eqref{eq:second_der_phi_partition_function_transversal_field}, we performed the following substitutions in Eq.~\eqref{eq:second_der_phi_partition_function}: $\epsilon\to(a,\sigma)$, $\zeta\to(b,-\sigma)$, $\gamma\to(d,\sigma)$ and $\delta\to(c,-\sigma)$. In the transversal particle-hole channel, another expression of the form $\chi^{\phi,-\sigma\sigma\sigma-\sigma}_{abdc}(z_1,z_2;z_4,z_3)$ is produced, but since it includes a four-point correlation function of the form $i\langle\tc\cc^{\dagger}_{d,\sigma}(z_4)\cc_{c,-\sigma}(z_3)\cc_{a,-\sigma}(z_1)\cc^{\dagger}_{b,\sigma}(z_2) \rangle_{\phi}$, it is equal to $0$ in the Hubbard model due to spin conservation. To match the four-point correlation function appearing in Eq.~\eqref{eq:four_point_self_G}, we need to perform at last the variable substitutions $(a,z_1)\to (l,z_1^+)$, $(b,z_2)\to (m,z_2)$, $(c,z_3)\to (l,z_1)$ and $(d,z_4)\to (l,z_1^{++})$. Doing so, the last term of Eq.~\eqref{eq:second_der_phi_partition_function_transversal_field} vanishes when the source field is turned off. Making the same variable substitutions in Eq.~\eqref{eq:bethe_Salpeter_equation} as done hitherto in Eq.~\eqref{eq:second_der_phi_partition_function}, we obtain
\begin{align}
\label{eq:bethe_Salpeter_equation_transversal_field}
&\chi^{\phi,\sigma-\sigma\sigma-\sigma}_{lml}(z_1^+,z_2; z_1^{++},z_1) = -i\gp_{l,\sigma}(z_1^+,z_1^{++})\gp_{lm,-\sigma}(z_1,z_2) \notag\\
&-\gp_{l\bar{a},\sigma}(z_1^+,\bar{z}_3)\Gamma^{\phi,\sigma-\sigma\bar{\sigma}^{\prime}\bar{\sigma}^{\prime\prime}}_{\bar{a}\bar{b}\bar{c}\bar{d}}(\bar{z}_3,\bar{z}_5;\bar{z}_6,\bar{z}_7)\notag\\
&\hspace{0.7cm}\times\chi^{\phi,\bar{\sigma}^{\prime}\bar{\sigma}^{\prime\prime}\sigma-\sigma}_{\bar{c}\bar{d}l}(\bar{z}_6,\bar{z}_7;z_1^{++},z_1)\gp_{\bar{b}m,-\sigma}(\bar{z}_5,z_2).
\end{align}
In Eq.~\eqref{eq:bethe_Salpeter_equation_transversal_field}, we used the spin selection rule forbidding antiparallel spins in Green's functions once $\phi\to0$. Next, we insert the result \eqref{eq:bethe_Salpeter_equation_transversal_field} into Eq.~\eqref{eq:second_der_phi_partition_function_transversal_field} to isolate the four-point correlation function and then multiply by $U(z_1)$ to recover something similar to Eq.~\eqref{eq:combination_bethe_Salpeter_two_particle_corr}, but now for the transversal channel. This yields an expression for the TPSC self-energy in the transversal channel
\begin{align}
\label{eq:self_transversal_derivation}
&\Sigma^{\phi,\text{trans.}}_{l\bar{b},-\sigma\bar{\sigma}^{\prime}}(z_1,\bar{z}_2)\mathcal{G}^{\phi}_{\bar{b}m,\bar{\sigma}^{\prime}-\sigma}(\bar{z}_2,z_2)\notag\\
&=iU(z_1)\gp_{l,-\sigma\sigma}(z_1,z_1^+)\gp_{lm,\sigma-\sigma}(z_1,z_2)\notag\\
&- iU(z_1)\gp_{l,\sigma}(z_1,z_1^{+})\gp_{lm,-\sigma}(z_1,z_2) \notag\\
&-U(z_1)\gp_{l\bar{a},\sigma}(z_1,\bar{z}_3)\Gamma^{\phi,\sigma-\sigma\bar{\sigma}^{\prime}\bar{\sigma}^{\prime\prime}}_{\bar{a}\bar{b}\bar{c}\bar{d}}(\bar{z}_3,\bar{z}_5;\bar{z}_6,\bar{z}_7)\notag\\
&\hspace{1.0cm}\times\chi^{\phi,\bar{\sigma}^{\prime}\bar{\sigma}^{\prime\prime}\sigma-\sigma}_{\bar{c}\bar{d}l}(\bar{z}_6,\bar{z}_7;z_1)\gp_{\bar{b}m,-\sigma}(\bar{z}_5,z_2).
\end{align} From Eq.~\eqref{eq:self_transversal_derivation}, the physical transversal component to the second-level TPSC self-energy reads
\begin{align}
\label{eq:self_transversal_isolated_from_derivation}
&\Sigma^{\text{trans.}}_{lm,-\sigma}(z_1,z_2) = U(z_1)n_{l,\sigma}(z_1)\delta^{\mathcal{C}}(z_1,z_2)\delta_{l,m}\notag\\
&-U(z_1)\mathcal{G}_{l\bar{a},\sigma}(z_1,\bar{z}_3)\Gamma^{\sigma-\sigma\sigma-\sigma}_{\bar{a}m\bar{c}\bar{d}}(\bar{z}_3,z_2;\bar{z}_6,\bar{z}_7)\notag\\
&\hspace{0.7cm}\times\chi^{\sigma-\sigma\sigma-\sigma}_{\bar{c}\bar{d}l}(\bar{z}_6,\bar{z}_7;z_1),
\end{align}
since $\chi^{\sigma\sigma-\sigma\sigma}=\chi^{\sigma-\sigma-\sigma\sigma}=0$. 

It is now time to have a closer look at the different components making up Eq.~\eqref{eq:self_transversal_isolated_from_derivation}, namely $\chi$ and $\Gamma$. To start with, we assume that the vertex appearing in Eq.~\eqref{eq:self_transversal_isolated_from_derivation} is fully local, as done in Sec.~\ref{ch:longitudinal_self} for the longitudinal component,
\begin{align}
\label{eq:off_diagonal_self_energy_functional_derivative}
&\Gamma^{\sigma-\sigma\sigma-\sigma}_{amcd}(z_3,z_2;z_6,z_7) = \Gamma^{\sigma-\sigma\sigma-\sigma}_{m}(z_2)\delta^{\mathcal{C}}(z_2,z_3)\notag\\
&\hspace{0.5cm}\times\delta^{\mathcal{C}}(z_2,z_6)\delta^{\mathcal{C}}(z_2^+,z_7)\delta_{l,m}\delta_{l,i}\delta_{l,j}.
\end{align} Next, we work out an expression for $\chi^{\sigma-\sigma\sigma-\sigma}$, using Eq.~\eqref{eq:second_der_phi_partition_function},
\begin{align}
\label{eq:chi_transverse_4_point_corr_expression}
&\chi^{\sigma-\sigma\sigma-\sigma}_{cdl}(z_6,z_7;z_1^{+},z_1)\notag\\
&= -i\left<\tc \cc^{\dagger}_{l,\sigma}(z_1^{+})\cc_{l,-\sigma}(z_1)\cc_{d,-\sigma}^{\dagger}(z_7)\cc_{c,\sigma}(z_6)\right>.
\end{align} Since it follows from Eq.~\eqref{eq:off_diagonal_self_energy_functional_derivative} that $z_7\to z_6^+$ and $d\to c$ in Eq.~\eqref{eq:chi_transverse_4_point_corr_expression}, we obtain
\begin{align}
\label{eq:spin_create_spin_annihilate_op}
&\chi^{\sigma-\sigma\sigma-\sigma}_{cl}(z_6,z_6^+;z_1^{+},z_1)= -i\langle\tc \hat{S}_{c,+}(z_6)\hat{S}_{l,-}(z_1)\rangle\notag\\
&= \chi_{cl;+-}(z_6,z_1),
\end{align} 
where $\hat{S}_{c,+/-}\equiv \frac{1}{2}\left(\hat{S}_{c,x}\pm i\hat{S}_{c,y}\right)$, such that Eq.~\eqref{eq:spin_create_spin_annihilate_op} can be expressed as
\begin{align}
\label{eq:spin_transverse_susceptibility_indentities}
&\chi_{cl;+-}(z_6,z_1)=-\frac{i}{4}\langle\tc\hat{S}_{c,x}(z_6)\hat{S}_{l,x}(z_1)\rangle\notag\\
&-\frac{i}{4}\langle\tc\hat{S}_{c,y}(z_6)\hat{S}_{l,y}(z_1)\rangle=-\frac{i}{2}\langle\tc\hat{S}_{c,z}(z_6)\hat{S}_{l,z}(z_1)\rangle. 
\end{align} Hence, from Eqs.~\eqref{eq:spin_transverse_susceptibility_indentities} and \eqref{eq:off_diagonal_self_energy_functional_derivative}, the transversal component \eqref{eq:self_transversal_isolated_from_derivation} becomes~\cite{senechal_bourbonnais_tremblay_2004}
\begin{align}
\label{eq:self_energy_approx_TPSC_locality_transversal}
&\Sigma^{\text{trans.}}_{lm,\sigma}(z_1,z_2) = U(z_1)n_{l,-\sigma}(z_1)\delta^{\mathcal{C}}(z_1,z_2)\delta_{l,m}\notag\\
&-\frac{U(z_1)}{2}\mathcal{G}_{lm,-\sigma}(z_1,z_2)\Gamma^{\sigma-\sigma\sigma-\sigma}_{m}(z_2)\chi^{\text{sp}}_{ml}(z_2;z_1).
\end{align}
The spin off-diagonal irreducible vertex $\Gamma^{\sigma-\sigma\sigma-\sigma}$ showing up in Eq.~\eqref{eq:self_energy_approx_TPSC_locality_transversal} will be specified in Sec.~\ref{ch:TPSC_ansatz} using the first-level TPSC approximations.

%%%%%%%%%%%%%%%%%%%%%%%%%%%%%%%%%%%% TPSC ansatz %%%%%%%%%%%%%%%%%%%%%%%%%%%%%%%%%%%%%%%%%%

\subsubsection{TPSC ansatz}
\label{ch:TPSC_ansatz}
To calculate the single- and two-particle correlation functions, TPSC employs an ansatz for the Luttinger-Ward functional $\Phi$ that approximates the local irreducible vertices in the particle-hole channel (transversal and longitudinal with respect to some generating field), namely the charge $\Gamma_{\text{ch}}$ and spin $\Gamma_{\text{sp}}$. The starting point is the following Luttinger-Ward functional~\cite{Vilk_1997}
\begin{align}
\label{eq:TPSC_Luttinger_Ward_functional}
&\Phi[\mathcal{G}] = \frac{1}{2}\int_{\mathcal{C}}\mathrm{d}z \ \sum_{\sigma}\mathcal{G}_{\sigma}(z,z^+)\Gamma_{\sigma\sigma}(z)\mathcal{G}_{\sigma}(z,z^+)\notag\\
&+\frac{1}{2}\int_{\mathcal{C}}\mathrm{d}z \ \sum_{\sigma}\mathcal{G}_{\sigma}(z,z^+)\Gamma_{\sigma-\sigma}(z)\mathcal{G}_{-\sigma}(z,z^+),
\end{align} where the quantities are defined on the Kadanoff-Baym contour, with arguments $z\in \mathcal{C}$. The integral can be decomposed into contour components according to the Langreth rules. From Eq.~\eqref{eq:TPSC_Luttinger_Ward_functional}, both the self-energy and the G-skeletonic irreducible vertices can be obtained. The first-level TPSC self-energy $\Sigma^{(0)}$ reads 
\begin{align}
\label{eq:TPSC_self_energy}
\Sigma^{(0)}_{\sigma}(z_2,z_3) = \frac{\delta\Phi[\mathcal{G}]}{\delta\mathcal{G}_{\sigma}(z_3,z_2)},
\end{align} 
which yields
\begin{align}
\label{eq:TPSC_self_energy_expanded}
\Sigma^{(0)}_{\sigma}(z_2,z_3) =& \,\, \Gamma_{\sigma\sigma}(z_3)\mathcal{G}_{\sigma}(z_3,z_3^+)\delta^{\mathcal{C}}(z_3^+,z_2)\notag\\
&+ \Gamma_{\sigma-\sigma}(z_3)\mathcal{G}_{-\sigma}(z_3,z_3^+)\delta^{\mathcal{C}}(z_3^+,z_2),
\end{align}
where the rotational spin symmetry $\Gamma_{\sigma-\sigma} = \Gamma_{-\sigma\sigma}$ was used. Since the $\Gamma$ factors are scalars, the first-level self-energy \eqref{eq:TPSC_self_energy_expanded} can be absorbed into a shift of the chemical potential $\mu_0$ when defining the lattice Green's function at the first level of approximation:
\begin{align}
\label{eq:first_level_approx_G}
&\left(i\partial_z + \mu_0 -  \Sigma_{\sigma}^{(0)}(z)\delta^{\mathcal{C}}(z,z^{\prime}) - \epsilon(\mathbf{k},z)\right)\mathcal{G}^{(0)}_{\mathbf{k},\sigma}(z,z^{\prime})\notag\\
&= \delta^{\mathcal{C}}(z,z^{\prime}).
\end{align} In essence, the Green's function at the first level of approximation 
is noninteracting. 

Let us now contrast Eq.~\eqref{eq:TPSC_self_energy_expanded} with the full expression describing the Hubbard self-energy \eqref{eq:four_point_self_G}. TPSC at the first level approximation corresponds to a Hartree-Fock factorization of Eq.~\eqref{eq:four_point_self_G}:
\begin{align}
\label{eq:first_approx_TPSC_two_particle}
&\Sigma^{\phi,(0)}_{l\bar{b},\sigma\bar{\sigma}^{\prime}}(z_1,\bar{z}_2)\mathcal{G}^{\phi,(0)}_{\bar{b}m,\bar{\sigma}^{\prime}\sigma}(\bar{z}_2,z_2)\notag\\
&\simeq A^{\phi}_{l,\sigma}(z_1)\biggl(\mathcal{G}^{\phi,(0)}_{l,-\sigma}(z_1,z_1^{+})\mathcal{G}^{\phi,(0)}_{lm,\sigma}(z_1,z_2)\notag\\
&\hspace{2.5cm}- \mathcal{G}^{\phi,(0)}_{l,\sigma-\sigma}(z_1,z_1^+)\mathcal{G}^{\phi,(0)}_{lm,-\sigma\sigma}(z_1,z_2)\biggr),
\end{align} 
where the kernel $A^{\phi}$ appearing in Eq.~\eqref{eq:first_approx_TPSC_two_particle} is defined as

\begin{align}
\label{eq:first_approx_TPSC_kernel_A}
A^{\phi}_{l,\sigma}(z_1)\equiv -iU(z_1)\frac{\left<\tc\hat{n}_{l,-\sigma}(z_1)\hat{n}_{l,\sigma}(z_1)\right>_{\phi}}{\left<\hat{n}_{l,-\sigma}(z_1)\right>_{\phi}\left<\hat{n}_{l,\sigma}(z_1)\right>_{\phi}}.
\end{align}
The kernel \eqref{eq:first_approx_TPSC_kernel_A} becomes exact in the local case where $z_2\to z_1^{+}$ and $m\to l$; one can indeed recover Eq.~\eqref{eq:four_point_self_G} given the definition of the local vertex $A^{\phi}_{l,\sigma}$. In Eq.~\eqref{eq:first_approx_TPSC_two_particle}, the source field is complete, \textit{i.e.} it contains both the diagonal (longitudinal) and off-diagonal (transversal) spin components. The first (second) term of Eq.~\eqref{eq:first_approx_TPSC_two_particle} results from the factorization of the longitudinal (transversal) four-point correlation function. From Eq.~\eqref{eq:first_approx_TPSC_two_particle}, because the transversal contribution vanishes when multiplying from the right by ${\mathcal{G}^{\phi}_{\sigma}}^{-1}$, the first-level longitudinal self-energy approximation reads
\begin{align}
\label{eq:first_approx_TPSC_two_particle_self}
\Sigma^{\phi,(0)}_{lm,\sigma}(z_1,z_2)&=A^{\phi}_{l,\sigma}(z_1)\mathcal{G}^{\phi}_{l,-\sigma}(z_1,z_1^+)\delta^{\mathcal{C}}(z_1,z_2)\delta_{l,m}\notag\\
&=iA^{\phi}_{l,\sigma}(z_1)n_{l,-\sigma}(z_1)\delta^{\mathcal{C}}(z_1,z_2)\delta_{l,m},
\end{align} such that 
\begin{align}
\label{eq:functional_derivative_sigma_first_approx}
&\frac{\delta\Sigma_{lm,\sigma}^{\phi,(0)}(z_1,z_2)}{\delta\mathcal{G}^{\phi,(0)}_{ij,-\sigma}(z_4,z_3)} = A^{\phi}_{l,\sigma}(z_1)\delta^{\mathcal{C}}(z_1,z_4)\delta^{\mathcal{C}}(z_1^+,z_3)\delta^{\mathcal{C}}(z_1,z_2)\delta_{l,i}\notag\\
&\times\delta_{l,j}\delta_{l,m}+\frac{\delta A^{\phi}_{l,\sigma}(z_1)}{\delta\mathcal{G}^{\phi}_{ij,-\sigma}(z_4,z_3)}n_{l,-\sigma}(z_1)\delta^{\mathcal{C}}(z_1,z_2)\delta_{l,m},
\end{align} 
and 
\begin{align}
\label{eq:functional_derivative_sigma_first_approx_2}
&\frac{\delta\Sigma_{lm,\sigma}^{\phi,(0)}(z_1,z_2)}{\delta\mathcal{G}^{\phi,(0)}_{ij,\sigma}(z_4,z_3)} = \frac{\delta A^{\phi}_{l,\sigma}(z_1)}{\delta\mathcal{G}^{\phi}_{ij,\sigma}(z_4,z_3)}n_{l,-\sigma}(z_1)\delta^{\mathcal{C}}(z_1,z_2)\delta_{l,m}.
\end{align} We have that $A^{\phi}_{l,\sigma}(z_1)=A^{\phi}_{l,-\sigma}(z_1)$. Now, since the irreducible vertex in the spin channel reads
\begin{align}
\label{eq:spin_irreducible_vertex_def}
&\Gamma^{\text{sp}}_{lmij}(z_1,z_2;z_4,z_3) \equiv \frac{\delta\Sigma^{\phi,(0)}_{\sigma}}{\delta\mathcal{G}^{\phi,(0)}_{-\sigma}}\biggr\rvert_{\phi\to 0} - \frac{\delta\Sigma^{\phi,(0)}_{\sigma}}{\delta\mathcal{G}^{\phi,(0)}_{\sigma}}\bigg\rvert_{\phi\to 0}\notag\\
&= A_{l,\sigma}(z_1)\delta^{\mathcal{C}}(z_1,z_4)\delta^{\mathcal{C}}(z_1^+,z_3)\delta^{\mathcal{C}}(z_1,z_2)\delta_{l,i}\delta_{l,j}\delta_{l,m},
\end{align} we can establish the following equivalence (within the TPSC approximation) between the local irreducible spin vertex and the double occupancy,
\begin{align}
\label{eq:equivalence_spin_irr_vertex_double_occupancy}
&\Gamma^{\text{sp}}_{lmij}(z_1,z_2;z_4,z_3)\notag\\
&= -iU(z_1)\frac{\left<\tc\hat{n}_{l,-\sigma}(z_1)\hat{n}_{l,\sigma}(z_1)\right>}{\left<\hat{n}_{l,-\sigma}(z_1)\right>\left<\hat{n}_{l,\sigma}(z_1)\right>}\notag\\
&\quad\times\delta^{\mathcal{C}}(z_1,z_4)\delta^{\mathcal{C}}(z_1^+,z_3)\delta^{\mathcal{C}}(z_1,z_2)\delta_{l,i}\delta_{l,j}\delta_{l,m}.
\end{align} 
The charge irreducible vertex is approximated in the same fashion as Eq.~\eqref{eq:spin_irreducible_vertex_def},
\begin{align}
\label{eq:charge_irreducible_vertex_def}
&\Gamma^{\text{ch}}_{lmij}(z_1,z_2;z_4,z_3) \equiv \frac{\delta\Sigma^{\phi,(0)}_{\sigma}}{\delta\mathcal{G}^{\phi,(0)}_{-\sigma}}\biggr\rvert_{\phi\to 0} + \frac{\delta\Sigma^{\phi,(0)}_{\sigma}}{\delta\mathcal{G}^{\phi,(0)}_{\sigma}}\bigg\rvert_{\phi\to 0}\notag\\
& = \Gamma_l^{\text{ch}}(z_1)\delta^{\mathcal{C}}(z_1,z_4)\delta^{\mathcal{C}}(z_1^+,z_3)\delta^{\mathcal{C}}(z_1,z_2)\delta_{l,i}\delta_{l,j}\delta_{l,m},
\end{align} where $\Gamma^{\text{ch}}$ has a different analytical expression from $\Gamma^{\text{sp}}$ and can be calculated from our knowledge of $\Gamma^{\text{sp}}$ using two-particle local sum-rules and Eq.~\eqref{eq:equivalence_spin_irr_vertex_double_occupancy}.

We next work out a useful expression for the vertex $\Gamma^{\sigma-\sigma\sigma-\sigma}$ appearing in Eq.~\eqref{eq:self_energy_approx_TPSC_locality_transversal}. To derive it, we need to calculate 
\begin{align*}
\Gamma^{\sigma-\sigma\sigma-\sigma}_{lmij}(z_1,z_2;z_4,z_3)=\frac{\delta\Sigma^{\phi,(0)}_{lm,\sigma-\sigma}(z_1,z_2)}{\delta\mathcal{G}^{\phi,(0)}_{ij,\sigma-\sigma}(z_4,z_3)}\bigg\rvert_{\phi\to 0},
\end{align*}
where $\Sigma^{\phi,(0)}_{\sigma-\sigma}$ can be extracted from Eq.~\eqref{eq:first_approx_TPSC_two_particle},
\begin{align}
\label{eq:transversal_channel_self_energy_get_vertex}
\Sigma^{\phi,(0)}_{lb,\sigma-\sigma}(z_1,z_2)=&\,\,iU(z_1)\frac{\left<\tc\hat{n}_{l,-\sigma}(z_1)\hat{n}_{l,\sigma}(z_1)\right>_{\phi}}{\left<\hat{n}_{l,-\sigma}(z_1)\right>_{\phi}\left<\hat{n}_{l,\sigma}(z_1)\right>_{\phi}}\notag\\
&\hspace{0cm}\times\mathcal{G}^{\phi,(0)}_{l,\sigma-\sigma}(z_1,z_1^+)\delta^{\mathcal{C}}(z_1,z_2)\delta_{lm}.
\end{align}
Hence, we obtain
\begin{align}
\label{eq:equivalence_transversal_irr_vertex_double_occupancy}
&\Gamma^{\sigma-\sigma\sigma-\sigma}_{lmij}(z_1,z_2;z_4.z_3) = iU(z_1)\frac{\left<\tc\hat{n}_{l,-\sigma}(z_1)\hat{n}_{l,\sigma}(z_1)\right>}{\left<\hat{n}_{l,-\sigma}(z_1)\right>\left<\hat{n}_{l,\sigma}(z_1)\right>}\notag\\
&\hspace{0.5cm}\times\delta^{\mathcal{C}}(z_1,z_4)\delta^{\mathcal{C}}(z_1^+,z_3)\delta^{\mathcal{C}}(z_1,z_2)\delta_{l,i}\delta_{l,j}\delta_{l,m}\notag\\
&=-\Gamma^{\text{sp}}_{lmij}(z_1,z_2;z_4.z_3).
\end{align}
Equation~\eqref{eq:equivalence_transversal_irr_vertex_double_occupancy} is inserted into Eq.~\eqref{eq:self_energy_approx_TPSC_locality_transversal} to replace $\Gamma^{\sigma-\sigma\sigma-\sigma}$. Gathering all the results stemming from the TPSC ansatz, we can express the total self-energy for the second-level approximation, which is an average of the longitudinal \eqref{eq:self_energy_approx_TPSC_locality_logitudinal} and the transversal \eqref{eq:self_energy_approx_TPSC_locality_transversal} components:
\begin{align}
\label{eq:self_energy_approx_TPSC_locality_total}
&\Sigma^{\text{TPSC},(1)}_{lm,\sigma}(z_1,z_2)\notag\\
&= U(z_1)n_{l,\downarrow}(z_1)\delta^{\mathcal{C}}(z_1,z_2)\delta_{l,m}+\frac{U(z_1)}{8}\mathcal{G}^{(0)}_{lm,\sigma}(z_1,z_2)\notag\\
&\quad\times\biggl[\Gamma_m^{\text{ch}}(z_2)\chi^{\text{ch}}_{ml}(z_2,z_1) + 3\Gamma_m^{\text{sp}}(z_2)\chi^{\text{sp}}_{ml}(z_2,z_1)\biggr].
\end{align}
The Fourier transform of Eq.~\eqref{eq:self_energy_approx_TPSC_locality_total} yields~\cite{Bergeron_2011_optical_cond}
\begin{align}
\label{eq:fft_total_self_energy}
&\int\mathrm{d}^D(\mathbf{r}_l-\mathbf{r}_m) \ e^{-i\mathbf{k}\cdot(\mathbf{r}_l-\mathbf{r}_m)}\Sigma^{\phi,(1)}_{lm,\sigma}(z_1,z_2)\notag\\
&= \Sigma^{\text{TPSC},(1)}_{\mathbf{k},\sigma}(z_1,z_2)\notag\\
&=U(z_1)n_{-\sigma}(z_1)\delta^{\mathcal{C}}(z_1,z_2)
+\frac{U(z_1)}{8}\int\mathrm{d}^D\mathbf{q} \ \mathcal{G}^{(0)}_{\mathbf{k}+\mathbf{q},\sigma}(z_1,z_2)\notag\\
&\quad\times\biggl[\Gamma^{\text{ch}}(z_2)\chi^{\text{ch}}_{\mathbf{q}}(z_2,z_1) + 3\Gamma^{\text{sp}}(z_2)\chi^{\text{sp}}_{\mathbf{q}}(z_2,z_1)\biggr].
\end{align}
The susceptibilities $\chi^{\text{ch/sp}}$ are functionals of $\mathcal{G}^{0}$ defined in Eq.~\eqref{eq:first_level_approx_G}. The steps which lead from the first-level approximation to the self-energy $\Sigma^{(0)}$ (Eq.~\eqref{eq:TPSC_self_energy_expanded}) to the second-level approximation $\Sigma^{(1)}$ (Eq.~\eqref{eq:fft_total_self_energy}) do not result in an approximation which is conserving in the Kadanoff-Baym sense, as was already pointed out in Ref.~\onlinecite{https://doi.org/10.1002/andp.202000399}. Nevertheless, in practice, the second-level approximation conserves energy rather well after a perturbation, for a large range of bare interactions $U$ and dopings $n$. Moreover, the fact that the second-level approximation to the TPSC self-energy \eqref{eq:fft_total_self_energy} reduces to the second-order lattice IPT self-energy~\cite{doi:10.1143/JPSJ.69.3912} in the limit of small $U$ makes it natural to combine TPSC within a DMFT scheme based on a weak-coupling impurity solver. %The latter makes it easy to avoid double-countings \textcolor{red}{[check]} since in the limit where $U/W$ is small, the second-level Luttinger-Ward functional that would generate $\Sigma^{\text{TPSC},(1)}$ can be approximated by that of the lattice second-order $\Sigma^{(2)}$ self-energy~\eqref{eq:nonequilibrium_quantum_many_body_physics:IPT:second_order_bubble}. 
%Hence, DMFT+TPSC is approximately conserving and it isn't enough to replace the local TPSC self-energy by the DMFT one to avoid double-counting.~\cite{PhysRevLett.118.246402_GW_Noneq} 
This nonequilibrium DMFT+TPSC scheme is explained in Sec.~\ref{ch:dmft_tpsc}.

%%%%%%%%%%%%%%%%%%%%%%%%%%%%%%%%%%%% Algorithm %%%%%%%%%%%%%%%%%%%%%%%%%%%%%%%%%%%%%%%%%%

%\paragraph{Algorithm}
\subsubsection{Algorithm}
\label{sec:tpsc_algorithm}

\begin{figure}[h!]
  \centering
    \includegraphics[width=1.0\columnwidth]{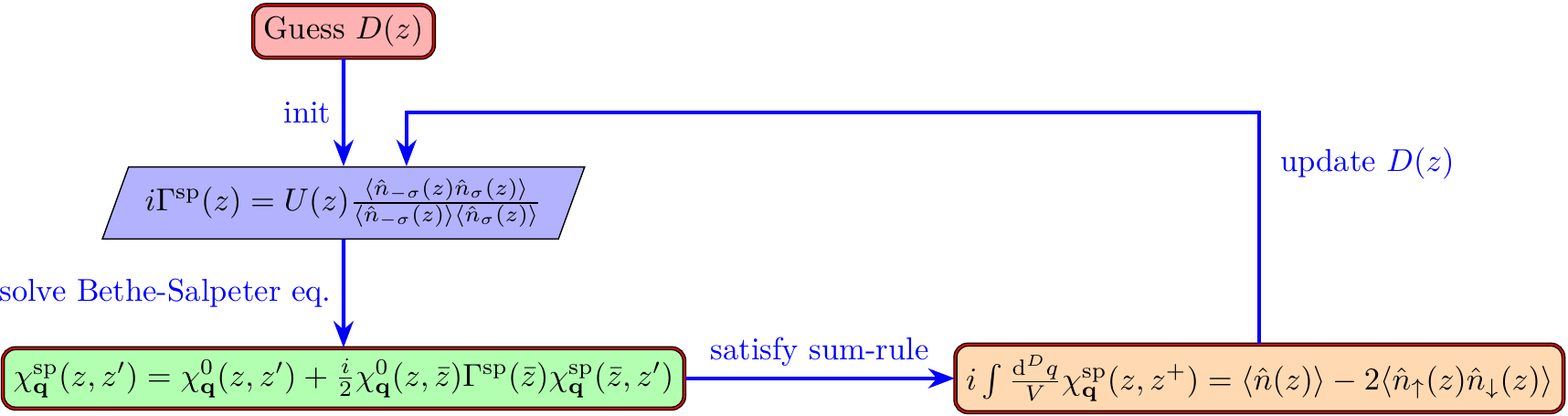}
      \caption{Flow chart describing the self-consistent determination of $D(z)$, $\chi^{\text{sp}}$ and $\Gamma^{\text{sp}}$ (alternative method). In the actual simulations, we modify the BSE as in Eq.~\eqref{eq:bethe_Salpeter_eq_approximated} and use the multidimensional root-finding method.
      }
  \label{fig:KB_sp_quantities}
\end{figure}

Our implementation of nonequilibrium TPSC contains the following steps: we first compute the noninteracting Green's function $\mathcal{G}^0$ that allows to calculate the noninteracting two-particle Green's function $\chi^0\equiv -2i\mathcal{G}^0\mathcal{G}^0$ and make an initial guess for the double occupancy $D(z)=\langle\hat{n}_{\uparrow,\sigma}(z)\hat{n}_{\downarrow,-\sigma}(z)\rangle$. Then, we simultaneously solve for $\chi^{\text{sp}}$ and $\Gamma^{\text{sp}}$ using the local spin two-particle sum-rule

\begin{align}
\label{eq:fluctuation_dissipation_two_particle}
&i\int\frac{\mathrm{d}^Dq}{\left(2\pi\right)^D} \ \chi^{\text{sp/ch}}_{\mathbf{q}}(z,z^{+})\notag\\
&\quad =n(z)+2(-1)^l\left<\hat{n}_{-\sigma}(z)\hat{n}_{\sigma}(z)\right> - (1-l)n(z)^2,
\end{align}
where $n=\left<\hat{n}_{\uparrow}+\hat{n}_{\downarrow}\right>$ is the density of particles, $l=0$ for charge (ch) and $l=1$ for spin (sp). This is done using a multidimensional root-finding method for a non-linear system of equations at each time step. Alternatively, as shown in the flow chart \ref{fig:KB_sp_quantities}, the spin quantities could be solved self-consistently until $D(z)$ converges. However, we make use of the multidimensional root-finding method due to its higher efficiency.

\begin{figure}[h!]
  \centering
    \includegraphics[width=1.0\columnwidth]{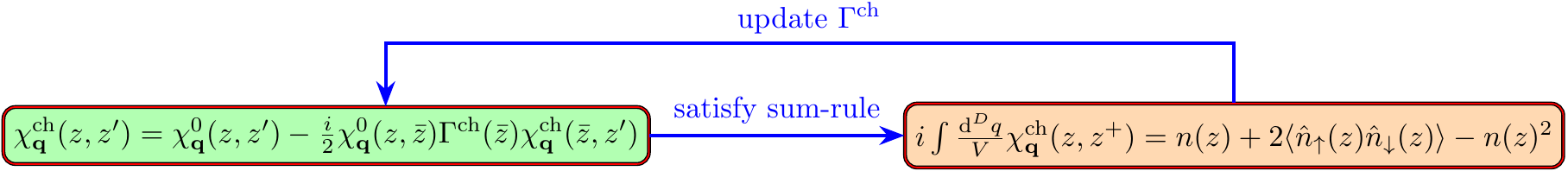}
      \caption{Flow chart describing the self-consistent determination of $\chi^{\text{ch}}$ and $\Gamma^{\text{ch}}$. In the actual simulations, we modify the BSE as in Eq.~\eqref{eq:bethe_Salpeter_eq_approximated}.}
  \label{fig:KB_ch_quantities}
\end{figure}

\noindent
The next step is to solve for the charge quantities $\chi^{\text{ch}}$ and $\Gamma^{\text{ch}}$. Again, a multidimensional root-finding method for a non-linear system of equations is used at each time step. The two equations which must be simultaneously solved are displayed in Fig.~\ref{fig:KB_ch_quantities}, which involves the charge two-particle sum-rule~\eqref{eq:fluctuation_dissipation_two_particle}.
\noindent

In order to satisfy the local sum-rules~\eqref{eq:fluctuation_dissipation_two_particle} out of equilibrium, we introduce and additional approximation, resulting in a modified (approximated) form of the Bethe-Salpeter equations written in the green panels of the flow charts \ref{fig:KB_sp_quantities} and \ref{fig:KB_ch_quantities}. The approximated form is
\begin{align}
\label{eq:bethe_Salpeter_eq_approximated}
&\chi_{\mathbf{q}}^{\text{sp/ch}}(z,z^{\prime}) = \chi^0_{\mathbf{q}}(z,z^{\prime})\notag\\ 
&\quad + (-1)^{l+1}\frac{i}{2} \Gamma^{\text{sp}/\text{ch}}(z)\chi_{\mathbf{q}}^0(z,\bar{z})\chi_{\mathbf{q}}^{\text{sp/ch}}(\bar{z},z^{\prime}),
\end{align}
where, once again, $l=0$ for charge (ch) and $l=1$ for spin (sp). The reason for the approximation \eqref{eq:bethe_Salpeter_eq_approximated} is explained in Appendix~\ref{appendice:ch:noneq_approx_TPSC_vertices}. As far as TPSC is concerned, the algorithm terminates once all the quantities in each channel have been solved and the self-energy

\begin{align}
\label{eq:tpsc_self_energy_alpha}
&\Sigma^{\text{TPSC},(1)}_{\mathbf{k},\sigma}[\alpha](z_1,z_2)\notag\\
&= U(z_1)n_{-\sigma}(z_1)\delta_{\mathcal{C}}(z_1,z_2) + \frac{U(z_1)}{8}\int\frac{\mathrm{d}^Dq}{(2\pi)^D}\alpha(z_2)\notag\\
&\!\!\times\!\!\biggl[3\Gamma^{\text{sp}}(z_2)\chi^{\text{sp}}_{\mathbf{q}}(z_2,z_1) + \Gamma^{\text{ch}}(z_2)\chi^{\text{ch}}_{\mathbf{q}}(z_2,z_1)\biggr]\mathcal{G}^{(0)}_{\mathbf{k}+\mathbf{q},\sigma}(z_1,z_2)
\end{align} has been computed. In Eq.~\eqref{eq:tpsc_self_energy_alpha}, the one-time variable $\alpha$ has been introduced into Eq.~\eqref{eq:fft_total_self_energy} to satisfy the sum-rule involving the double occupancy appearing in Eqs.~\eqref{eq:first_approx_TPSC_two_particle} and \eqref{eq:first_approx_TPSC_kernel_A} which is needed for solving the spin quantities (see Fig.~\ref{fig:KB_sp_quantities}),
\begin{align}
\label{eq:double_occupancy_sum_rule_alpha}
&\frac{-i}{2}\int\frac{\mathrm{d}^Dk}{(2\pi)^D} \ \left[\Sigma_{\mathbf{k},\bar{\sigma}}^{\text{TPSC},(1)}[\alpha](z_1,\bar{z})\mathcal{G}^{(1)}_{\mathbf{k},\bar{\sigma}}[\Sigma^{\text{TPSC},(1)}](\bar{z},z_1^{+})\right]\notag\\
& \quad = U(z_1)\langle \hat n_{-\sigma}(z_1)\hat n_{\sigma}(z_1) \rangle.
\end{align}
This extra renormalization of the irreducible vertices is necessary in order to get physically sound results after parameter quenches in the Hubbard model~\eqref{eq:Hubbard_model_intro}. 

The variant TPSC+GG reintroduces the Green's function $\mathcal{G}^{(1)}[\Sigma^{\text{TPSC},(1)}]$ computed with Eq.~\eqref{eq:tpsc_self_energy_alpha} into the noninteracting bubble $\chi^0$ and repeats the subroutines described in Figures~\ref{fig:KB_sp_quantities} and \ref{fig:KB_ch_quantities} until overall convergence.

%%%%%%%%%%%%%%%%%%%%%%%%%%%%%%%%%%%% DMFT+TPSC %%%%%%%%%%%%%%%%%%%%%%%%%%%%%%%%%%%%%%%%%%

\subsection{Nonequilibrium DMFT+TPSC}
\label{ch:dmft_tpsc}

\subsubsection{General remarks}

\begin{figure}[h!]
  \centering
    \includegraphics[width=1.0\columnwidth]{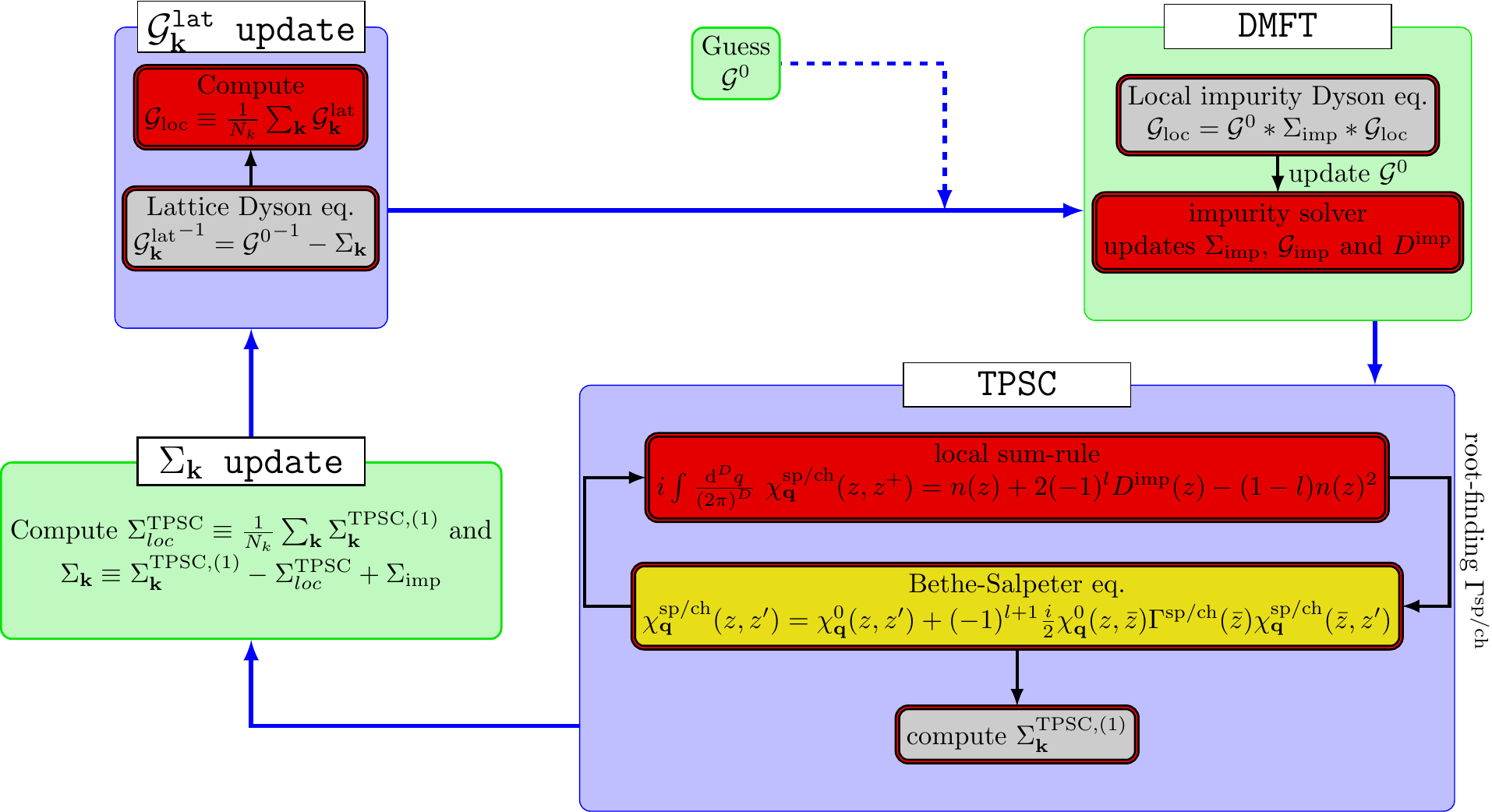}
      \caption{Flow chart describing the self-consistent DMFT+TPSC procedure. In the actual calculations, the Bethe-Salpeter equations inside the yellow panel are approximated by Eq.~\eqref{eq:bethe_Salpeter_eq_approximated}.
      }
  \label{fig:DMFT_flow_chart}
\end{figure}

Similar in spirit to established schemes like GW+DMFT\cite{Biermann2003,Ayral2013} or FLEX+DMFT,\cite{Huang2015,Kitatani2015} the combination of DMFT (introduced in Sec.~\ref{subsec:DMFT}) and TPSC (introduced in Sec.~\ref{subsec:TPSC_variants}) can be accomplished by replacing the local TPSC self-energy component with the DMFT one in a self-consistent manner in order to better capture the local correlations. The resulting self-energy reads $\Sigma^{\text{DMFT+TPSC}}_{ij}=\Sigma^{\text{imp}}\delta_{ij}+\Sigma^{\text{TPSC},(1)}(1-\delta_{ij})$, with $i,j$ lattice site indices, and thus incorporates the effects of local and nonlocal correlations on the spin and charge degrees of freedom. These correlations feed back into the DMFT calculations within a self-consistency loop. In the following subsection, we describe the algorithmic procedure that defines nonequilibrium DMFT+TPSC. The full scheme is illustrated as a flow chart in Fig.~\ref{fig:DMFT_flow_chart}.

%\paragraph{Algorithm} 
\subsubsection{Algorithm}

To start the DMFT+TPSC procedure, one must guess an initial Weiss Green's function \eqref{eq:PM:Weiss_Green_hyb} (e.g.~local Green's function of the noninteracting lattice) that enters the impurity solver described in Sec.~\ref{subsubsec:IPT}. The impurity solver computes an impurity self-energy, denoted by $\Sigma_{\text{imp}}[\mathcal{G}_0]$ in this section, that renormalizes and broadens the energy spectrum of the impurity electrons. Then, the impurity double occupancy %$D^{\text{imp}}$
\begin{align}
\label{eq:impurity_double_occupancy}
D^{\text{imp}}(z) =& \,\frac{-i}{2U(z)}\text{Tr}\left[\Sigma^{\text{imp}}_{\sigma}(z,\bar{z})\mathcal{G}^{\text{imp}}_{\sigma}(\bar{z},z)\right]^<\notag\\
&+ \frac{1}{4}\sum_{\sigma}n_{\sigma}(z)n_{-\sigma}(z),
\end{align}
is used instead of that extracted from the ansatz~\eqref{eq:equivalence_spin_irr_vertex_double_occupancy}, which is employed in TPSC and TPSC+GG. $D^{\text{imp}}$ determines both the spin and charge irreducible vertices according to Figs.~\ref{fig:KB_sp_quantities} and \ref{fig:KB_ch_quantities}, respectively, making use of the respective local sum-rules~\eqref{eq:fluctuation_dissipation_two_particle}. This time, the susceptibilities defined through the Bethe-Salpeter equation~\eqref{eq:charge_spin_susceptibility_TPSC_definition_developed_2} are slightly different, in that the ``bare" two-particle Green's function $\chi^0$ is defined as

\begin{align}
\label{eq:noninteracting_two_particle_Gfunc}
\chi^0_{\mathbf{q}}(z,z^{\prime}) = -2i\int\frac{\mathrm{d}^Dk}{(2\pi)^D} \ \mathcal{G}_{\mathbf{k}}(z,z^{\prime})\mathcal{G}_{\mathbf{k}+\mathbf{q}}(z,z^{\prime}),
\end{align}
where the lattice Green's function $\mathcal{G_{\mathbf{k}}}$ is obtained from Eq.~\eqref{eq:PM:projected_green_function_impurity} and contains the local impurity self-energy. Then, the momentum-dependent TPSC self-energy can be calculated using Eq.~\eqref{eq:fft_total_self_energy} (with $\mathcal{G}^{(0)}$ replaced with $\mathcal{G}$). We finally replace the local self-energy component of $\Sigma^{\text{TPSC},(1)}_{\mathbf{k}}$,
\begin{align}
\Sigma^{\text{TPSC},(1)}_{\text{loc},\sigma}(z,z^{\prime})\equiv\frac{1}{N_k}\sum_{\mathbf{k}}\Sigma_{\mathbf{k},\sigma}^{\text{TPSC},(1)}(z,z^{\prime}),
\end{align}
by the impurity self-energy $\Sigma^{\sigma}_{\text{imp}}$. The DMFT+TPSC self-energy with improved local correlations thus reads

\begin{align}
\label{eq:sigma_DMFT_TPSC}
&\Sigma^{(1)}_{\mathbf{k},\sigma}(z,z^{\prime})\notag\\
&\equiv\Sigma_{\mathbf{k},\sigma}^{\text{TPSC},(1)}(z,z^{\prime})-\Sigma^{\text{TPSC},(1)}_{\text{loc},\sigma}(z,z^{\prime})+\Sigma^{\text{imp}}_{\sigma}(z,z^{\prime}),
\end{align}
and the improved lattice Green's function $\mathcal{G}^{\text{lat},(1)}_{\mathbf{k}}$ with $\Sigma_{\mathbf{k}}$~from Eq.~\eqref{eq:sigma_DMFT_TPSC} is defined 
as the solution of the Dyson equation
\begin{align}
\label{eq:lattice_G_definition_improved}
&\left[i\partial_z + \mu - \epsilon(\mathbf{k})-\Sigma_{\text{imp}}^{\delta,\sigma}(z)\right]\mathcal{G}^{\text{lat},(1)}_{\mathbf{k},\sigma}(z,z^{\prime})\notag\\
&\hspace{0.7cm}-\Sigma^{(1)}_{\mathbf{k},\sigma}(z,\bar{z})\mathcal{G}^{\text{lat},(1)}_{\mathbf{k},\sigma}(\bar{z},z^{\prime}) = \delta^{\mathcal{C}}(z,z^{\prime}).
\end{align}
Once the improved lattice Green's function \eqref{eq:lattice_G_definition_improved} is known, the lattice average $\mathcal{G}_{\text{loc}}^{\sigma}(z,z^{\prime})\equiv\frac{1}{N_k}\sum_{\mathbf{k}}\mathcal{G}^{\text{lat},(1)}_{\mathbf{k},\sigma}(z,z^{\prime})$ is calculated and identified with the impurity Green's function. Finally, by solving the Volterra equation \eqref{eq:PM:impurity_self_energy}, the Weiss Green's function can be updated and reinserted into the impurity solver. The whole process is repeated until the scheme converges.

Apart from looking at the energy conservation during the time propagation of the (undriven) DMFT+TPSC solution, the comparison between the DMFT double occupancy $D^{\text{imp}}$~\eqref{eq:impurity_double_occupancy} and the one extracted from the lattice quantities
\begin{align}
\label{eq:tr_sk_Gk_TPSC}
D^{\text{TPSC},(1)}(z) =& \, \frac{-i}{2U(z)} \text{Tr}\left[\Sigma^{(1)}_{\mathbf{k},\sigma}(z,\bar{z})\mathcal{G}^{\text{lat},(1)}_{\mathbf{k},\sigma}(\bar{z},z)\right]^< \notag\\
&\hspace{0.0cm}+ \frac{1}{4}\sum_{\sigma}n_{\sigma}(z)n_{-\sigma}(z),
\end{align}
with $\Sigma^{(1)}_{\mathbf{k}}$ defined in Eq.~\eqref{eq:sigma_DMFT_TPSC} and $\mathcal{G}^{\text{lat},(1)}_{\mathbf{k}}$ defined in Eq.~\eqref{eq:lattice_G_definition_improved}, turns out to be a good consistency check for the method. If the difference between $D^{\text{imp}}(z)$ and $D^{\text{TPSC},(1)}(z)$ becomes too large, the results become unreliable. Note that in our single-band model, Eq.~\eqref{eq:tr_sk_Gk_TPSC} can be obtained by Fourier-transforming Eq.~\eqref{eq:four_point_self_G}.

Similarly to TPSC and TPSC+GG, which employ the sum-rule \eqref{eq:double_occupancy_sum_rule_alpha} to obtain a consistent result for the double occupation, DMFT+TPSC can be modified by enforcing that the impurity double occupancy $D^{\text{imp}}$~\eqref{eq:impurity_double_occupancy} be equal to that computed from the lattice quantities obtained from TPSC~\eqref{eq:tr_sk_Gk_TPSC}:
\begin{align}
\label{eq:double_occupancy_sum_rule_alpha_DMFT_TPSC}
&\text{Tr}\left[\Sigma^{(1)}_{\mathbf{k},\sigma}[\alpha](z,\bar{z})\mathcal{G}^{\text{lat},(1)}_{\mathbf{k},\sigma}(\bar{z},z)\right]^<\notag\\
&\hspace{2.0cm} \equiv \text{Tr}\left[\Sigma^{\text{imp}}_{\sigma}(z,\bar{z})\mathcal{G}^{\text{imp}}_{\sigma}(\bar{z},z)\right]^<,
\end{align}
with 

\begin{align}
\label{eq:sigma_DMFT_TPSC_alpha}
&\Sigma^{(1)}_{\mathbf{k},\sigma}[\alpha](z,z^{\prime})\notag\\
&=\Sigma_{\mathbf{k},\sigma}^{\text{TPSC},(1)}[\alpha](z,z^{\prime})-\Sigma^{\text{TPSC},(1)}_{\text{loc},\sigma}[\alpha](z,z^{\prime})+\Sigma^{\text{imp}}_{\sigma}(z,z^{\prime}),
\end{align}
or, alternatively,
\begin{align}
\label{eq:sigma_DMFT_TPSC_alpha_2}
&\Sigma^{(1)}_{\mathbf{k},\sigma}[\alpha](z,z^{\prime})\notag\\
&=\Sigma_{\mathbf{k},\sigma}^{\text{TPSC},(1)}(z,z^{\prime})-\alpha(z)\Sigma^{\text{TPSC},(1)}_{\text{loc},\sigma}(z,z^{\prime})+\Sigma^{\text{imp}}_{\sigma}(z,z^{\prime}),
\end{align}
where $\alpha$, in the case of Eq.~\eqref{eq:sigma_DMFT_TPSC_alpha}, serves a similar purpose as in Eq.~\eqref{eq:double_occupancy_sum_rule_alpha}, in that it renormalizes further the irreducible vertices in Eq.~\eqref{eq:tpsc_self_energy_alpha} so as to fulfil Eq.~\eqref{eq:double_occupancy_sum_rule_alpha_DMFT_TPSC}. In Eq.~\eqref{eq:sigma_DMFT_TPSC_alpha_2}, the parameter $\alpha$ can be seen as a time-dependent correction to the hybridization function appearing in the DMFT self-consistency (Eq.~\eqref{eq:DMFT:DMFT_action}). These \textit{modified} DMFT+TPSC methods are coined DMFT+TPSC$\alpha$. It turns out, however, that neither the lattice self-energy \eqref{eq:sigma_DMFT_TPSC_alpha} nor the one defined in Eq.~\eqref{eq:sigma_DMFT_TPSC_alpha_2} leads to a stable nonequilibrium evolution. Thus, DMFT+TPSC$\alpha$ will only be discussed in equilibrium set-ups, making use of Eq.~\eqref{eq:sigma_DMFT_TPSC_alpha}.

%%%%%%%%%%%%%%%%%%%%%%%%%%%%%%%%%%%%%%%%%%%%%%%% Results %%%%%%%%%%%%%%%%%%%%%%%%%%%%%%%%%%%%%%%%%%%%%%%%%%%%%%%%%%%%%%%%%%%
%%    

\subsection{Summary of the different schemes}

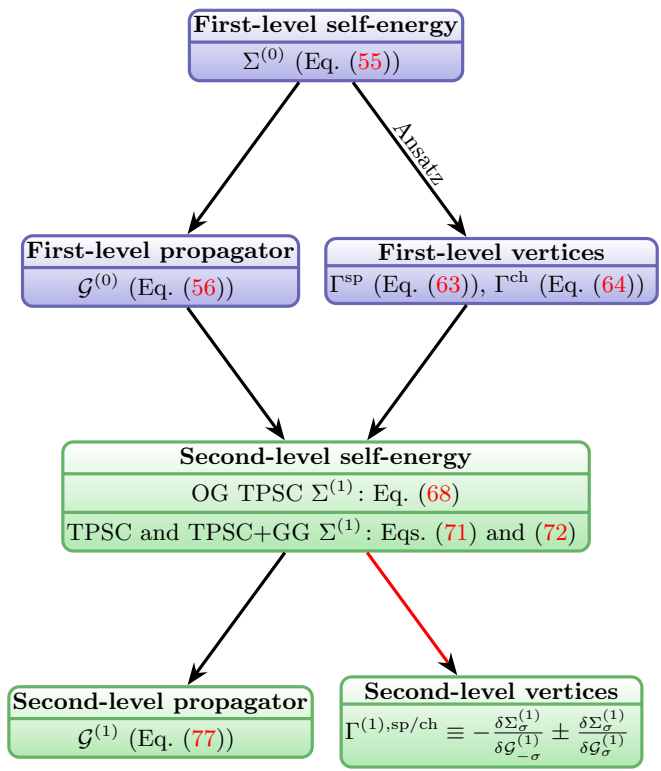
\begin{figure}[h!]
\begin{center}
\begin{tikzpicture}[
    %grow=right,
    level 1/.style={sibling distance=5.0cm, level distance=4.5cm, align=center},
    level 2/.style={sibling distance=5.0cm, level distance=4.5cm, align=center},
    edge from parent/.style={very thick,draw=blue!50!black!90,
        shorten >=5pt, shorten <=5pt,->},
    edge from parent path={(\tikzparentnode.south) -- (\tikzchildnode.north)},
    kant/.style={text width=2cm, text centered, sloped},
    every node/.style={text ragged, inner sep=.5mm, align=center},
    punkt/.style={rectangle, rounded corners, shade, top color=white,
    bottom color=blue!70!black!30, draw=blue!50!black!60, very
    thick },
    punkt2/.style={rectangle, rounded corners, shade, top color=white,
    bottom color=green!70!black!30, draw=green!50!black!60, very
    thick }
    ]

\begin{scope}
    \node[punkt] [rectangle split, rectangle split, rectangle split parts=2, text ragged] (A) at (0,0) {
            \textbf{First-level self-energy}
                  \nodepart{second}
            $\Sigma^{(0)}$~(Eq.~\eqref{eq:TPSC_self_energy_expanded})
        };
    \node[punkt] [rectangle split, rectangle split, rectangle split parts=2,
         text ragged] (B) at (2.2cm,-3cm) {
            \textbf{First-level vertices}
                  \nodepart{second}
            $\Gamma^{\text{sp}}$~(Eq.~\eqref{eq:equivalence_spin_irr_vertex_double_occupancy}), $\Gamma^{\text{ch}}$~(Eq.~\eqref{eq:charge_irreducible_vertex_def})
        };
   \node[punkt] [rectangle split, rectangle split, rectangle split parts=2,
         text ragged] (C) at (-2.2cm,-3cm) {
            \textbf{First-level propagator}
                  \nodepart{second}
            $\mathcal{G}^{(0)}$~(Eq.~\eqref{eq:first_level_approx_G})
        };
    
     \node[punkt2] [rectangle split, rectangle split, rectangle split parts=3] (D) at (0,-6cm) {
                \textbf{Second-level self-energy}
                \nodepart{second}
                $\text{OG TPSC} \ \Sigma^{(1)}\colon \text{Eq.~\eqref{eq:fft_total_self_energy}}$
                \nodepart{third}
                $\text{TPSC and TPSC+GG} \ \Sigma^{(1)}\colon \text{Eqs.~\eqref{eq:tpsc_self_energy_alpha} and \eqref{eq:double_occupancy_sum_rule_alpha}}$
        };

     \node[punkt2] [rectangle split, rectangle split, rectangle split parts=2] (E) at (2.2cm,-9cm) {
                \textbf{Second-level vertices}
                \nodepart{second}
                $\Gamma^{(1),\text{sp/ch}}\equiv -\frac{\delta\Sigma_{\sigma}^{(1)}}{\delta\mathcal{G}_{-\sigma}^{(1)}}\pm \frac{\delta\Sigma_{\sigma}^{(1)}}{\delta\mathcal{G}_{\sigma}^{(1)}}$
        };
        
     \node[punkt2] [rectangle split, rectangle split, rectangle split parts=2] (F) at (-2.2cm,-9cm) {
                \textbf{Second-level propagator}
                \nodepart{second}
                $\mathcal{G}^{(1)}$~(Eq.~\eqref{eq:lattice_G_definition_improved})
        };
\end{scope}

\begin{scope}[>={Stealth[black]}]
    \path [->,draw=black,very thick] (A) edge[text width=1.5cm, text centered, anchor=south, sloped] node {\small Ansatz} (B);
    \path [->,draw=black,very thick] (A) edge[text width=1.5cm, text centered, anchor=south, sloped] node {} (C);
    \path [->,draw=black,very thick] (C) edge node {} (D);
    \path [->,draw=black,very thick] (B) edge node {} (D);
    \path [->,draw=red,very thick] (D) edge[text width=2cm, text centered, sloped] node {} (E);
    \path [->,draw=black,very thick] (D) edge[text width=1.5cm, text centered, anchor=south, sloped] node {} (F);
\end{scope}

\end{tikzpicture}
\end{center}
\caption{Flow graph showing the connections between the two levels of TPSC, namely the first- (blue boxes) and second-level (green boxes) approximations. The red line shows that second-level irreducible vertices could in principle be obtained from the second-level self-energy $\Sigma^{(1)}$. The $\alpha$ renormalization of the vertices introduced via Eq.~\eqref{eq:double_occupancy_sum_rule_alpha} modifies the irreducible vertices such that the two levels of the approximation become consistent. 
}
\label{fig:graph_second_level_first_level}
\end{figure}

In order to clarify the similarities and differences between the methods considered in this paper, we summarize the key characteristics of the methods in Table~\ref{table:nonequilibrium_methods_laid_out}. Moreover, the graph in Fig.~\ref{fig:graph_second_level_first_level} illustrates the connection between the first- and second-level approximations.

\begin{table}[h!]
\centering
\resizebox{\columnwidth}{!}{
 \begin{tabular}{||c|c|c|c||} 
 \hline
 \backslashbox{\phantom{=}}{} & Self-consistent & $D$ consistency & $\Sigma^{(1)}_{\mathbf{k}}$\\
 \hline
OG TPSC & $\text{\sffamily X}$ & $\text{\sffamily X}$ & Eq.~\eqref{eq:fft_total_self_energy} \\
 \hline
TPSC & $\text{\sffamily X}$ &  $\checkmark$ & Eqs.~\eqref{eq:tpsc_self_energy_alpha} \& \eqref{eq:double_occupancy_sum_rule_alpha} \\
 \hline
TPSC+GG & $\checkmark$ & $\checkmark$ & Eqs.~\eqref{eq:tpsc_self_energy_alpha} \& \eqref{eq:double_occupancy_sum_rule_alpha} \\
 \hline
DMFT+TPSC & $\checkmark$ & $\text{\sffamily X}$ & Eqs.~\eqref{eq:sigma_DMFT_TPSC} \\
 \hline
DMFT+TPSC$\alpha$ & $\checkmark$ & $\checkmark$ & Eqs.~\eqref{eq:sigma_DMFT_TPSC_alpha} \& \eqref{eq:double_occupancy_sum_rule_alpha_DMFT_TPSC} \\[1ex]
 \hline
\end{tabular}
}
\caption{Properties of the TPSC variants considered in this study. Checkmarks ($\checkmark$) indicate that a method is endowed with the corresponding characteristic, while the x-marks ($\text{\sffamily X}$) mean the opposite. In the last column, we list the equations defining the lattice self-energy.}
\label{table:nonequilibrium_methods_laid_out}
\end{table}

The first column of Table~\ref{table:nonequilibrium_methods_laid_out} titled ``Self-consistent'' specifies which methods are self-consistent, \textit{i.e.} feed back the interacting lattice Green's functions into a self-consistency loop until convergence. The methods without this characteristic compute the self-energy and related quantities in a ``one-shot'' fashion. The second column titled ``$D$ consistency'' indicates which methods make use of a parameter $\alpha$ to enforce consistency between the double occupancies obtained from local and lattice quantities. For example, in the case of TPSC and TPSC+GG, the sum-rule~\eqref{eq:double_occupancy_sum_rule_alpha} ensures that the double occupancy obtained within the first-level approximation from Eq.~\eqref{eq:equivalence_spin_irr_vertex_double_occupancy} is equal to that calculated from the second-level quantities $\Sigma^{(1)}_{\mathbf{k}}$ and $\mathcal{G}^{(1)}$. Indeed, in a fully %conserving 
consistent 
scheme, the double occupancy appearing in Eq.~\eqref{eq:equivalence_spin_irr_vertex_double_occupancy}, which is extracted from the first-level approximation self-energy $\Sigma^{(0)}_{\mathbf{k}}$~\eqref{eq:transversal_channel_self_energy_get_vertex}, should be equal to that obtained from the second-level single-particle quantities $\Sigma^{(1)}_{\mathbf{k}}$ and $\mathcal{G}^{(1)}_{\mathbf{k}}$ (Eq.~\eqref{eq:double_occupancy_sum_rule_alpha}). Finally, the last column of Table~\ref{table:nonequilibrium_methods_laid_out} refers to the second-level self-energies featuring in each method, together with the extra sum-rule they need to satisfy if the method is ``$D$-consistent''.

%---------------- TREE GRAPH ------------------

%---------------- END TREE GRAPH ------------------

\section{Results}
\label{sec:results}

\subsection{General remarks}

We first test TPSC, TPSC+GG and DMFT+TPSC as introduced in Sec.~\ref{ch:dmft_tpsc} by studying equilibrium lattice models and comparing some results with data published in the literature.\cite{PhysRevX.11.011058} In Sec.~\ref{sec:results:equilibrium}, we benchmark our results against Diagrammatic Monte Carlo (DiagMC)\cite{PhysRevLett.81.2514,VanHoucke2010} and compare our implementations with TPSC in its original formulation, coined from now on ``OG TPSC''.\cite{tpsc_1997} Then, TPSC, TPSC+GG and DMFT+TPSC are used to compute various equilibrium properties of the cubic lattice Hubbard model. In Sec.~\ref{sec:results:Nonequilibrium}, we present the nonequilibrium applications. We simulate ramps in one of the hopping terms to induce a dimensional crossover from a square to a cubic lattice and analyze the corresponding spin and charge dynamics. 
%In Sec.~\ref{sec:results:Nonequilibrium}, we furthermore study the prethermalization of the occupation function near the Fermi level.

\subsection{Equilibrium}
\label{sec:results:equilibrium}

\subsubsection{Benchmarks against DiagMC}

\begin{figure}[t]
\begin{minipage}[h]{\columnwidth}
\begin{center}
\includegraphics[width=\columnwidth]{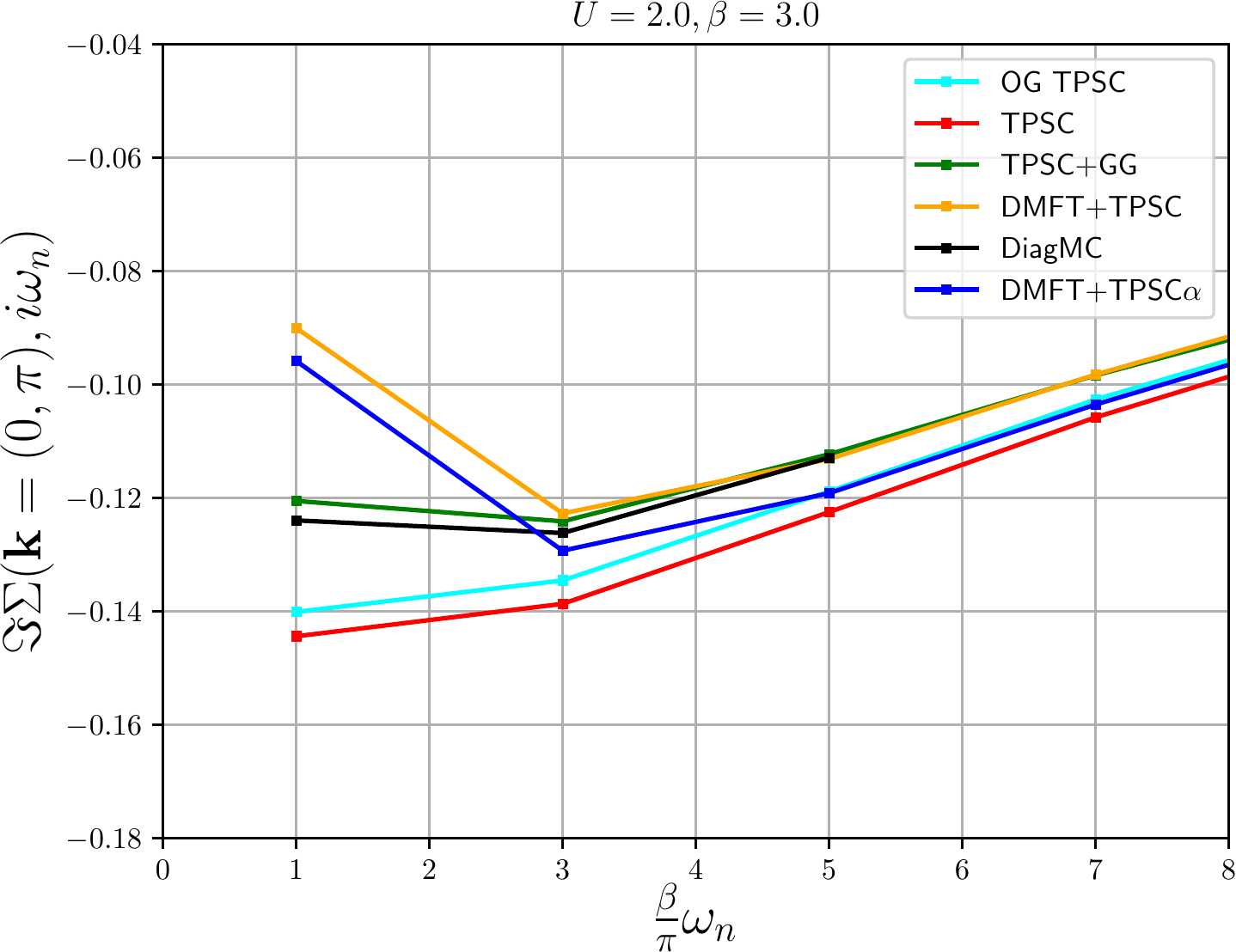}
\end{center} 
\end{minipage}
\hfill
\vspace{0.1 cm}
\begin{minipage}[h]{\columnwidth}
\begin{center}
\includegraphics[width=\columnwidth]{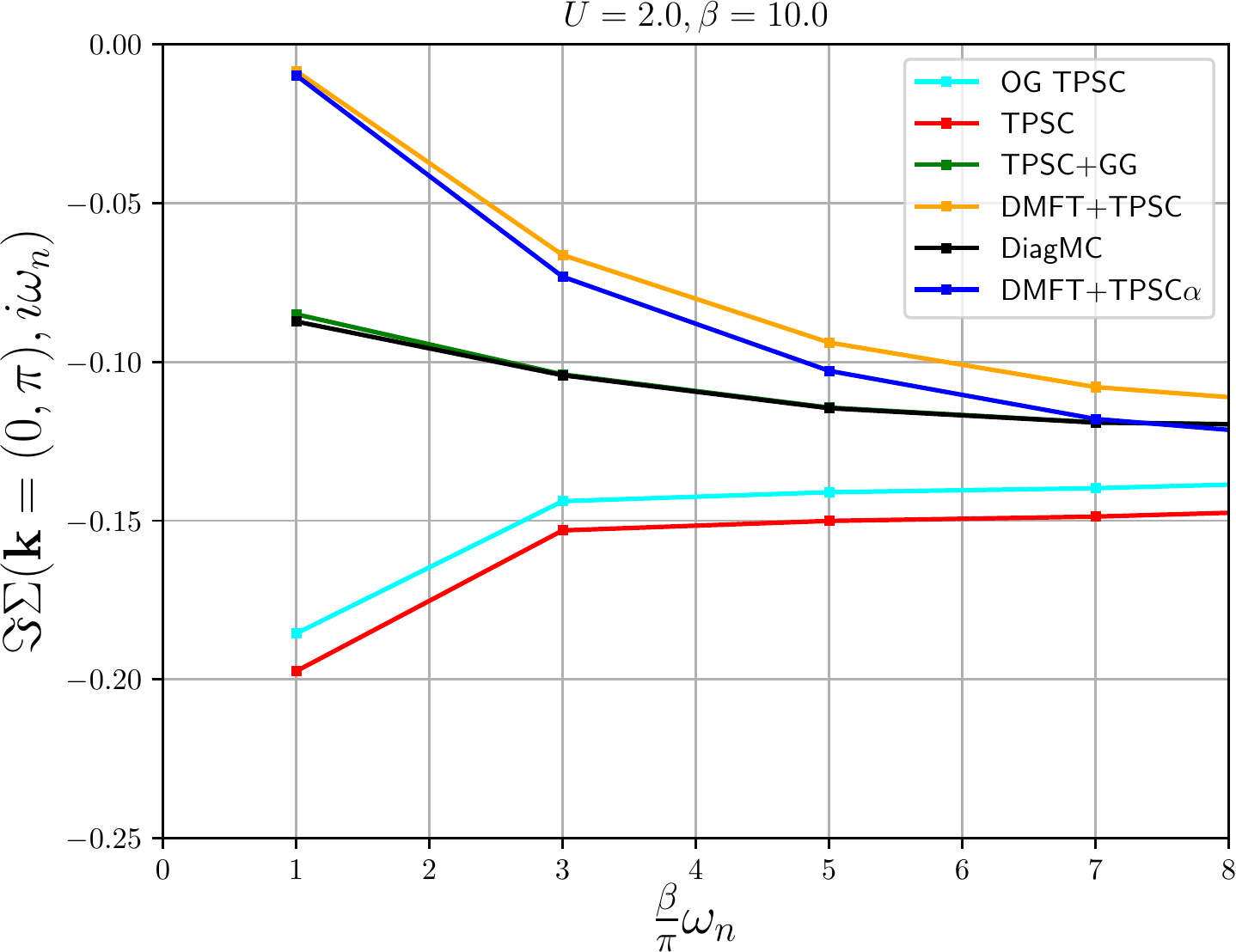}
\end{center}
\end{minipage}
\caption{Imaginary part of the Matsubara self-energy at the antinode ($\mathbf{k}=(0,\pi)$) for the half-filled Hubbard model at $U=2$. Results for $T=0.33$ (top subplot) and $T=0.1$ (bottom subplot) are shown for the various methods indicated in the legend. %The first several Matsubara frequencies are shown. 
This figure can be compared with the ``TPSC'' panel in Fig.~10 of Ref.~\onlinecite{PhysRevX.11.011058}.}
\label{fig:self_energy_antinode_matsubara}
\end{figure}

To understand how well the different methods capture nonlocal correlations, we first focus on the 2D square lattice Hubbard model. The first Matsubara frequencies of the self-energy at the antinode $\Sigma^{(1)}(\mathbf{k}=(0,\pi);i\omega_n)$ are plotted for $U=2$ in Fig.~\ref{fig:self_energy_antinode_matsubara} for the original TPSC formulation (OG TPSC), TPSC, TPSC+GG, DMFT+TPSC, DMFT+TPSC$\alpha$ and DiagMC. The TPSC and TPSC+GG schemes used here were introduced in Ref.~\onlinecite{https://doi.org/10.48550/arxiv.2205.13813}, while OG TPSC corresponds to the variant introduced in Ref.~\onlinecite{tpsc_1997}. The DiagMC results are taken from Ref.~\onlinecite{PhysRevX.11.011058}. The top subplot shows results for $T=0.33$ ($\beta=3$) and the bottom subplot for $T=0.1$ ($\beta=10$). As a reminder, we note that OG TPSC does not ensure consistency in the double occupancy between the first- and second-level TPSC approximations, \textit{i.e.} no $\alpha$ parameter is used. Comparing the results of Fig.~\ref{fig:self_energy_antinode_matsubara} with the ``TPSC'' panel in Fig.~10 of Ref.~\onlinecite{PhysRevX.11.011058}, which in our notation corresponds to OG TPSC, one can notice that TPSC+GG (green curves) improves the self-energy substantially so that it almost overlaps with the numerically exact result from the DiagMC method (black curves). DMFT+TPSC (orange curves) and DMFT+TPSC$\alpha$ also show a good agreement at $T=0.33$ with TPSC+GG and DiagMC. In the DMFT+TPSC schemes, the antinodal self-energy follows very closely that of TPSC+GG and DiagMC, except for the lowest Matsubara frequency, which reveals a too metallic behavior in this weak-coupling regime. The TPSC self-energy, on the other hand, systematically overestimates the self-energy (red curves). This result is rescaled, with respect to the result of OG TPSC (cyan curves), by the introduction of the parameter $\alpha$ (see Eq.~\eqref{eq:tpsc_self_energy_alpha}), which worsens the agreement with DiagMC. However, since TPSC+GG also uses the parameter $\alpha$ and agrees very well with DiagMC, the lack of self-consistency seems to be the main problem. At the lower temperature $T=0.1$, shown in the bottom panel of Fig.~\ref{fig:self_energy_antinode_matsubara}, TPSC+GG is clearly the most accurate of the TPSC variants, and again remarkably on top of the exact DiagMC result.  While DMFT+TPSC and DMFT+TPSC$\alpha$ underestimates the antinodal self-energy, it follows qualitatively the trend of the TPSC+GG and DiagMC results, while this is not the case for both TPSC and OG TPSC which bend in the opposite direction at lower Matsubara frequencies and hence overestimate the pseudogap tendency.  
Furthermore, the DMFT+TPSC schemes and TPSC+GG allow one to access lower temperature results by alleviating the convergence problems that limit the applicability of TPSC and OG TPSC in the vicinity of $T_x$ (crossover temperature to the renormalized classical regime). It is also worth mentioning that the non-self-consistent TPSC+DMFT scheme introduced in Ref.~\onlinecite{https://doi.org/10.48550/arxiv.2211.01919} matches the DiagMC data well, although less accurately than TPSC+GG.

As we will see in Sec.~\ref{sec:results:Nonequilibrium}, even though TPSC+GG looks most convincing in the benchmark of Fig.~\ref{fig:self_energy_antinode_matsubara}, this is not anymore the case out of equilibrium when evaluating local quantities such as the impurity double occupancy~\eqref{eq:impurity_double_occupancy}, 
although we lack exact benchmarks in this case. 
% \textcolor{blue}{[do we have exact benchmarks there, or does this statement somehow assume that DMFT is reliable? REP: This partly is based on the double occupancies calculated in Figs. 17 and 18.]}, and the same holds for the comparison between OG TPSC and TPSC. 

\subsubsection{Spin and charge vertices}

\begin{figure}[t]
  \centering
    \includegraphics[width=1.0\columnwidth]{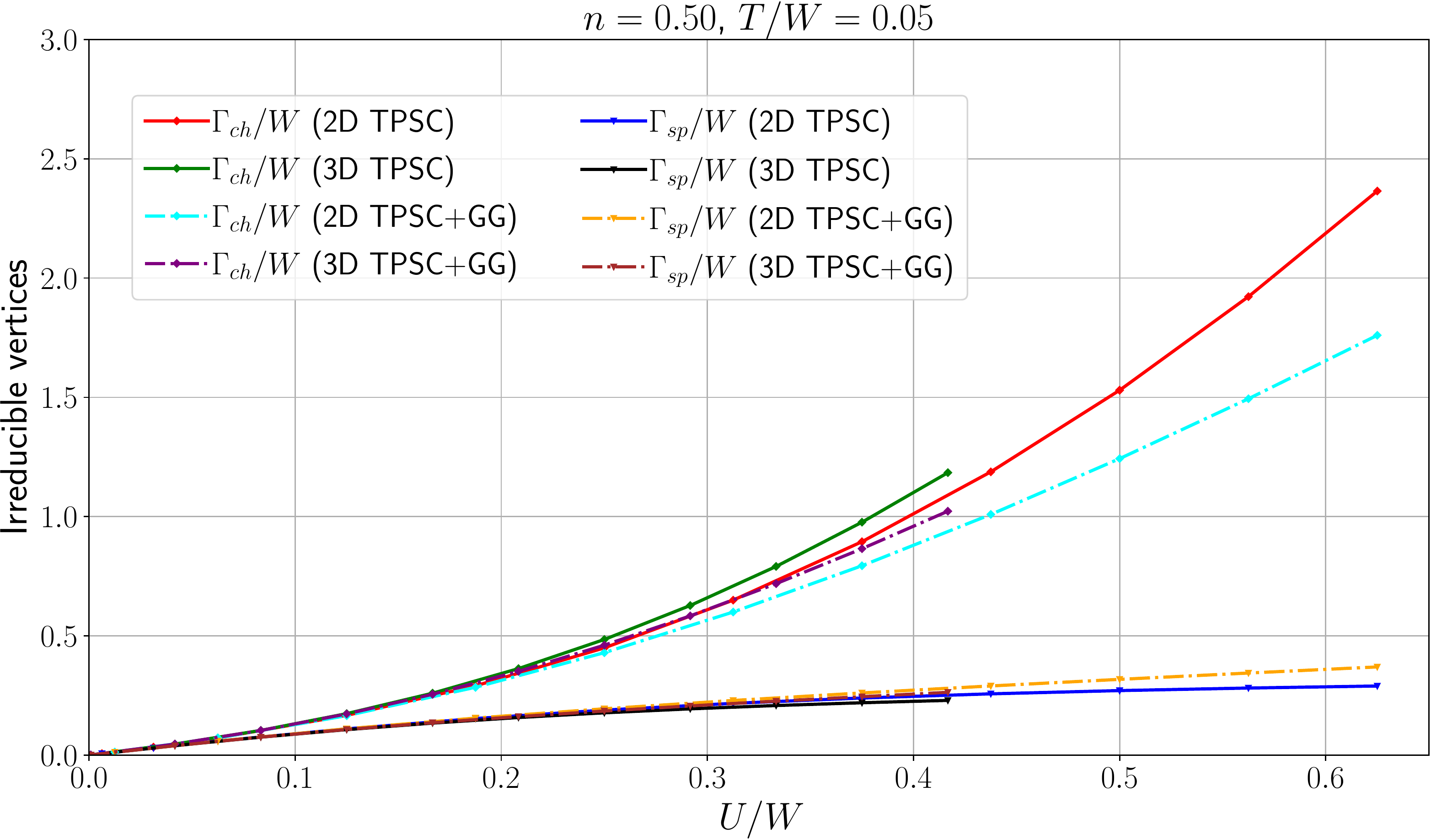}
      \caption{Bandwidth-renormalized spin and charge irreducible vertices as a function of normalized bare interaction for the nearest-neighbor square (2D) and cubic (3D) lattices for TPSC (bold lines) and TPSC+GG (dashed lines). The dimensionless temperature is $T/W=0.05$ and we consider half-filled systems. These data are taken from Ref.~\onlinecite{https://doi.org/10.48550/arxiv.2205.13813}.
}
  \label{fig:usp_uch_vs_u_n_0p5}
\end{figure}

TPSC gives access to self-consistently computed spin and charge vertices, which exhibit a distinct $U$ dependence. In 3D, the separation between the charge and spin vertices, renormalized by the bandwidth $W$, grows a bit faster with $U/W$ than in 2D, as shown in Fig.~\ref{fig:usp_uch_vs_u_n_0p5}.\cite{https://doi.org/10.48550/arxiv.2205.13813} The  distinction between $\Gamma^{\text{ch}}$ and $\Gamma^{\text{sp}}$ is more pronounced in TPSC compared to TPSC+GG, for both dimensions considered. Corresponding results without rescaling of the vertices and of the interaction by $W$ can be found in Ref.~\onlinecite{https://doi.org/10.48550/arxiv.2205.13813}. 

\begin{figure}[t]
  \centering
    \includegraphics[width=1.0\linewidth]{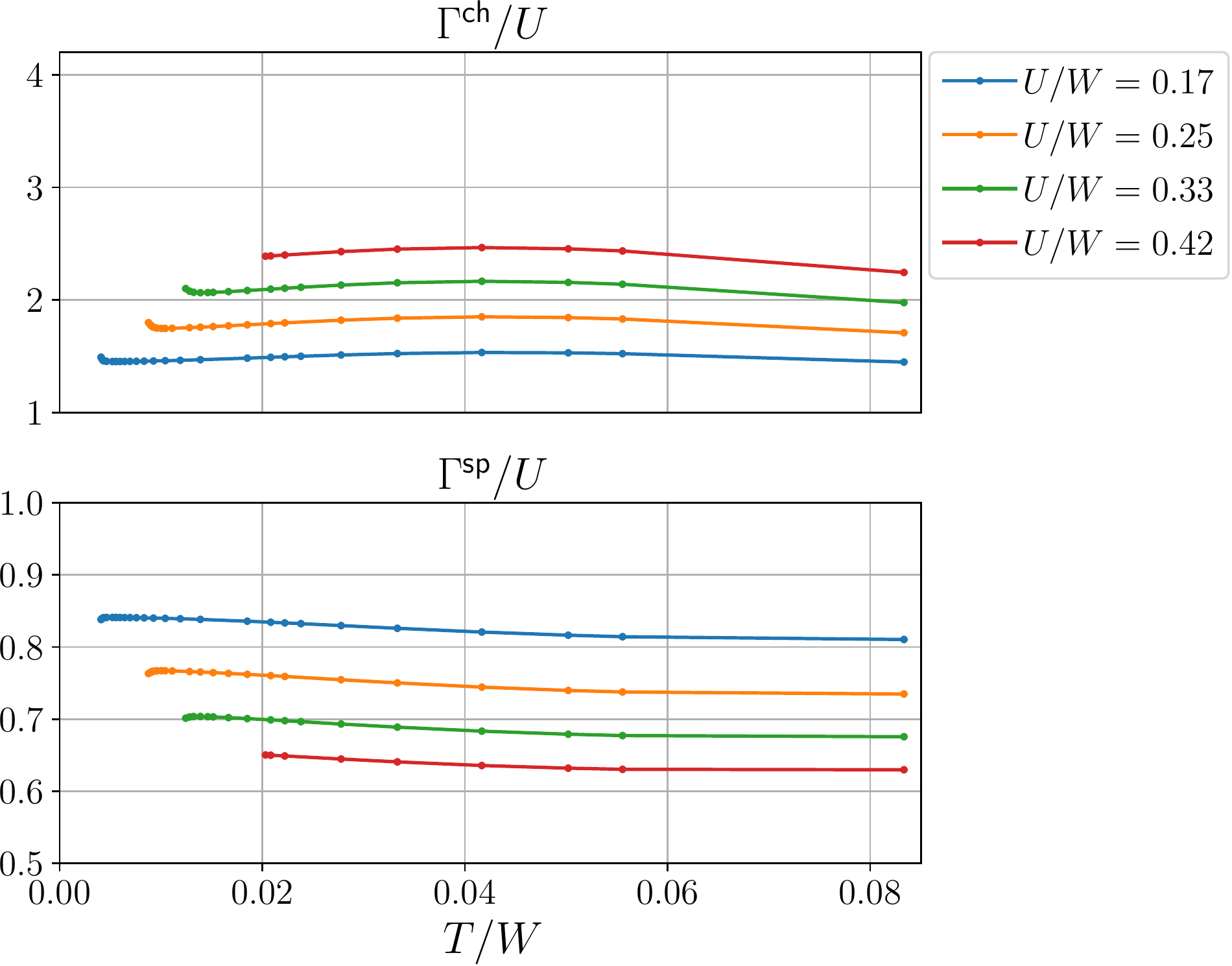}
      \caption{$\Gamma^{\text{ch}}$ (top panel) and $\Gamma^{\text{sp}}$ (bottom panel) as a function of $T/W$ for $U=2,3,4,5$ in the half-filled 3D Hubbard model, calculated with TPSC+GG. The values of the vertices are normalized by $U$ for presentation reasons.
      }
  \label{fig:Uch_Usp_vs_beta_3D_tpsc_GG}
\end{figure}

\begin{figure}[t]
  \centering
    \includegraphics[width=1.0\linewidth]{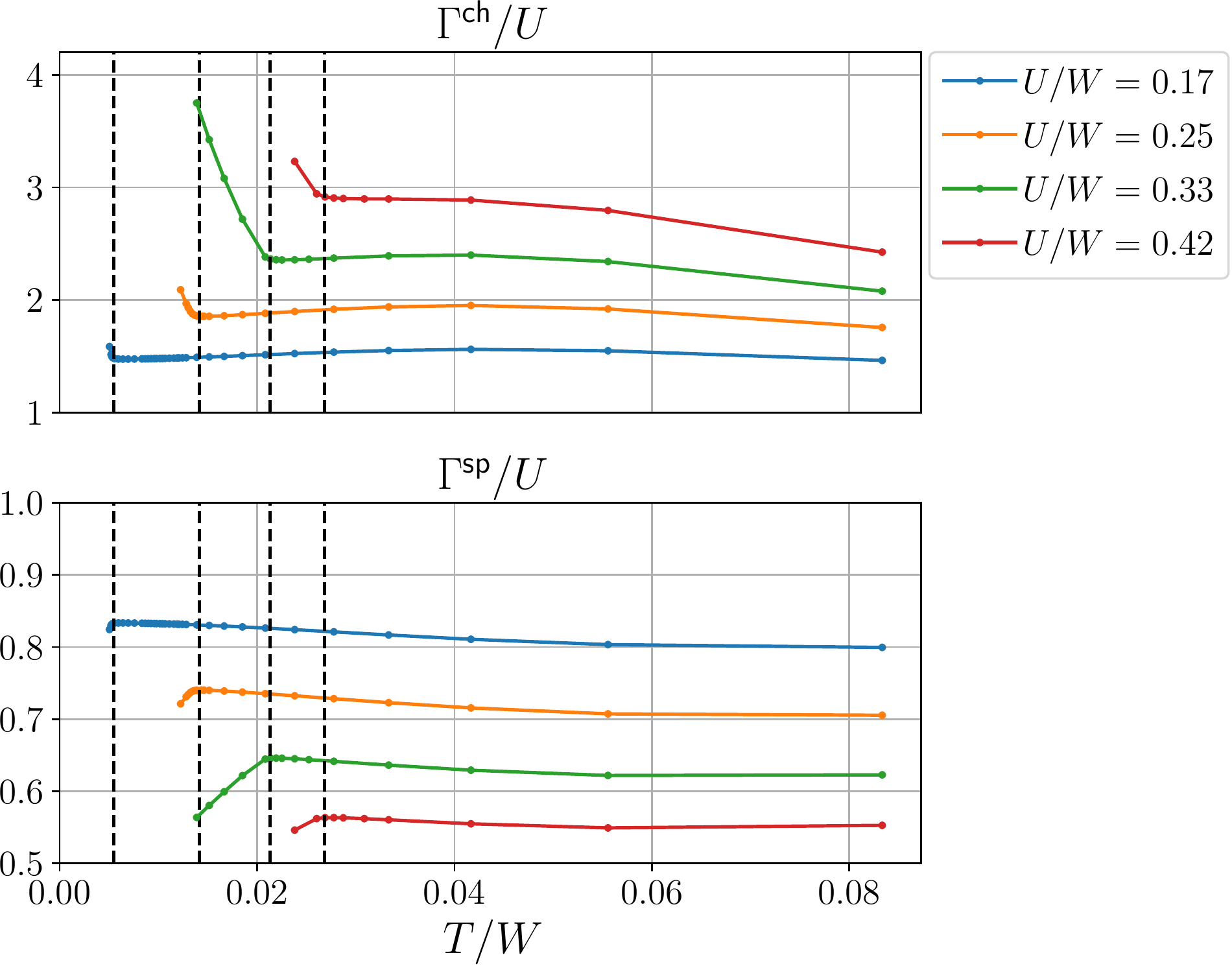}
      \caption{$\Gamma^{\text{ch}}$ (top panel) and $\Gamma^{\text{sp}}$ (bottom panel) as a function of $T/W$ for $U=2,3,4,5$ in the half-filled 3D Hubbard model, calculated with TPSC. The values of the vertices are normalized by $U$.
     }
  \label{fig:Uch_Usp_vs_beta_3D_tpsc}
\end{figure}

In Fig.~\ref{fig:Uch_Usp_vs_beta_3D_tpsc_GG}, the temperature dependence of the vertices calculated with TPSC+GG for various interaction strengths is plotted for the cubic lattice, while in Fig.~\ref{fig:Uch_Usp_vs_beta_3D_tpsc}, the TPSC results are shown for the same model parameters. These plots illustrate how the effective charge and spin interactions evolve when the renormalized classical regime is approached in the two methods. The vertical dotted lines in Fig.~\ref{fig:Uch_Usp_vs_beta_3D_tpsc} indicate the temperatures where $\Gamma^{\text{sp}}$ bends down and these temperatures will be later linked to a sharp upturn in the static spin susceptibility.\footnote{These vertical lines roughly mark the temperatures below which the temperature dependence of the spin vertex is no longer linear, \textit{i.e.}, starts deviating from a straight line that fits the neighboring points at higher $T$.} There is no significant $T$ dependence of the spin and charge vertices in TPSC+GG at intermediate temperatures. In TPSC+GG only a hint of an upturn in $\Gamma^{\text{ch}}$ can be resolved near $T_x$, due to a convergence slowdown at low temperatures, while in the case of TPSC a much more pronounced up-turn can be observed. The sharp downturn of $\Gamma^{\text{sp}}$ close to the renormalized classical regime Fig.~\ref{fig:Uch_Usp_vs_beta_3D_tpsc} is due to the suppression of the double occupancy extracted from the ansatz~\eqref{eq:equivalence_spin_irr_vertex_double_occupancy}. 

\begin{figure}[t]
  \centering
    \includegraphics[width=1.0\linewidth]{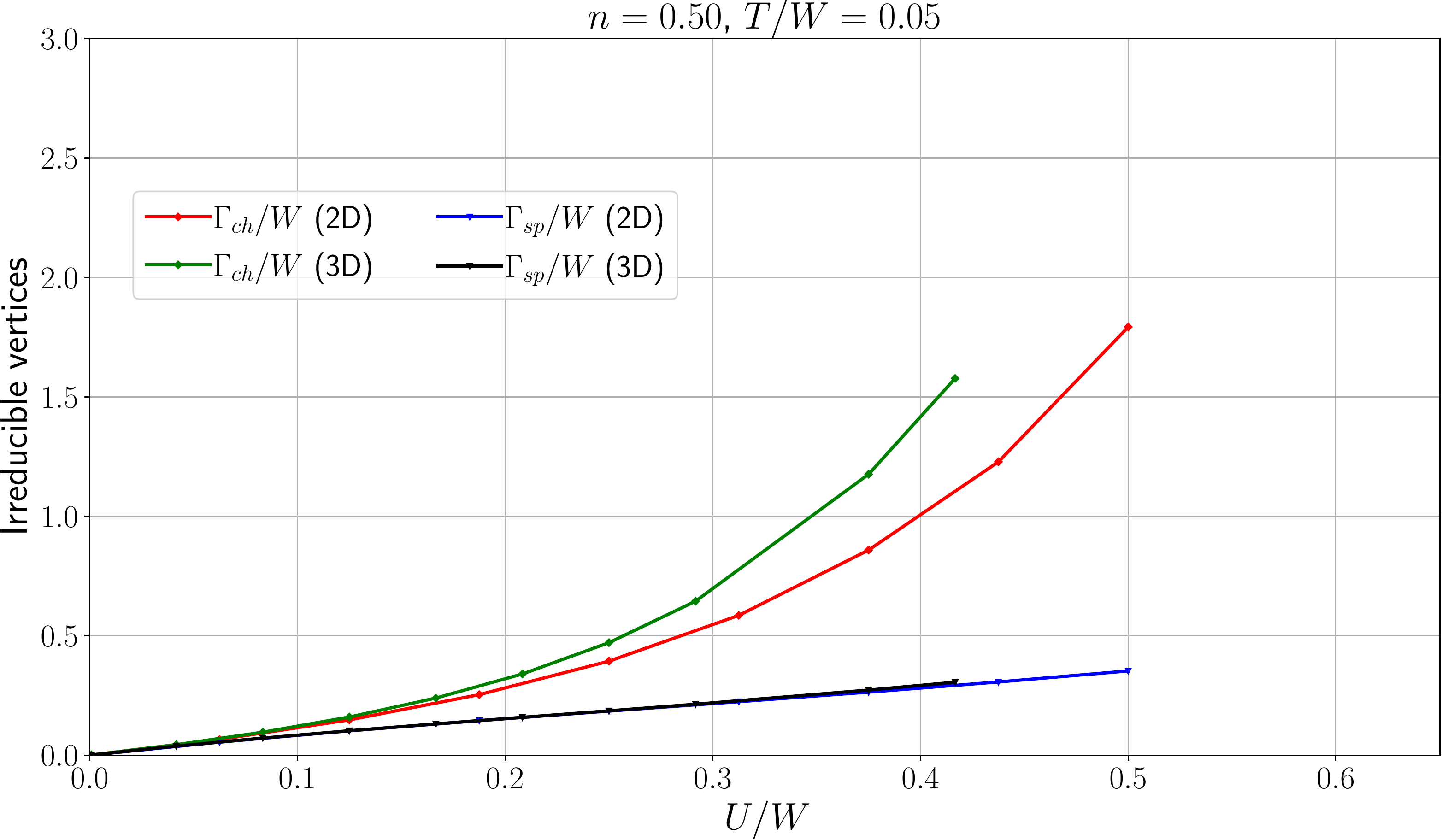}
      \caption{Dimensionless spin and charge irreducible vertices as a function of normalized bare interaction for the square and cubic lattices, calculated with DMFT+TPSC. The dimensionless temperature is $T/W=0.05$ and the systems are half-filled.
}
  \label{fig:usp_uch_vs_u_n_0p5_dmft_tpsc}
\end{figure}

The local two-particle irreducible spin and charge vertices can also be computed within DMFT+TPSC. Throughout this work, the weak-coupling impurity solver introduced in Sec.~\ref{subsubsec:IPT} is used to treat the local impurity interactions. At half-filling, the second-order IPT self-energy is used, unless mentioned otherwise, in which case the self-energy diagrams up to the third-order are considered. In Fig.~\ref{fig:usp_uch_vs_u_n_0p5_dmft_tpsc}, the irreducible vertices are plotted as a function of the normalized bare interaction parameter $U/W$ at normalized temperature $T/W=0.05$. These can be compared with the TPSC and TPSC+GG results for 2D and 3D in Fig.~\ref{fig:usp_uch_vs_u_n_0p5}, which are very similar. $\Gamma^{\text{ch}}$ and $\Gamma^{\text{sp}}$ drift apart with increasing $U/W$, and as mentioned before this is more pronounced in 3D than in 2D. In DMFT+TPSC, both $\Gamma^{\text{ch}}$ and $\Gamma^{\text{sp}}$ have larger values than in TPSC or TPSC+GG at a given $U/W$. Because the IPT impurity solver is reliable only in the weak-coupling regime, the range of interactions shown is limited to $U/W=0.5$.

\begin{figure}[t]
  \centering
    \includegraphics[width=1.0\linewidth]{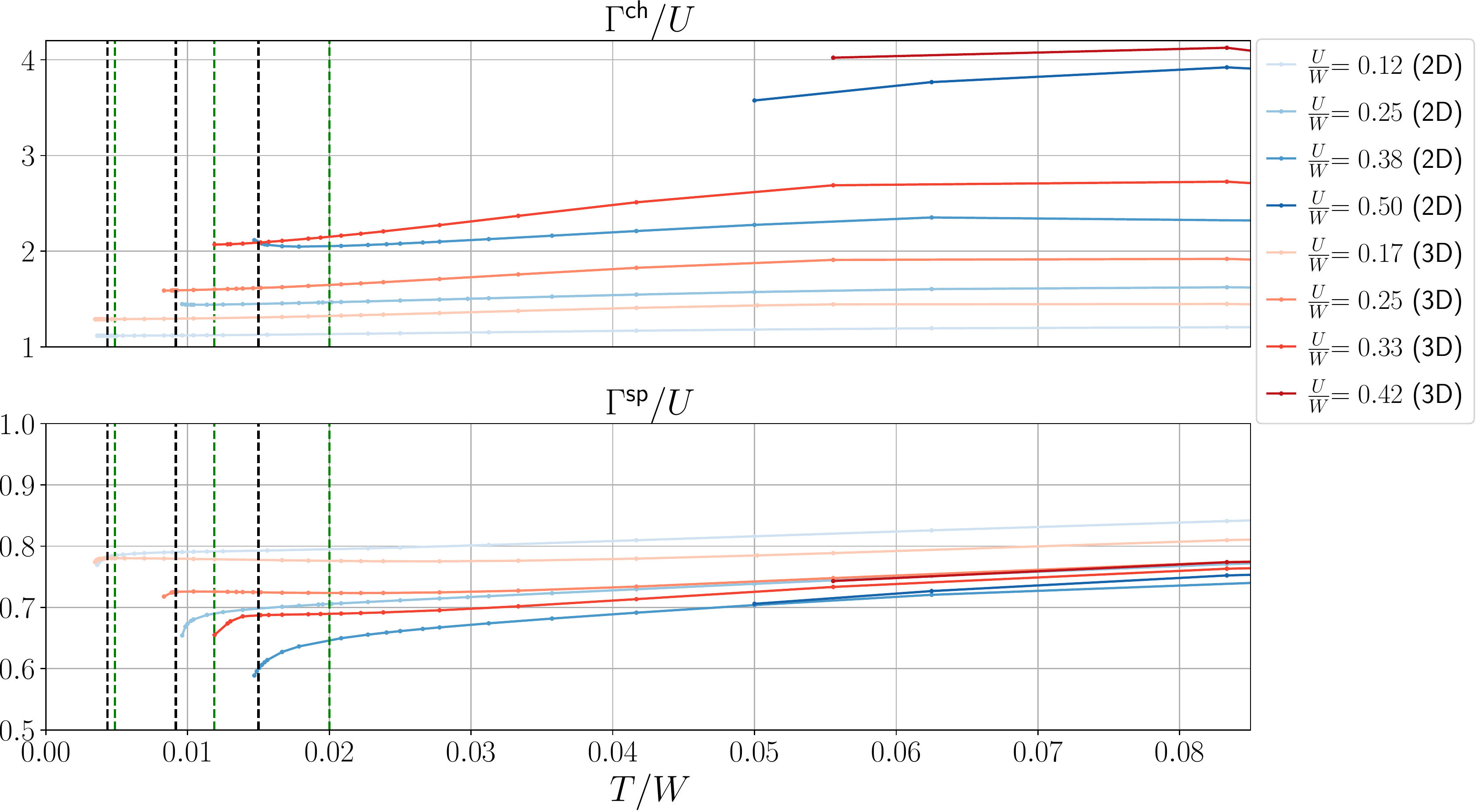}
      \caption{$\Gamma^{\text{ch}}$ (top panel) and $\Gamma^{\text{sp}}$ (bottom panel) as a function of $T$ for $U=2,3,4,5$ in the 3D half-filled nearest-neighbor Hubbard model. The values of the vertices are normalized by $U$ and were obtained using DMFT+TPSC. At lower temperatures for $U=5$, the DMFT solution could not be converged.
      %\textcolor{blue}{[how did you choose the dashed lines? they don't seem to match the "sharp" downturn.] REP: I have stressed in the blue sentence above that the vertical lines pin down the temperatures where the T-dependence of the spin vertex is no longer linear, i.e. it deviates from a straight line. Maybe I should adjust the ``amount of deviation'' to a larger number?}
      }
  \label{fig:Uch_Usp_vs_beta_3D_dmft_tpsc}
\end{figure} 

The DMFT+TPSC irreducible vertices $\Gamma^{\text{ch}}$ (top panel) and $\Gamma^{\text{sp}}$ (bottom panel) are plotted in Fig.~\ref{fig:Uch_Usp_vs_beta_3D_dmft_tpsc} as a function of temperature for the half-filled 3D Hubbard model with $U=\{2,3,4,5\}$. For a better comparison with the TPSC+GG and TPSC results, we use here the same $y$ axis range as in Figs.~\ref{fig:Uch_Usp_vs_beta_3D_tpsc_GG} and \ref{fig:Uch_Usp_vs_beta_3D_tpsc}. Again, the vertical lines in Fig.~\ref{fig:Uch_Usp_vs_beta_3D_dmft_tpsc} indicate the temperatures where $\Gamma^{\text{sp}}$ bends down, and these will be related to an upturn in the static spin susceptibility. Contrary to the TPSC+GG and TPSC temperature dependence of $\Gamma^{\text{ch}}$, the charge vertex gets significantly reduced as temperature is lowered, but it starts from higher values at high $T$. On the other hand, $\Gamma^{\text{sp}}$ almost saturates at lower temperatures in DMFT+TPSC, and then sharply drops in the renormalized classical regime near $T_x$. In contrast to TPSC, the rapid decrease of the spin irreducible vertex $\Gamma^{\text{sp}}$ (concomitant with a drop in the double occupancy) in DMFT+TPSC does not coincide with a shooting up of $\Gamma^{\text{ch}}$ (compare Figs.~\ref{fig:Uch_Usp_vs_beta_3D_tpsc} and \ref{fig:Uch_Usp_vs_beta_3D_dmft_tpsc}).

\subsubsection{Spin susceptibility}

\begin{figure}[t]
\begin{minipage}[h]{1.0\columnwidth}
\begin{center}
    \includegraphics[width=1.0\columnwidth]{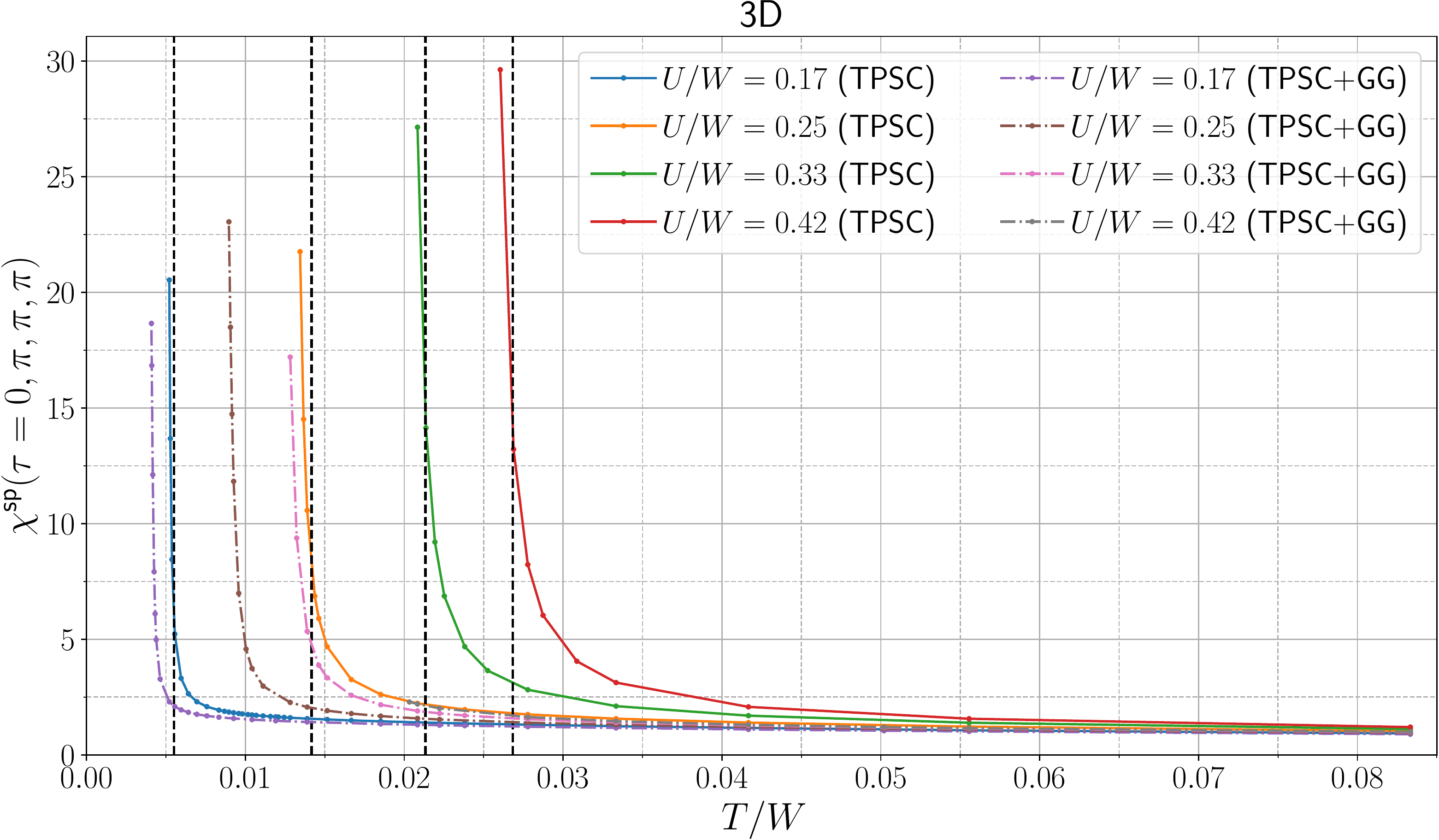}
\end{center}
\end{minipage}
\hfill
\vspace{0.1cm}
\begin{minipage}[h]{1.0\columnwidth}
\begin{center}
   \includegraphics[width=1.0\linewidth]{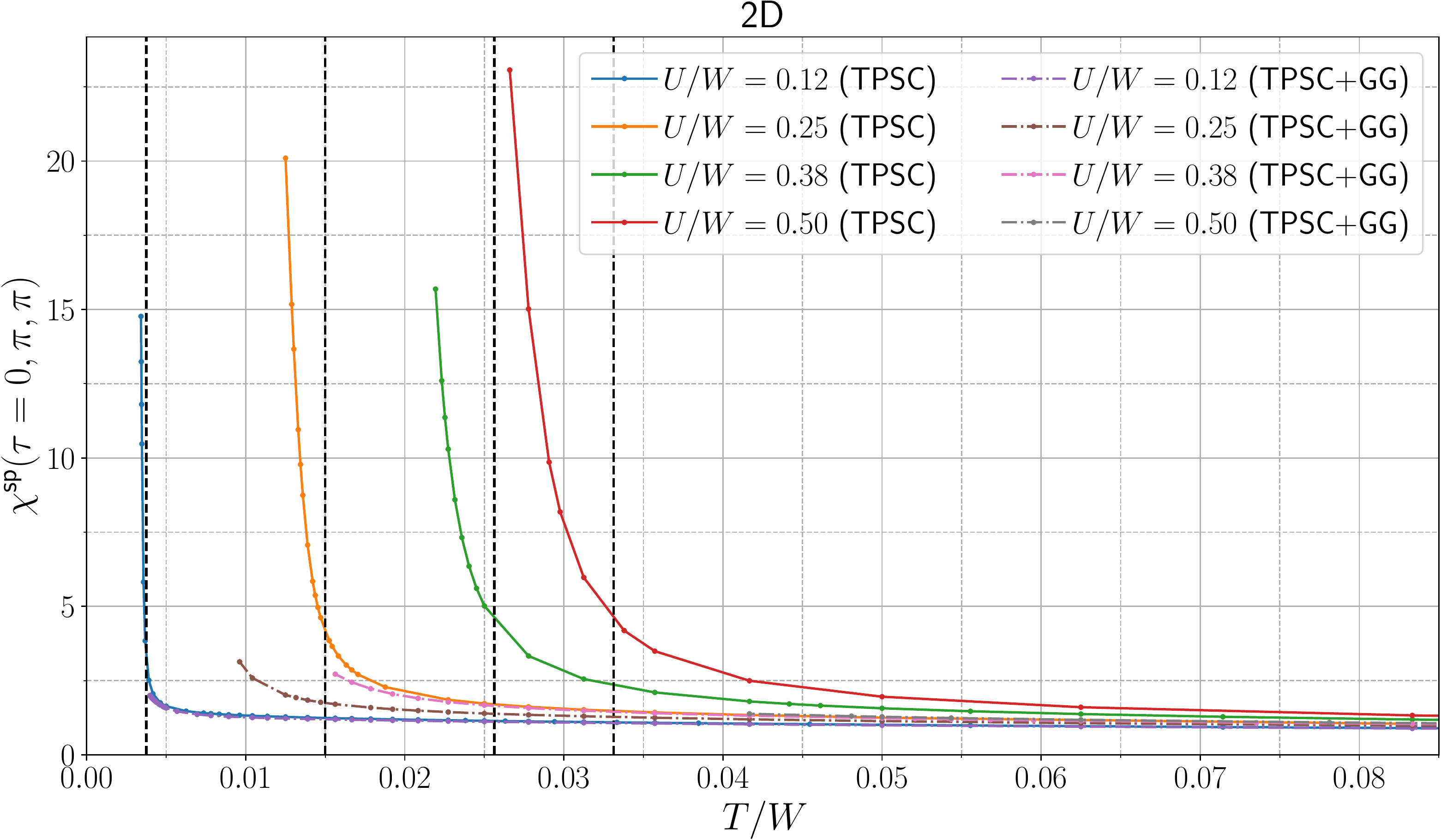}
\end{center}
\end{minipage}
\caption{Static spin susceptibility of the 3D (top subplot) and 2D (bottom subplot) models at momentum $\mathbf{k}_{\pi}$ as a function of temperature for the interactions $U=2,3,4,5$ and half-filling. Results are shown for TPSC (bold lines) and TPSC+GG (dashed lines). 
 The vertical lines coincide with those in Fig.~\ref{fig:Uch_Usp_vs_beta_3D_tpsc}.
 }
\label{fig:static_sus_vs_T_TPSC}
\end{figure}

In Fig.~\ref{fig:static_sus_vs_T_TPSC}, the static spin susceptibility at half-filling is plotted for both TPSC and TPSC+GG in 2D (bottom subplot) and 3D (top subplot). It shows the growth of the static spin correlations as temperature is lowered. The up-turn in $\chi^{\text{sp}}(\tau=0,\mathbf{k}_{\pi})$ marks the temperature crossover $T_x$ to the renormalized classical regime. Increasing the interaction $U$ displaces the up-turn to higher temperatures in both TPSC and TPSC+GG. However, in TPSC+GG, for the same interaction value, the estimated crossover temperature $T_x$ is consistently lower than that extracted from the TPSC static susceptibility. In 3D, the shooting-up of the static spin susceptibility at $\mathbf{k}_{\pi}$ at low temperature in TPSC coincides with the up-turn of $\Gamma^{\text{ch}}$, \textit{cf.} Figs.~\ref{fig:Uch_Usp_vs_beta_3D_tpsc} and \ref{fig:static_sus_vs_T_TPSC} (top subplot), as becomes clear from the vertical dashed lines which are at the same temperatures in both figures. 

In the bottom subplot of Fig.~\ref{fig:static_sus_vs_T_TPSC}, the static spin susceptibility $\chi^{\text{sp}}(\tau=0,\mathbf{k}_{\pi})$ is plotted for the 2D model. For equal interaction strengths $U$ (without normalizing $U$ by $W$), the up-turns in the static susceptibility happen at slightly lower temperatures when increasing the dimension, except at $U\geq 4$. As a consequence, a larger temperature range is accessible in 3D compared to 2D at weak coupling, since $T_x$ is lowered in 3D. In 3D, the TPSC+GG results of the static susceptibility are qualitatively more similar to TPSC than it is the case in 2D, where only the beginning of the up-turn is numerically accessible.~\cite{https://doi.org/10.48550/arxiv.2205.13813} This might be an indication that TPSC is more accurate in 3D.

\begin{figure}[t]
  \centering
    \includegraphics[width=1.0\linewidth]{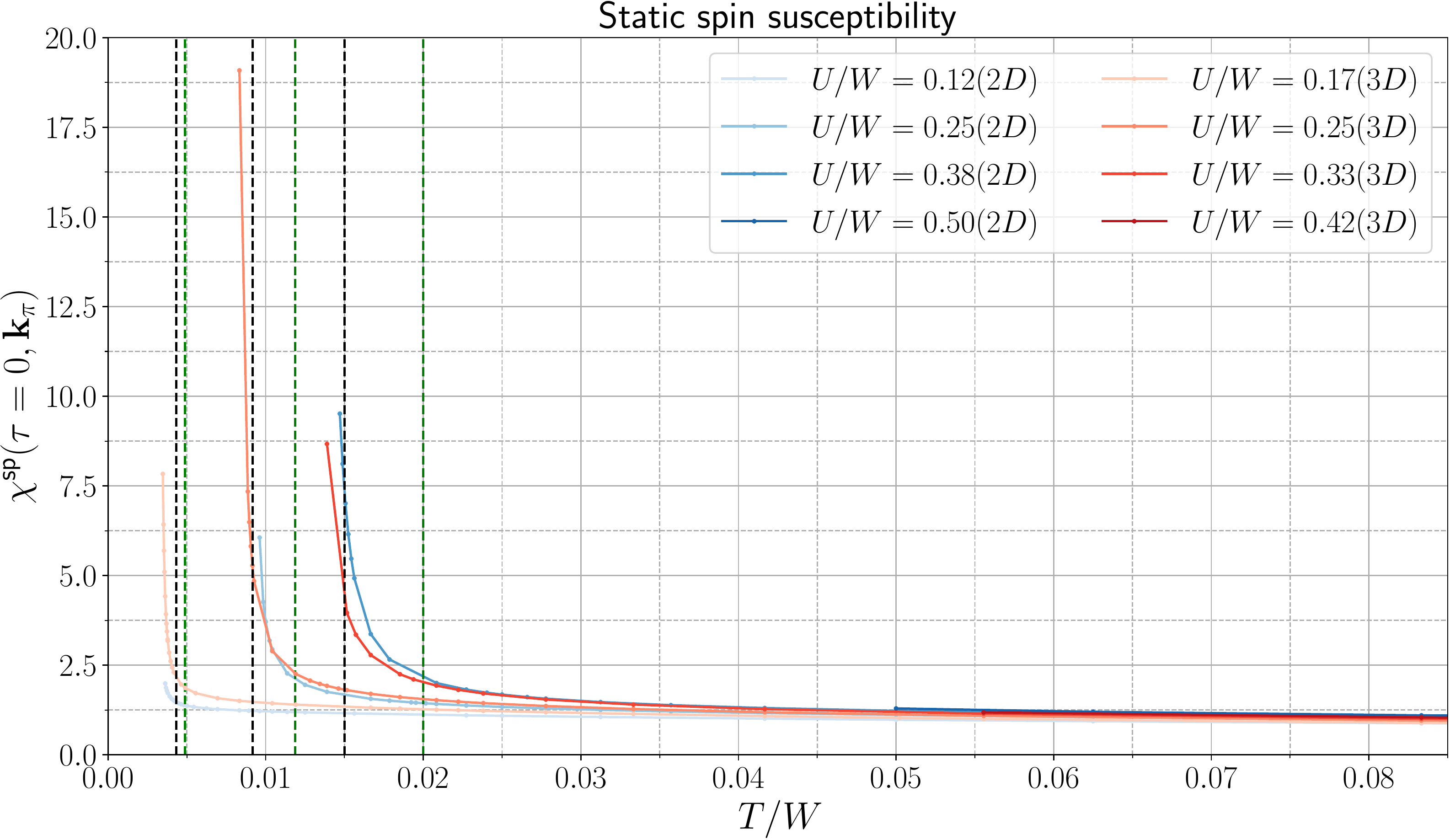}
      \caption{Static spin susceptibility of the half-filled 2D and 3D model at momentum $\mathbf{k}_{\pi}$ as a function of temperature for interactions $U=1,2,3,4$ (2D) and $U=2,3,4,5$ (3D), obtained with DMFT+TPSC. The vertical lines coincide with those in Fig.~\ref{fig:Uch_Usp_vs_beta_3D_dmft_tpsc}.
      }
  \label{fig:static_spin_sus_fermi_surface_dmft_tpsc_3D}
\end{figure}

To demonstrate that DMFT+TPSC still captures the growth of the AFM correlations with decreasing temperature at various interactions, the 2D and 3D static spin susceptibilities are plotted for DMFT+TPSC in Fig.~\ref{fig:static_spin_sus_fermi_surface_dmft_tpsc_3D}. These results can be compared directly to Fig.~\ref{fig:static_sus_vs_T_TPSC} for TPSC and TPSC+GG. It is obvious that the same qualitative behavior of the static spin response is observed also in the presence of the DMFT correction: with increasing interaction strength, the up-turn in the static spin susceptibility is shifted to higher temperatures. Furthermore, the relative change in the $T$ value of the up-turns increases as $U$ is decreased. 
(Remember that since TPSC and its variants make use of the spin rotational symmetry in the derivation, these methods can only describe the growth of spin correlations, but not the spontaneous symmetry-breaking.)
Similarly to TPSC+GG, the up-turns at fixed $U$ in DMFT+TPSC occur at lower temperatures when compared to TPSC. 

\begin{figure}[h!]
  \centering
    \includegraphics[width=1.0\linewidth]{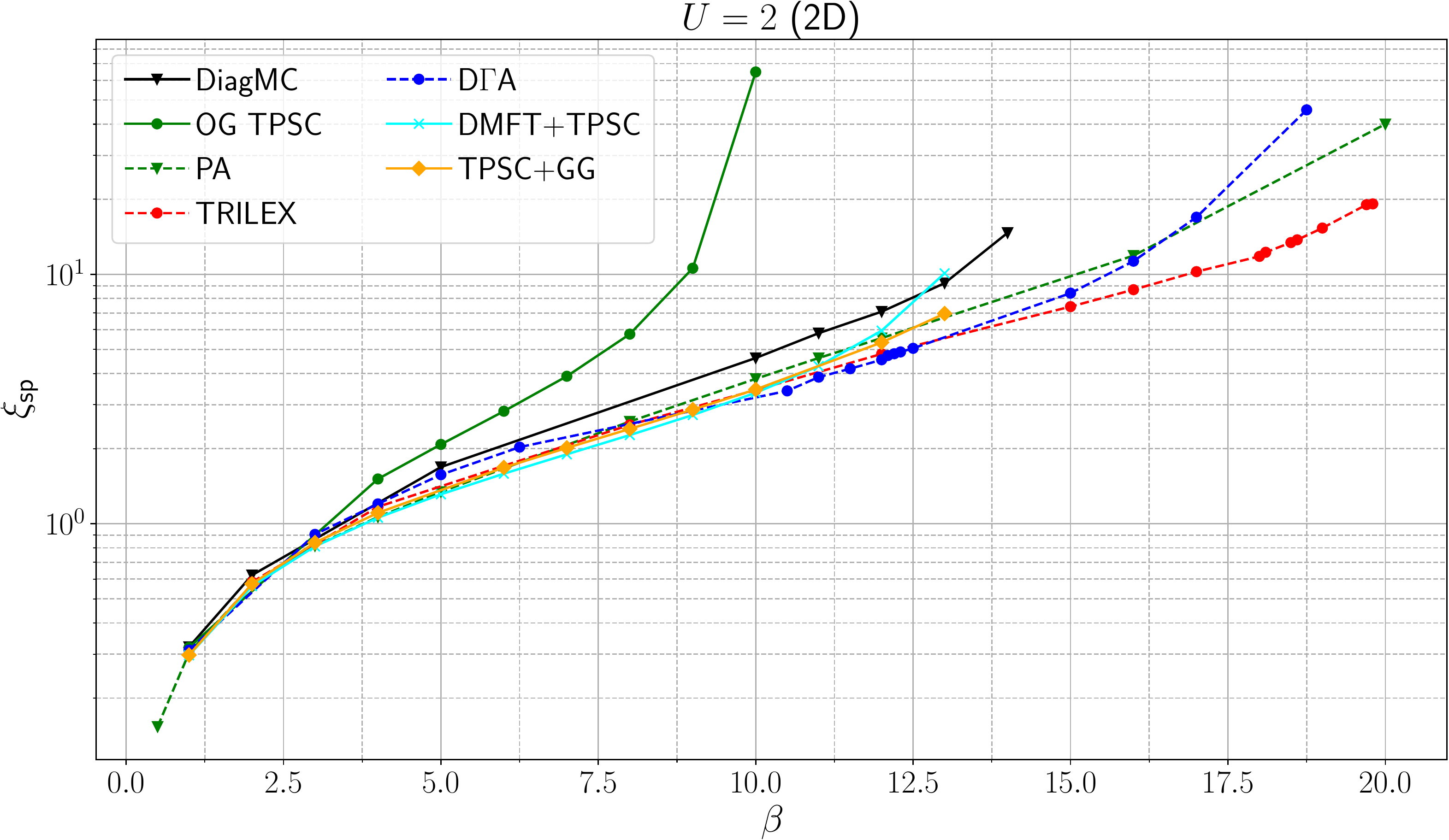}
      \caption{$\xi_{{\text{sp}}}$ as a function of $\beta=1/T$ for $U=2$ in the half-filled 2D Hubbard model. The $y$-axis uses a logarithmic scale. The methods compared are ``OG TPSC'' (green circles, called TPSC in Refs.~\onlinecite{PhysRevX.11.011058,https://doi.org/10.48550/arxiv.2211.01919}), TPSC+GG (orange diamonds), DMFT+TPSC (cyan crosses), D$\Gamma$A (blue circles), DiagMC (black triangles), TRILEX (red circles) and PA (green triangles). The data calculated using TRILEX, DiagMC, OG TPSC, D$\Gamma$A and PA were taken from Ref.~\onlinecite{PhysRevX.11.011058}. The $3^{\text{rd}}$-order IPT impurity solver is used in DMFT+TPSC (see Sec.~\ref{ch:3rd_order_IPT}).}
  \label{fig:corr_len_comparison_eq}
\end{figure}

A different way of quantifying the growth of the spin correlations is to plot the antiferromagnetic correlation length $\xi_{\text{sp}}$ as a function of inverse temperature. In Fig.~\ref{fig:corr_len_comparison_eq}, $\xi_{\text{sp}}$ is shown for the half-filled 2D square lattice Hubbard model at constant interaction $U=2$. Several methods are compared against each other, namely OG TPSC, TPSC+GG, DMFT+TPSC, D$\Gamma$A,\cite{PhysRevB.75.045118_dynamical_vertex_approx} DiagMC,\cite{PhysRevLett.81.2514,Kozik_2010} TRILEX~\cite{PhysRevB.92.115109,PhysRevB.93.235124} and the Parquet Approximation (PA).\cite{doi:10.1063/1.1704062,doi:10.1063/1.1704064} The correlation length $\xi_{\text{sp}}$ is extracted from the Ornstein-Zernicke fit of the momentum-dependent static spin susceptibility $\chi^{\text{sp}}_{\mathbf{q}-\mathbf{Q}}(iq_{n}=0)$ in the vicinity of the AFM scattering wave vector $\mathbf{Q}$:
\begin{align*}
%\label{eq:ornstein_zernicke_spectral_fit}
\chi^{\text{sp}}_{\mathbf{q}-\mathbf{Q}}(iq_{n}=0) \approx \frac{A}{(\mathbf{q}-\mathbf{Q})^{2} + \xi_{\text{sp}}^{-2}},
\end{align*}
where $\mathbf{Q}=\mathbf{k}_{\pi}$ ($\mathbf{k}_{\pi}=(\pi,\pi)$ in 2D) at half-filling and $A$ is some weight of the order of $1$. It is clear from Fig.~\ref{fig:corr_len_comparison_eq} that the original formulation of TPSC (OG TPSC) overestimates the growth of spin correlations as the temperature is decreased, \textit{i.e.}, $T_{x}$ is much higher than the values estimated by the other more accurate methods. The latter predict similar correlation lengths in the temperature range up to $\beta\simeq 12$. In particular, both TPSC+GG and DMFT+TPSC follow very closely the $\xi_{\text{sp}}$ results obtained from TRILEX, PA and D$\Gamma$A. Thus, TPSC+GG and DMFT+TPSC both correct the overestimation of the spin correlations of OG TPSC and this is reflected also in the antinodal self-energy at the Fermi surface, where TPSC+GG and DMFT+TPSC agree quite well with DiagMC, especially in the case of TPSC+GG (Fig.~\ref{fig:self_energy_antinode_matsubara}).

\subsubsection{Double occupancy}

\begin{figure}[h!]
  \centering
    \includegraphics[width=1.0\linewidth]{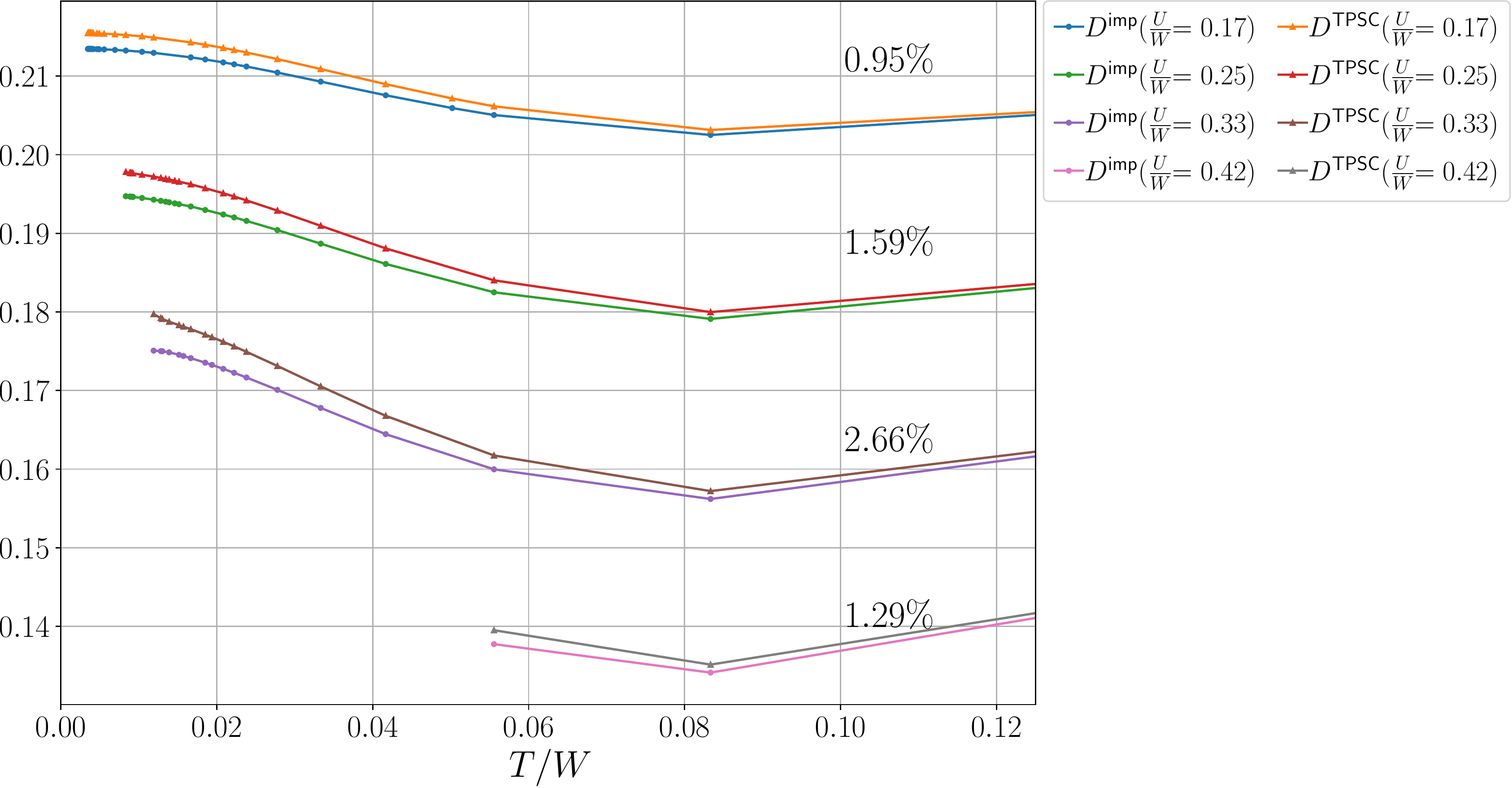}
      \caption{Double occupancies $D^{\text{imp}}$ (Eq.~\eqref{eq:impurity_double_occupancy}) and $D^{\text{TPSC}}$ (Eq.~\eqref{eq:tr_sk_Gk_TPSC}) as a function of temperature for several interactions $U$ in the half-filled 3D Hubbard model. The annotated percentages denote the largest absolute variation relative to $D^{\text{imp}}$.}
  \label{fig:Ds_dmft_tpsc_3D}
\end{figure}

In DMFT+TPSC, there are local Green's functions and self-energies of the auxiliary Anderson impurity model, \textit{i.e.} $\mathcal{G}^{\text{imp}}$, $\Sigma^{\text{imp}}$, and corresponding functions defined on the lattice, \textit{i.e.} $\mathcal{G}^{\text{TPSC}}$, $\Sigma^{\text{TPSC}}$. With these quantities, we can calculate a double occupancy for the impurity $D^{\text{imp}}$ via Eq.~\eqref{eq:impurity_double_occupancy} and a double occupancy on the lattice $D^{\text{TPSC}}$ via Eq.~\eqref{eq:tr_sk_Gk_TPSC}. 
In Fig.~\ref{fig:Ds_dmft_tpsc_3D} we plot both estimates for the 3D model. The lower the temperature and the larger the interaction, the larger the deviation between $D^{\text{imp}}$ and $D^{\text{TPSC}}$ becomes. The largest deviation for each interaction is displayed in the figure as an absolute relative percentage with respect to $D^{\text{imp}}$. Overall, the deviations are quite small (below $3\%$). The deviations are larger in the 2D model, but the same qualitative trend in $U$ and $T$ is observed (not shown). At larger temperature the double occupancies flex upwards since they approach $D=0.25$ as $T\to\infty$ at half-filling.

\subsection{Nonequilibrium}
\label{sec:results:Nonequilibrium}

\subsubsection{General remarks}

We now switch to the real-time dynamics of perturbed correlated lattice systems, as described by the different TPSC variants. In Fig.~\ref{fig:self_energy_antinode_matsubara}, it was shown by comparing to DiagMC that the equilibrium self-energy at the antinodal point of the Fermi surface calculated with TPSC+GG and DMFT+TPSC was improved substantially, compared to TPSC, especially at higher temperatures. One might thus naively expect that these two methods also provide the best description of the nonequilibrium dynamics. However, as shown below, the incorporation of the DMFT local self-energy has substantial effects on the time evolution and cures some anomalies of our (approximate) TPSC+GG implementation.

\subsubsection{Interaction ramps}

We first investigate the double occupancy following an interaction ramp from $U=0\to 1$ in the 2D Hubbard model at half-filling, which is the most challenging filling for TPSC.~\cite{tpsc_1997,https://doi.org/10.48550/arxiv.2211.01919} Besides the various TPSC-based methods, we consider second-order lattice perturbation theory, $\Sigma^{(2)}$,~\cite{Tsuji2014} which employs the self-energy
\begin{align*}
%\label{eq:lattice_ipt_compared_with_tpsc}
&\Sigma^{(2)}_{\mathbf{k},\sigma}(z_1,z_2) = U(z_1)U(z_2)\int\frac{\mathrm{d}^Dq\mathrm{d}^Dk^{\prime}}{(2\pi)^{2D}}\notag\\
&\hspace{0.0cm}\times\mathcal{G}^0_{\mathbf{k}+\mathbf{q},\sigma}(z_1,z_2)\mathcal{G}^0_{\mathbf{k}^{\prime}+\mathbf{q},-\sigma}(z_2,z_1^+)\mathcal{G}_{\mathbf{k}^{\prime}+\mathbf{q},-\sigma}^0(z_1,{z_2}^+)
\end{align*}
in the lattice Dyson equation~\eqref{eq:lattice_G_definition_improved}. This scheme should provide useful reference data in the weak-coupling regime $U\ll W$. OG TPSC refers to the original formulation of TPSC that utilizes the self-energy $\Sigma_{\mathbf{k}}\to \Sigma^{(1),\text{TPSC}}_{\mathbf{k}}$~\eqref{eq:fft_total_self_energy}. In the case of TPSC+GG, the self-energy $\Sigma_{\mathbf{k}}$ used is laid out in Eq.~\eqref{eq:tpsc_self_energy_alpha}. DMFT employs the third-order IPT as impurity solver~(see Sec.~\ref{subsubsec:IPT}), so that the local self-energy becomes $\Sigma^{(3)}_{\text{imp}}$, while DMFT+TPSC uses the momentum-dependent $\Sigma_{\mathbf{k}}$ defined in Eq.~\eqref{eq:sigma_DMFT_TPSC}. We remind the reader that OG TPSC does not enforce the sum-rule~\eqref{eq:double_occupancy_sum_rule_alpha}, \textit{i.e.} it does not include the time-dependent parameter $\alpha$ that forces the double occupancy calculated from the TPSC ansatz Eq.~\eqref{eq:equivalence_spin_irr_vertex_double_occupancy} to be the same as that computed from the trace over lattice TPSC quantities~\eqref{eq:double_occupancy_sum_rule_alpha}.

In this paper, the interaction ramp $\Delta U$ is described by the error function 
\begin{align}
\label{eq:ramp_profile_function_erf}
\Delta U(t) = \pm\left(\frac{U_{\text{f}}-U_{\text{i}}}{2}\right)\text{erf}(\gamma t + \delta) + \left(\frac{U_{\text{f}}+U_{\text{i}}}{2}\right),
\end{align}
where $U_{\text{i}}$ corresponds to the initial interaction value and $U_{\text{f}}$ to the final one, $\gamma$ controls the steepness of the inflection of the curve and $\delta$ its position on the time axis. A global minus sign appears in Eq.~\eqref{eq:ramp_profile_function_erf} in the case of a down ramp ($U_f<U_i$). The same form is also used for the lattice hopping ramps ($U\to t^{\text{hop}}_z$).

\begin{figure}[t]
\includegraphics[width=\linewidth]{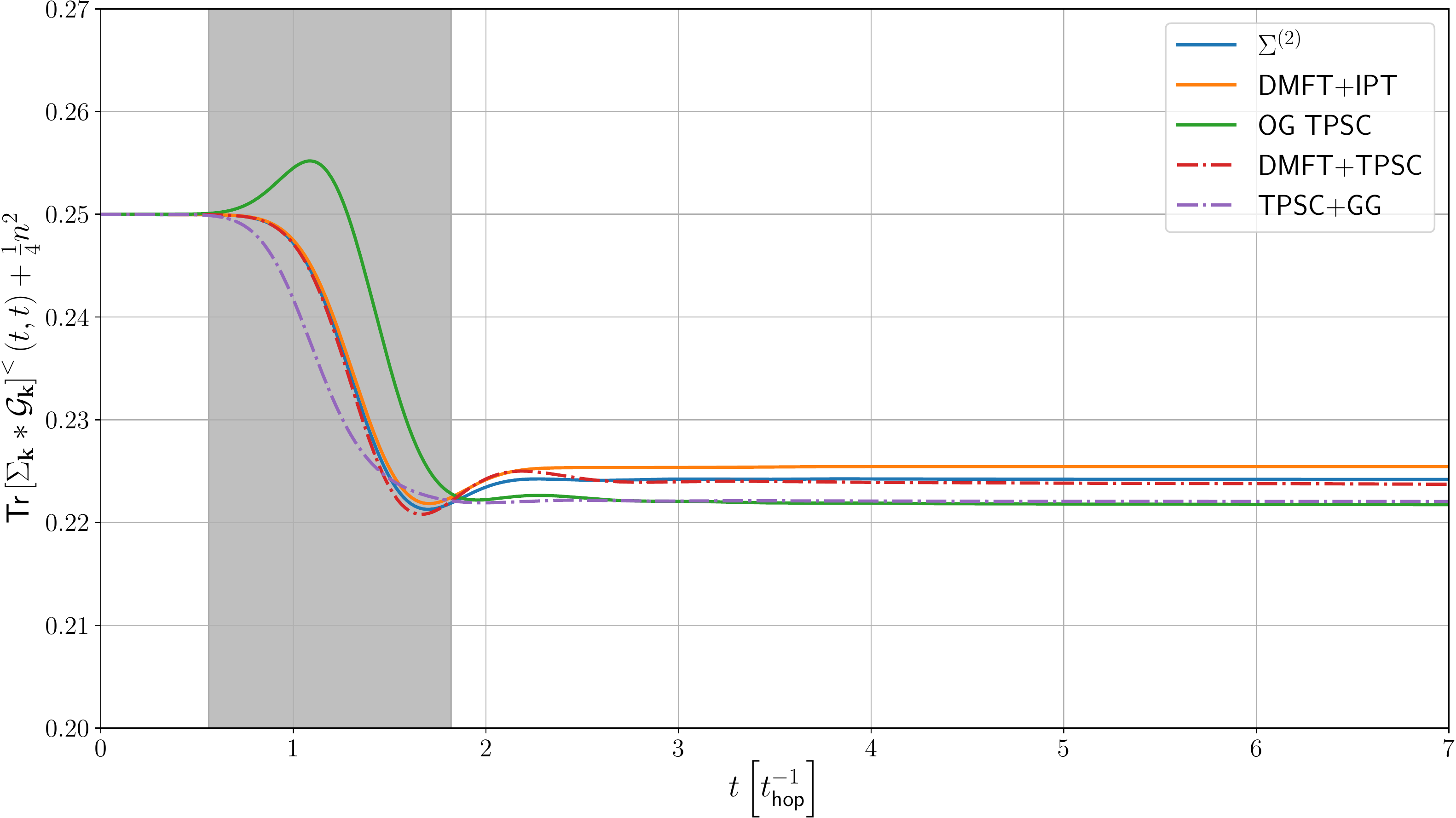}
\caption{Double occupancy of the 2D Hubbard model calculated from the lattice quantities, Eq.~\eqref{eq:tr_sk_Gk_TPSC}, for $\Sigma^{(2)}$, DMFT, OG TPSC, DMFT+TPSC and TPSC+GG. The interaction is ramped from $U=0$ to $U=1$ in the time interval indicated by the grey shading and the initial temperature is $T=0.2$.
} 
\label{fig:DMFT_TPSC_tr_sigmak_Gk_0_1}
\end{figure}

Figure~\ref{fig:DMFT_TPSC_tr_sigmak_Gk_0_1} plots the double occupancy calculated from the lattice quantities (Eq.~\eqref{eq:tr_sk_Gk_TPSC}) for an interaction ramp with parameters $\gamma=3.5$ and $\delta=2.45$ in Eq.~\eqref{eq:ramp_profile_function_erf}. The double occupancies $D$ computed by DMFT+IPT and $\Sigma^{(2)}$ follow each other quite closely, both featuring a dip at the end of the interaction ramp, succeeded by a fast thermalization. OG TPSC, with the approximate solution \eqref{eq:bethe_Salpeter_eq_approximated} of the BSE, however predicts a qualitatively different transient behavior of this local quantity: it yields an (unphysical) increase of the double occupancy at the beginning of the interaction ramp and no dip at the end of the ramp. Furthermore, the thermalized value of the double occupancy is lower than the value predicted by the other methods. DMFT+TPSC agrees rather well at all times with the results from DMFT+IPT and $\Sigma^{(2)}$. 

\begin{figure}[t]
\includegraphics[width=\linewidth]{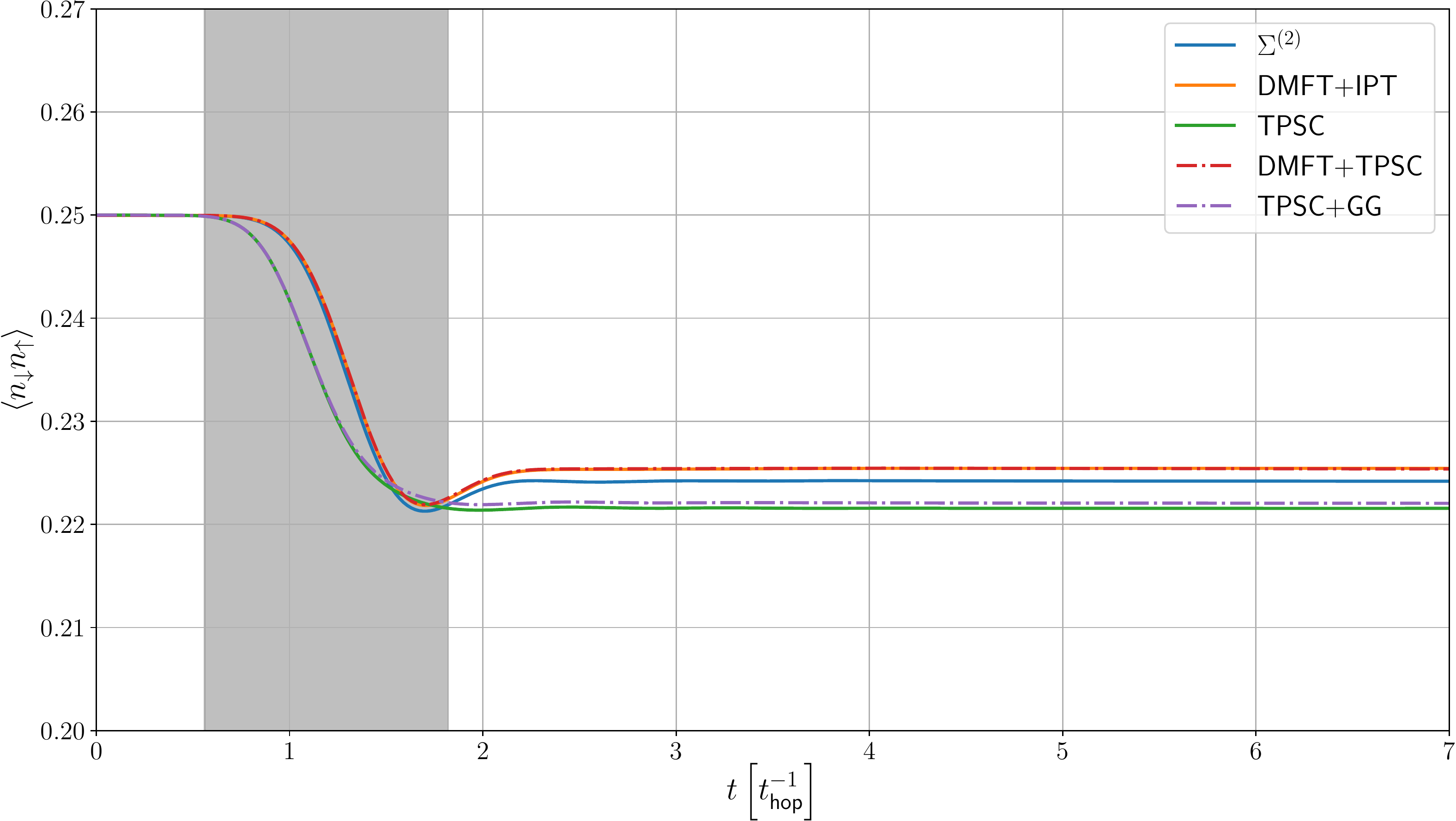}
\caption{Double occupancies calculated using the impurity quantities~\eqref{eq:impurity_double_occupancy} in the cases of DMFT+IPT and DMFT+TPSC. In the case of TPSC and TPSC+GG, the double occupancy taken from Eq.~\eqref{eq:equivalence_spin_irr_vertex_double_occupancy} is shown. The result for $\Sigma^{(2)}$ as well as the parameters are the same as in Fig.~\ref{fig:DMFT_TPSC_tr_sigmak_Gk_0_1}.
} 
\label{fig:DMFT_TPSC_Dloc_0_1}
\end{figure}

One way to correct the transient anomalies of OG TPSC is to resort to the sum-rule \eqref{eq:double_occupancy_sum_rule_alpha} and employ the TPSC second-level approximation~\eqref{eq:tpsc_self_energy_alpha}, \textit{i.e.} switch to TPSC (or TPSC+GG if there is self-consistency). In these schemes, the double occupancy does not show a transient increase at the start of the up-ramp and there is no ambiguity in the definition of the double occupancy, since $D$ obtained from the ansatz is equal to $D$ calculated from the lattice quantities by construction (Eq.~\eqref{eq:double_occupancy_sum_rule_alpha}). The effect of this correction is illustrated in Fig.~\ref{fig:DMFT_TPSC_Dloc_0_1} along with the same result for $\Sigma^{(2)}$ as in Fig.~\ref{fig:DMFT_TPSC_tr_sigmak_Gk_0_1}. While the unphysical increase in the double occupation no longer appears, there is no minimum at the end of the ramp and -- most prominently -- a time-shift in the response to the interaction ramp appears, compared to the other methods. Some of these discrepancies may be related to the fact that we approximately solve the Bethe-Salpeter equations by using Eq.~\eqref{eq:bethe_Salpeter_eq_approximated}.

Note that for DMFT+IPT and DMFT+TPSC the double occupancies illustrated in Fig.~\ref{fig:DMFT_TPSC_Dloc_0_1} are obtained from the impurity quantities using \eqref{eq:impurity_double_occupancy}. In the case of DMFT+IPT this gives the same result as in Figs.~\ref{fig:DMFT_TPSC_tr_sigmak_Gk_0_1}, while there is a small difference for DMFT+TPSC, which employs a momentum-dependent self-energy. However, the difference between the DMFT+TPSC data of Figs.~\ref{fig:DMFT_TPSC_tr_sigmak_Gk_0_1} and \ref{fig:DMFT_TPSC_Dloc_0_1} is only about $1$\%.

\begin{figure}[t]
\includegraphics[width=\linewidth]{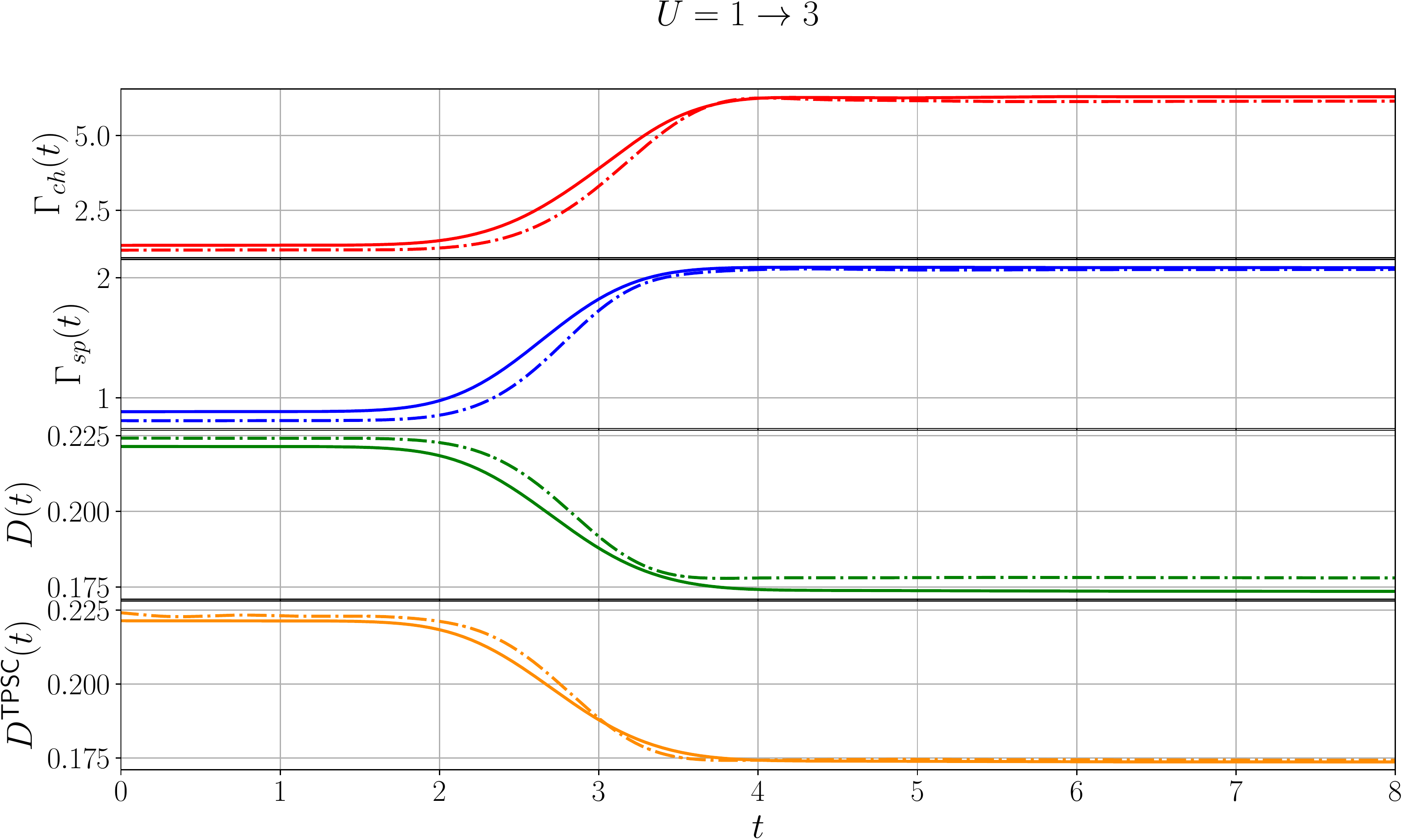}
\caption{Local DMFT+TPSC (dashed lines) and TPSC+GG (solid lines) quantities in the 2D Hubbard model for the ramp from $U=1$ to $U=3$ at initial temperature $T=0.33$. The charge irreducible vertex (top panel), spin irreducible vertex (second panel from top), $D^{\text{imp}}$ (third panel from top) and $D^{\text{TPSC}}$ (bottom panel) are plotted for a time window of $\Delta t=8$.
} 
\label{fig:DMFT_TPSC_local_qties_U_1_3}
\end{figure}

We next consider an interaction ramp from $U=1$ to $U=3$ with the ramp profile corresponding to the parameters $\gamma=1.5$ and $\delta=0.675$ in Eq.~\eqref{eq:ramp_profile_function_erf}. The initial temperature is $T=0.33$, the model is still the half-filled 2D Hubbard model, and we focus on the results from DMFT+TPSC. In Fig.~\ref{fig:DMFT_TPSC_local_qties_U_1_3}, the local irreducible vertices $\Gamma^{\text{ch}}$ (top panel) and $\Gamma^{\text{sp}}$ (second panel from top), the impurity double occupancy $D^{\text{imp}}$ (Eq.~\eqref{eq:impurity_double_occupancy}, third panel from top) and lattice double occupancy $D^{\text{TPSC}}$ (Eq.~\eqref{eq:tr_sk_Gk_TPSC}, bottom panel) are displayed over a time window of $\Delta t = 8$. After the ramp, $\Gamma^{\text{ch}}$ thermalizes to $6.10$ and $\Gamma^{\text{sp}}$ to $2.05$ in DMFT+TPSC (dashed lines). These values are close to those obtained with TPSC+GG for the same ramp (solid lines), which are $\Gamma^{\text{ch}}\simeq 6.01$ and $\Gamma^{\text{sp}}\simeq 2.05$. The same holds for the local double occupancies, which are calculated from Eq.~\eqref{eq:equivalence_spin_irr_vertex_double_occupancy} in TPSC+GG and from Eq.~\eqref{eq:impurity_double_occupancy} in DMFT+TPSC: for TPSC+GG, the double occupancy reaches $D=0.172$, while the value is $D^{\text{imp}}=0.177$ for DMFT+TPSC (green curves). The thermalized value of the lattice double occupancy $D^{\text{TPSC}}$ (Eq.~\eqref{eq:tr_sk_Gk_TPSC}) is $0.174$ (orange curve), which is quite close to that of TPSC+GG. The double occupancies $D^{\text{TPSC}}$ and $D^{\text{imp}}$ overlap almost perfectly. Moreover, given that the interaction ramp used in Fig.~\ref{fig:DMFT_TPSC_local_qties_U_1_3} is slower than that used in Figs.~\ref{fig:DMFT_TPSC_tr_sigmak_Gk_0_1} and \ref{fig:DMFT_TPSC_Dloc_0_1}, no transient dips in the double occupancies are observed near the end of the ramp. Notice that the response of the charge vertex $\Gamma^{\text{ch}}$ to the ramp (top panel of Fig.~\ref{fig:DMFT_TPSC_local_qties_U_1_3}) is delayed compared to that of the spin vertex $\Gamma^{\text{sp}}$ (second top panel of Fig.~\ref{fig:DMFT_TPSC_local_qties_U_1_3}), as was previously reported in the case of TPSC and TPSC+GG,\cite{https://doi.org/10.48550/arxiv.2205.13813} which in contrast to DMFT+TPSC makes use of the ansatz \eqref{eq:equivalence_spin_irr_vertex_double_occupancy} to connect $D$ and $\Gamma^{\text{sp}}$.

\begin{figure}[t]
\includegraphics[width=\linewidth]{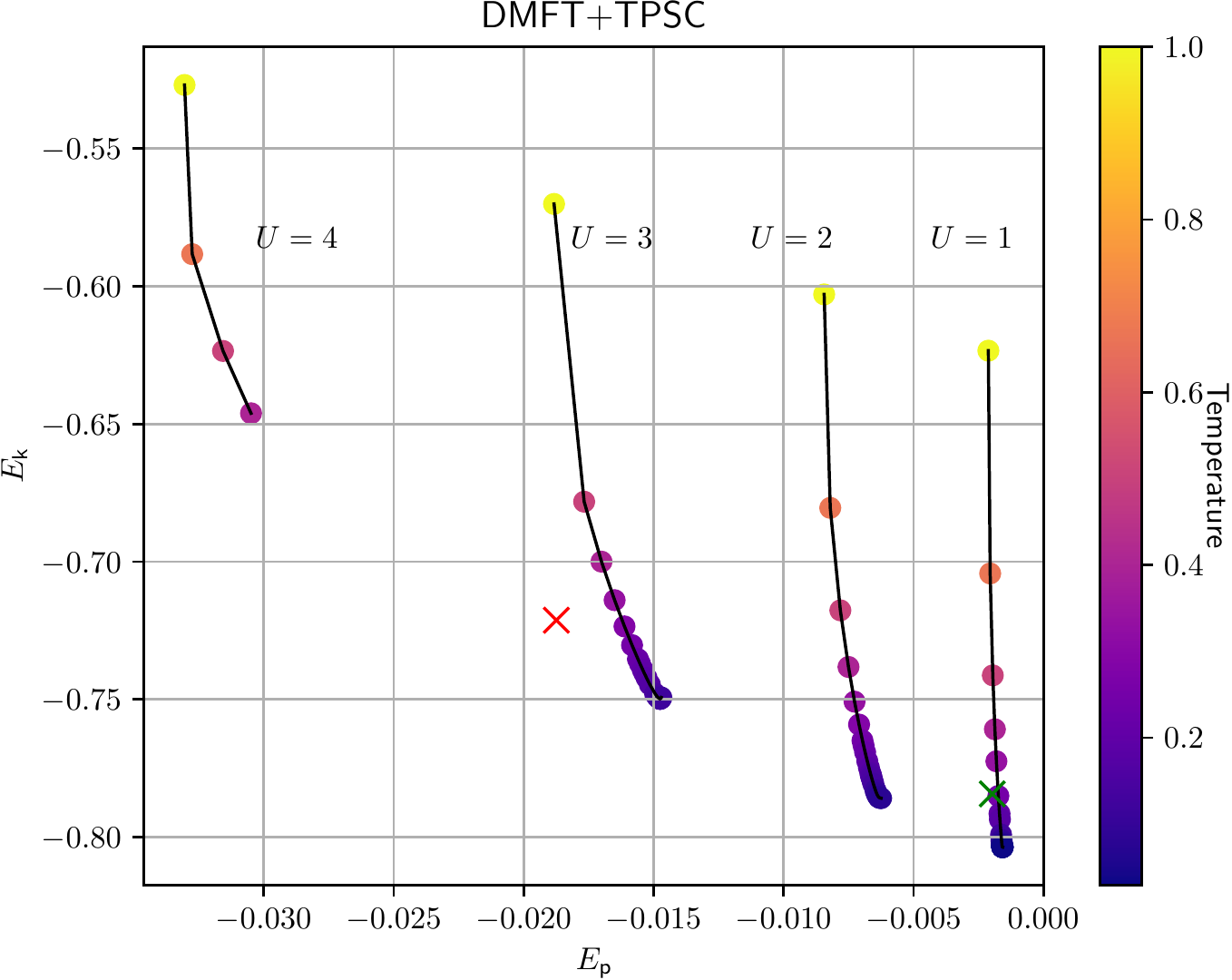}
\caption{Color plot illustrating the relation between the potential energy $E_{\text{p}}$ ($x$-axis), the kinetic energy $E_{\text{k}}$ ($y$-axis) and the corresponding equilibrium temperature for DMFT+TPSC and $U=1,2,3,4$ (see annotations). The 2D square lattice Hubbard model is used. The red (green) cross marks the post-ramp state, obtained from the interaction ramp shown in Fig.~\ref{fig:DMFT_TPSC_local_qties_U_1_3} (Fig.~\ref{fig:DMFT_TPSC_tr_sigmak_Gk_0_1}).
} 
\label{fig:DMFT_TPSC_energy_plane_Us}
\end{figure}

A drawback of the DMFT+TPSC implementation which does not enforce the equivalence of $D^{\text{TPSC}}$ (Eq.~\eqref{eq:tr_sk_Gk_TPSC}) and $D^{\text{imp}}$ (Eq.~\eqref{eq:impurity_double_occupancy}) is that there is no unambiguous way to determine the potential energy and hence the thermalized temperature from the total energy after the ramp. 
In the following analysis, we calculate the total energy from the lattice quantities $\Sigma_{\mathbf{k}}$ (Eq.~\eqref{eq:sigma_DMFT_TPSC}) and $\mathcal{G}_{\mathbf{k}}^{\text{lat}}$  (Eq.~\eqref{eq:lattice_G_definition_improved}). Then
the kinetic energy of the system is $E_\text{k}(t)=\frac{-i}{N_{\mathbf{k}}}\sum_{\mathbf{k}}\epsilon_{\mathbf{k}}\mathcal{G}^{<}_{\mathbf{k}}(t,t)$, while the potential energy is $E_{\text{p}}(t)=\frac{-i}{N_k}\sum_{\mathbf{k}}\int_{\mathcal{C}}\mathrm{d}z \left[\Sigma_{\mathbf{k}}(t,z)\mathcal{G}_{\mathbf{k}}(z,t)\right]^<$, which gives the total energy of the lattice electrons $E_{\text{tot}}(t)=E_{\text{k}}(t)+E_{\text{p}}(t)$.  A temperature of $T_{\text{therm}}\simeq 0.32$ is obtained for the $U=1\to 3$ ramp used in Fig.~\ref{fig:DMFT_TPSC_local_qties_U_1_3}. In Fig.~\ref{fig:DMFT_TPSC_energy_plane_Us}, the corresponding total energy in the post-ramp state is marked by a red cross in the energy plane and compared to results calculated in equilibrium (colored dots). The green cross shows the DMFT+TPSC total lattice energy after the interaction ramp $U=0\to 1$ presented in Fig.~\ref{fig:DMFT_TPSC_tr_sigmak_Gk_0_1}. One can notice that the red cross is quite far from the thermal reference points for $U=3$, corresponding to the post-ramp value of the interaction, meaning that the state after the ramp is not a thermalized state (even though there seems to be little evolution in physical observables). This is surprising, since a trapping in nonthermal states is generically expected for weak interactions, but not in the intermediate coupling regime.\cite{Moeckel2008,Eckstein2009}

From Fig.~\ref{fig:DMFT_TPSC_energy_plane_Us}, different effective temperatures could be defined based on the potential energy $E_{\text{p}}$ or the kinetic energy $E_{\text{k}}$. The temperature extracted from $E_{\text{p}}$ is $T_{\text{therm}}(E_{\text{p}})\simeq 0.93$, whereas that extracted from $E_{\text{k}}$ is $T_{\text{therm}}(E_{\text{k}})\simeq 0.28$.
This unexpected trapping in a nonthermal state may be related to the fact that $U=3$ is close to the regime where the weak-coupling impurity solver breaks down.\cite{Tsuji_2013b} At the weaker post-ramp interaction $U=1$, the energy is almost compatible with a thermalized state, since the green cross practically falls on the $U=1$ line of thermalized states. Here, the discrepancy to the thermalized value may indeed be the result of slow thermalization. 

\begin{figure}[t]
\includegraphics[width=\linewidth]{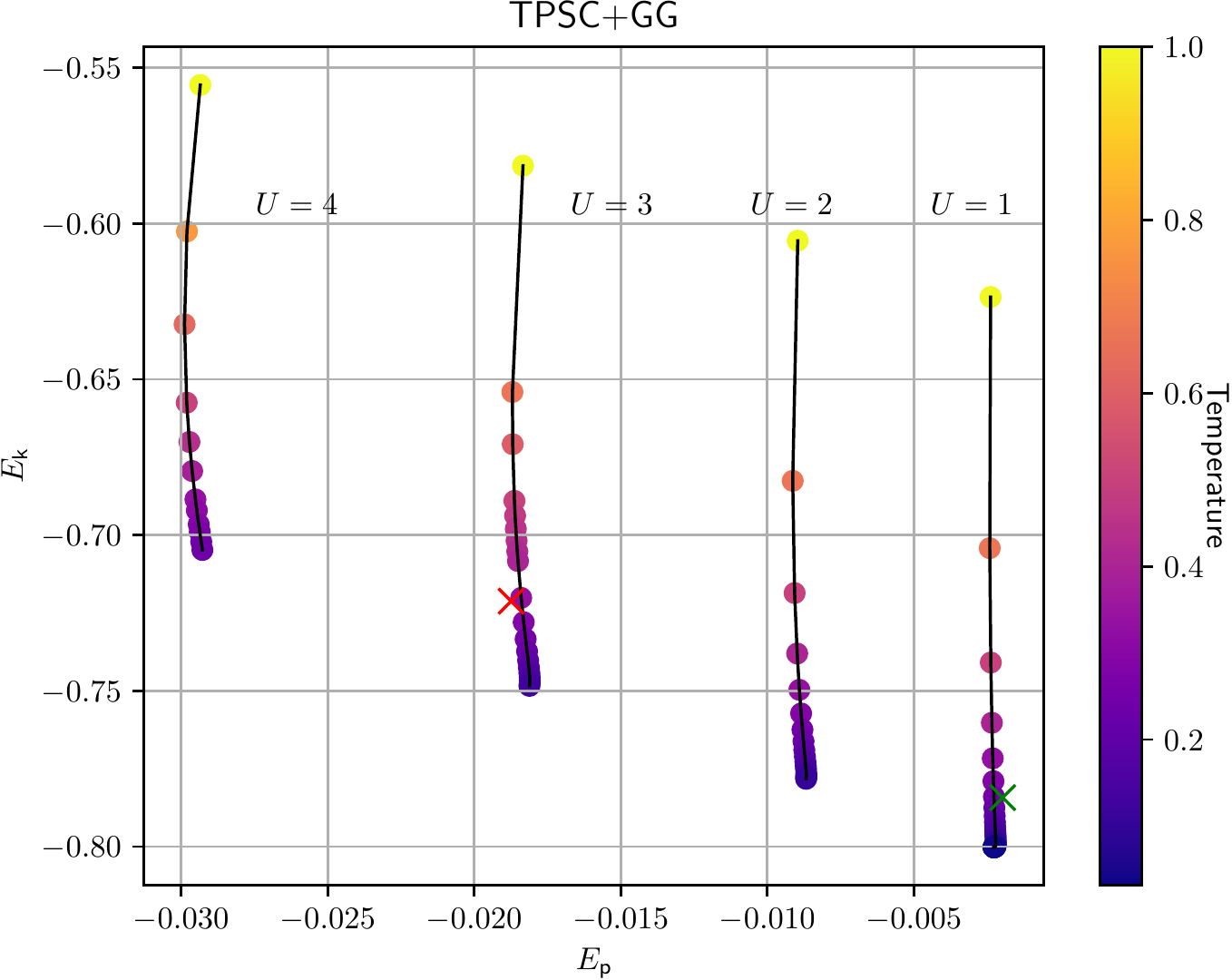}
\caption{Color plot analogous to Fig.~\ref{fig:DMFT_TPSC_energy_plane_Us}, but for TPSC+GG. The red (green) cross marks the post-ramp state obtained from the interaction ramp shown in Fig.~\ref{fig:DMFT_TPSC_local_qties_U_1_3} (Fig.~\ref{fig:DMFT_TPSC_tr_sigmak_Gk_0_1}).
} 
\label{fig:TPSC_GG_energy_plane_Us}
\end{figure}

For comparison, we show in Fig.~\ref{fig:TPSC_GG_energy_plane_Us} the same type of analysis as in Fig.~\ref{fig:DMFT_TPSC_energy_plane_Us}, but for TPSC+GG. This time, the red (green) cross corresponds to the TPSC+GG post-ramp state for the ramp shown in Fig.~\ref{fig:DMFT_TPSC_local_qties_U_1_3} (Fig.~\ref{fig:DMFT_TPSC_tr_sigmak_Gk_0_1}). This figure clearly demonstrates that within TPSC+GG the system approximately thermalizes after an interaction ramp, even at $U=3$. The problem of trapping or unexpectedly slow thermalization at intermediate $U$ is thus much reduced in TPSC+GG, compared to DMFT+TPSC with the bare IPT impurity solver.

One way to address the issue of non-unique double occupations and potential energies is to introduce a parameter $\alpha$ that enforces the equivalence between the impurity $D^{\text{imp}}$~\eqref{eq:impurity_double_occupancy} and the lattice $D^{\text{TPSC}}$~\eqref{eq:tr_sk_Gk_TPSC}, as indicated in Eq.~\eqref{eq:double_occupancy_sum_rule_alpha_DMFT_TPSC}. This extra sum-rule promotes DMFT+TPSC to DMFT+TPSC$\alpha$. This scheme, however, only works well in equilibrium, as already mentioned, and it does not solve problems originating from the bare IPT solver. 

\subsubsection{Dimensional crossover}

\begin{figure}[t]
\includegraphics[width=\linewidth]{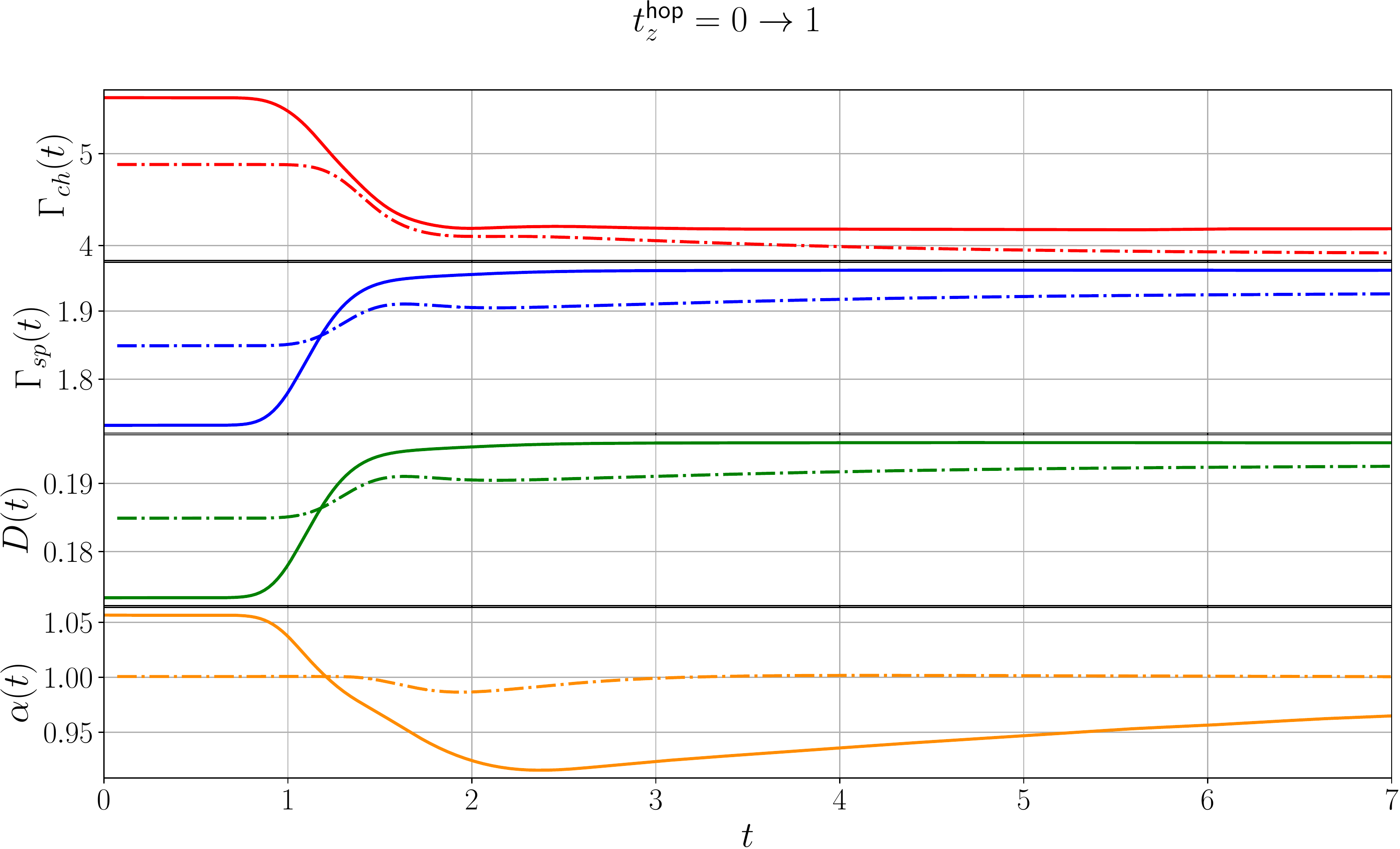}
\caption{Local TPSC (solid lines) and TPSC+GG (dashed lines) quantities in a dimensional ramp from a square lattice to a cubic lattice corresponding to a ramp from $t^{\text{hop}}_z=0$ to $t^{\text{hop}}_z=1$ in the dispersion relation \eqref{eq:dispersion_relation}. The initial temperature is $T=0.2$ and the constant interaction is $U=2.5$. The charge irreducible vertex (top panel), spin irreducible vertex (second panel from top), $D^{\text{imp}}$ (third panel from top) and $\alpha$ (bottom panel) are plotted for a time window of $\Delta t=7$.} 
\label{fig:TPSC_local_qties_tp_0_1}
\end{figure}

We next consider lattice hopping ramps to test the performance of TPSC, TPSC+GG and DMFT+TPSC in dimensions $\ge 2$. In these ramps, we switch on the hopping $t^{\text{hop}}_z$ in the direction perpendicular to the plane, and thus induce a transition from the 2D Hubbard model  ($t^{\text{hop}}_z=0$) to the 3D model ($t^{\text{hop}}_z=1$). Figure~\ref{fig:TPSC_local_qties_tp_0_1} shows TPSC (solid lines) and TPSC+GG (dashed lines) results of such a ramp for the constant interaction $U=2.5$ and initial temperature $T=0.2$. As the dimension is increased, $\Gamma^{\text{ch}}$ decreases while the double occupation increases. This makes sense, since the bandwidth $W$ increases from $8t^{\text{hop}}$ (square lattice) to $12t^{\text{hop}}$ (cubic lattice) and hence the correlation strength is reduced. On the other hand, the spin irreducible vertex $\Gamma^{\text{sp}}$ varies in the opposite direction (see second panel from the top), since $D$ increases and $\Gamma^{\text{sp}}$ and $D$ are related via the ansatz~\eqref{eq:equivalence_spin_irr_vertex_double_occupancy}. As a result, the spin and charge vertices become more similar, which is the expected result if $U/W$ decreases. The parameter $\alpha$, which enforces consistency between the different evaluations of the double occupancy, relaxes slowly since it is strongly affected by the $\mathbf{k}$-dependent thermalization of the (convolved) single-particle quantities. Overall, TPSC admits larger variations of the quantities with faster thermalization compared to TPSC+GG.

By construction, nonequilibrium TPSC and its variants rely to a much larger extent on the conservation of the potential energy $E_{\text{p}}$ than on the kinetic energy $E_{\text{k}}$, because the local irreducible vertices are strongly dependent on the double occupancy $D$ (see for instance Eqs.~\eqref{eq:equivalence_spin_irr_vertex_double_occupancy}, \eqref{eq:fluctuation_dissipation_two_particle} or \eqref{eq:double_occupancy_sum_rule_alpha}). When the total energy drifts after the ramp, which happens for too large and/or too fast ramps, especially for TPSC following a lattice hopping ramp like depicted in Fig.~\ref{fig:TPSC_local_qties_tp_0_1}, this drift is mainly caused by $E_{\text{k}}$. Therefore, as long as $E_{\text{p}}$ is stable after the ramps, which is the case in most situations, the TPSC quantities such as $\Gamma^{\text{sp/ch}}$ and $D$ will stabilize at some value. One particularly useful observation is that even if $E_{\text{k}}$ drifts, thermalized temperatures can be assigned within TPSC frameworks by matching the post-ramp values of the local quantities ($\Gamma^{\text{sp/ch}}$ and $D$) with those calculated at equilibrium for the same post-ramp value: the $T_{\text{therm}}$ values thereby extracted for each local quantities are almost exactly the same,\footnote{This is however not the case in DMFT+TPSC.} \textit{i.e.} $T_{\text{therm}}(\Gamma^{\text{ch}})=T_{\text{therm}}(\Gamma^{\text{sp}})=T_{\text{therm}}(D)$. Since $E_{\text{k}}$ is calculated with $\mathcal{G}_{\mathbf{k}}$, the only meaningful kinetic energy is that of the lattice. When calculating the thermalized temperature of the system after the $t^{\text{hop}}_{z}$ ramp in Fig.~\ref{fig:TPSC_local_qties_tp_0_1}, one finds that the variation from the initial temperature ($T=0.2$) is negligible in TPSC. Hence, the thermalized values of the local quantities depicted in Fig.~\ref{fig:TPSC_local_qties_tp_0_1} are those, at equilibrium, of a cubic lattice at $U=2.5$ and $T_{\text{therm}}\simeq0.204$. On the other hand, the thermalized temperature calculated from TPSC+GG would be much higher, that is $T_{\text{therm}}\simeq1.06$. Note that the system heats up much more in TPSC+GG as well when ramping the interaction, compared to TPSC.\cite{https://doi.org/10.48550/arxiv.2205.13813} The way the thermalized temperature is computed after a lattice hopping ramp is the same as the one explained for $U$-ramps (Eq.~\eqref{eq:ramp_profile_function_erf}), with the exception that equilibrium results are calculated with the post-ramp $t^{\text{hop}}_{z}$ ($U$ is fixed). 

\begin{figure}[h!]
\includegraphics[width=1\linewidth]{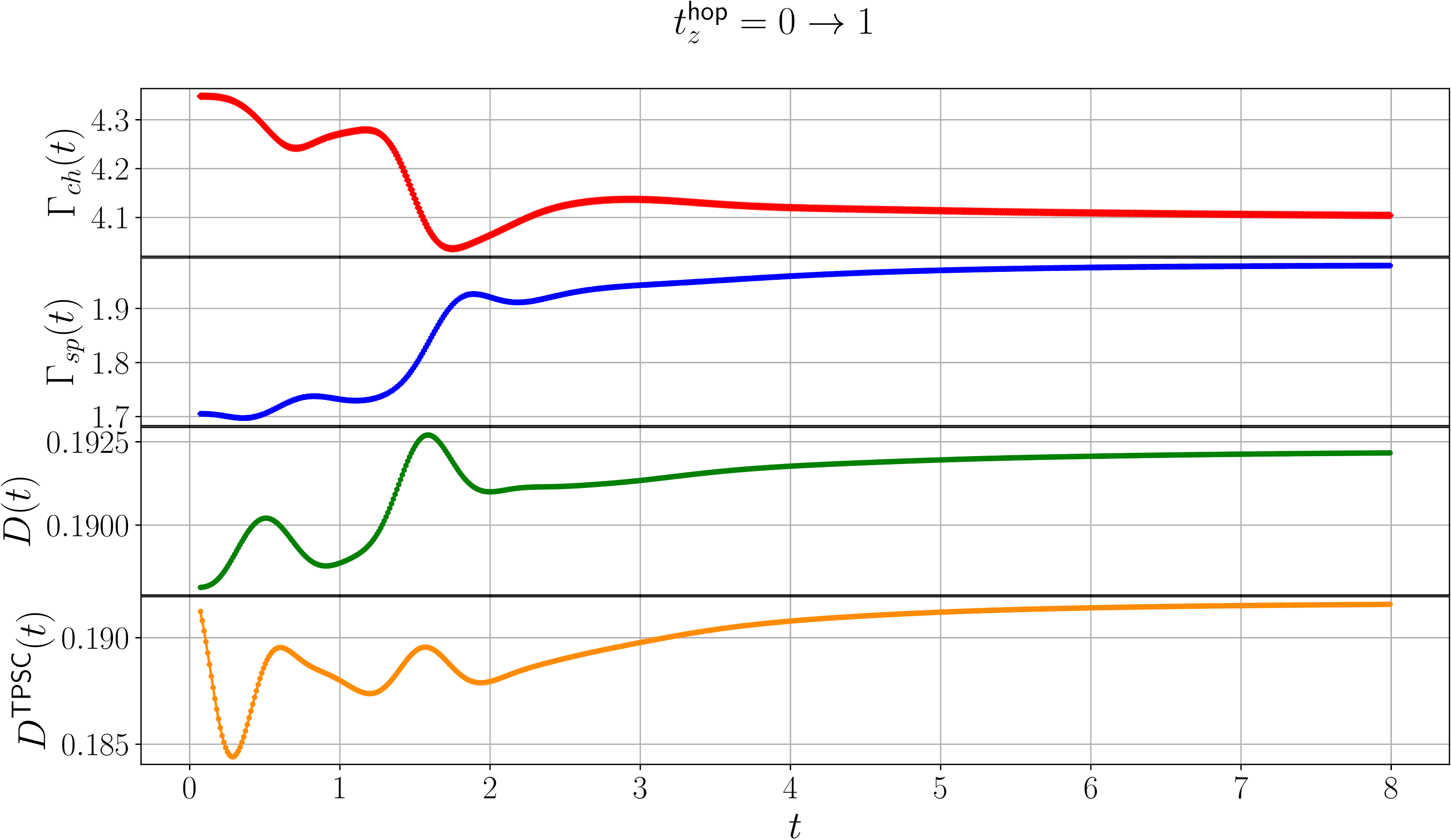} 
\caption{Local DMFT+TPSC quantities in the dimensional ramp from $t^{\text{hop}}_z=0$ to $t^{\text{hop}}_z=1$ in the single-band nearest-neighbor Hubbard model for $U=2.5$ at initial temperature $T=0.2$. The charge irreducible vertex (top panel), spin irreducible vertex (second panel from top), $D^{\text{imp}}$ (third panel from top) and $D^{\text{TPSC}}$ (bottom panel) are plotted for a time window of $\Delta t=8$.}
\label{fig:up_quench_loc_dmft_tpsc_tp_0_1_U_2p5}
\end{figure}

The analogous results to Fig.~\ref{fig:TPSC_local_qties_tp_0_1} but for DMFT+TPSC are shown in Fig.~\ref{fig:up_quench_loc_dmft_tpsc_tp_0_1_U_2p5}.
The overall trend follows that of Fig.~\ref{fig:TPSC_local_qties_tp_0_1}, in that $\Gamma^{\text{ch}}$ is reduced and $\Gamma^{\text{sp}}$ increased as the dimensionality is increased from 2D to 3D. Also the double occupancy $D^\text{imp}$ increases, although significantly less than what is observed in TPSC (Fig.~\ref{fig:TPSC_local_qties_tp_0_1}), while $D^\text{TPSC}$ even shows a transient reduction. The main qualitative difference for this particular set-up however is that the DMFT+TPSC results exhibit prominent humps -- one located at $t\simeq 0.7$ and the other at $t\simeq 1.7$ -- in all the local quantities in Fig.~\ref{fig:up_quench_loc_dmft_tpsc_tp_0_1_U_2p5} and that there is a slower approach to the thermalized state. The lattice hopping ramp stops around the time of the second hump. The minima in the charge vertex correlate with maxima in $\Gamma^{\text{sp}}$ as well as in the double occupancies.

%\subsubsection{Prethermalization}
%
%It was shown in Ref.~\onlinecite{PhysRevB.104.245127_non_eq_piton_simard} that the RPA-type $\pi$-ton vertex corrections drove the prethermalization phase following up the interaction ramps. In both TPSC and TPSC+GG, even though the local vertices thermalize rather fast (within a few inverse $t^{\text{hop}}$), the $\mathbf{k}$-defined susceptibilities associated with their respective vertex (charge and spin) take much longer to thermalize.~\cite{https://doi.org/10.48550/arxiv.2205.13813} In DMFT+TPSC, based off Fig.~\ref{fig:up_quench_loc_dmft_tpsc_tp_0_1_U_2p5}, one can clearly notice that the thermalization of the local vertices and double occupancies is delayed compared to TPSC and TPSC+GG. Thereof, it looks like the continuous feedback between the lattice degrees of freedom and impurity degrees of freedom is related to the slow-down of the thermalization, since all the quantities continue to change well past the end of the ramp at $t\simeq 1.7$, even if the total energy is constant (not shown).
%
%\textcolor{red}{[it could be interesting to add here a section about prethermalization. for this, one would have to study the evolution of the occupations $n_{k+}-n_{k-}$, where $k+$ and $k-$ are symmetric w. r. t. the Fermi surface. we could compare this for TPSC, DMFT, and DMFT+TPSC]}

\subsubsection{Momentum-resolved spectra}

\begin{figure}[h!]
\begin{minipage}[h]{1.0\linewidth}
\begin{center}
\includegraphics[width=1\linewidth]{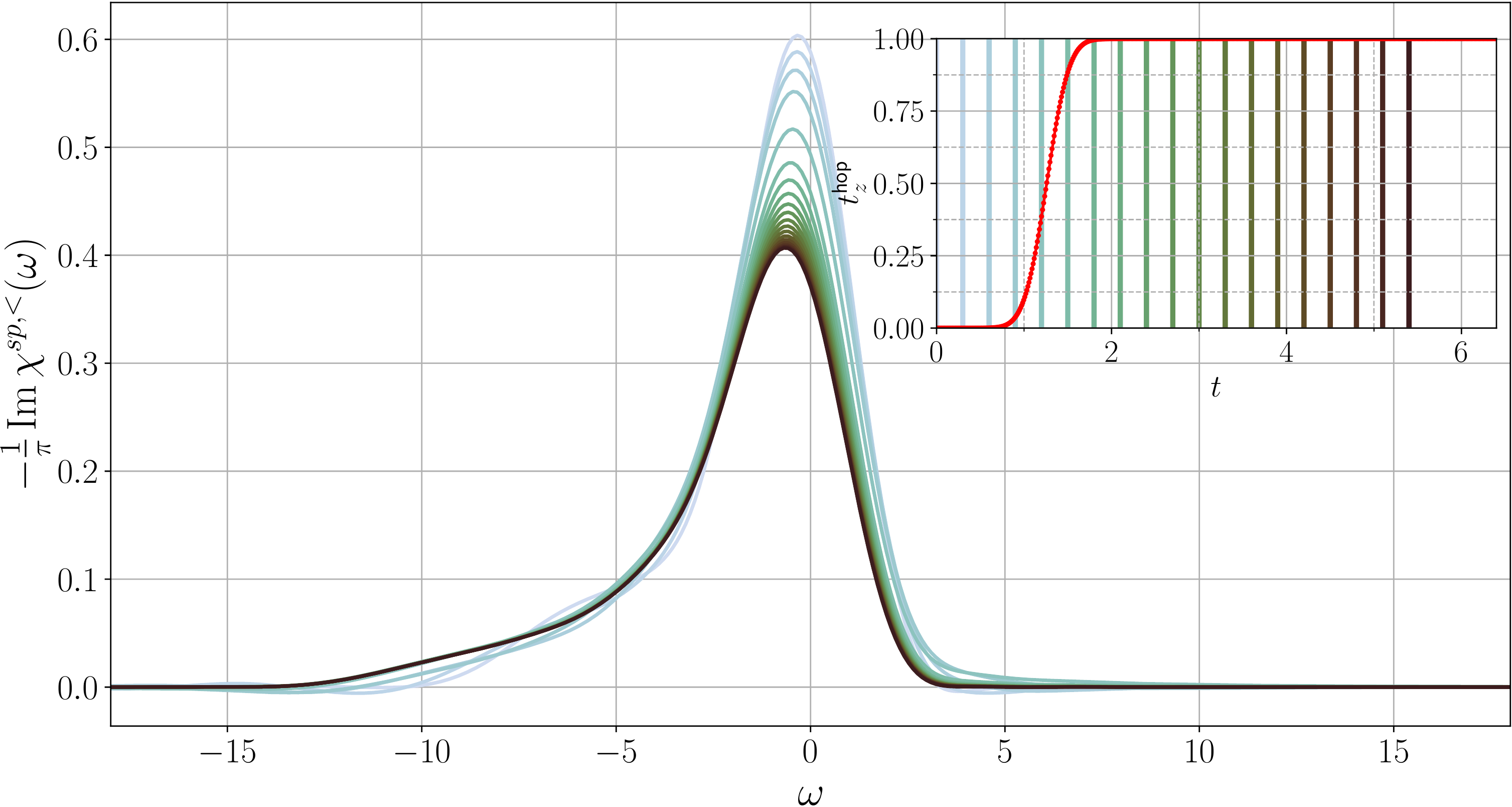} 
\end{center} 
\end{minipage}
\vspace{0.1 cm}
\begin{minipage}[h]{1.0\linewidth}
\begin{center}
\includegraphics[width=1\linewidth]{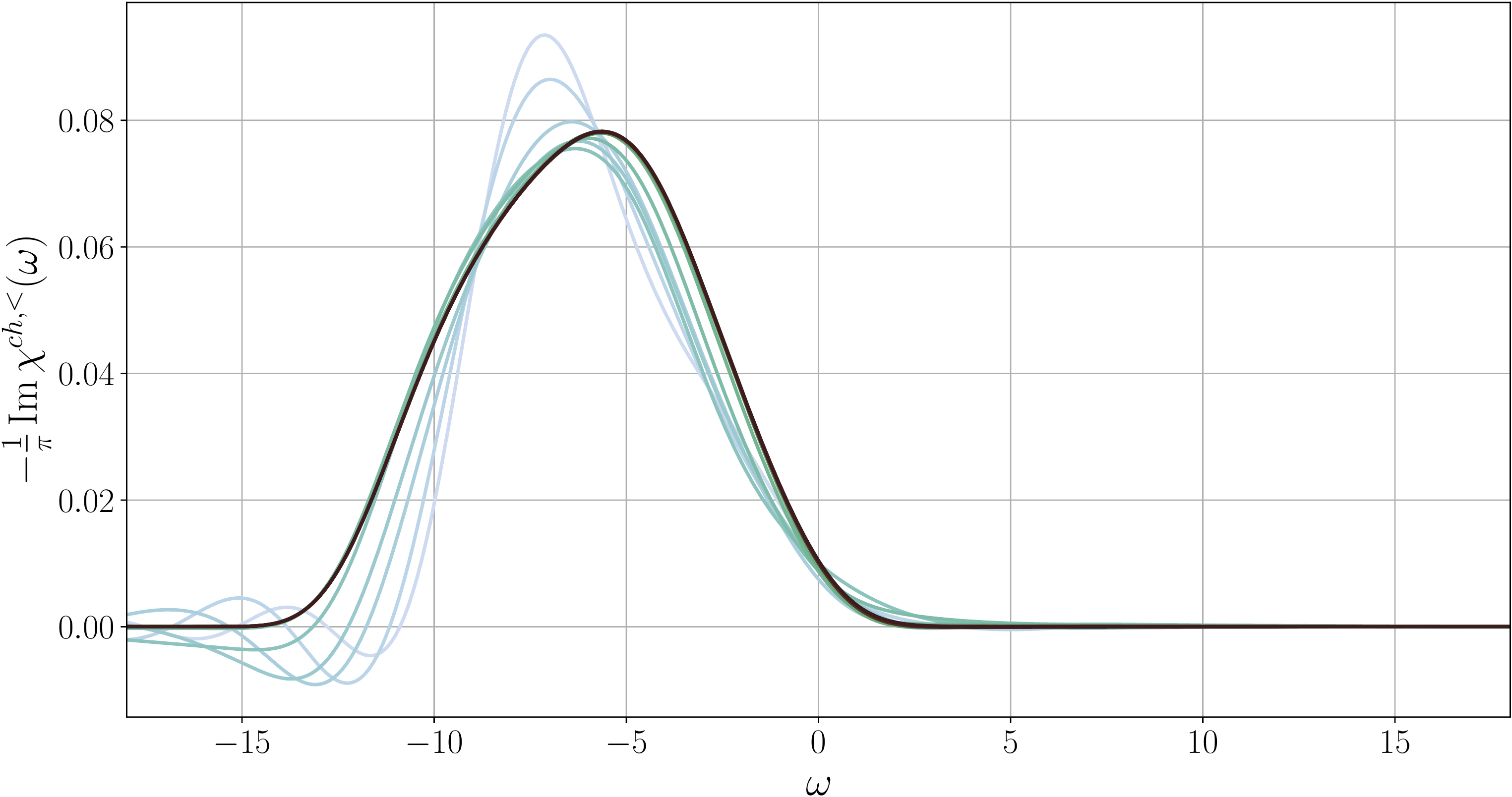} 
\end{center}
\end{minipage}
\caption{Imaginary parts of the \textit{lesser} component of the spin (top subplot) and charge (bottom subplot) susceptibilities for momentum $\mathbf{k}_{\pi}$ and TPSC. The initial temperature is $T=0.2$ and the interaction is $U=2.5$. The inset shows the profile of the perpendicular hopping ramps $t^{\text{hop}}_z$ with the vertical bars representing the times for which the spectra are calculated. The time window for the Fourier transformation is $\Delta t=2.5$.
}
\label{fig:lesser_spin_and_charge_sus_time_evolution_2D_3D_TPSC}
\end{figure}

Next, the time evolution of the spin and charge susceptibilities is illustrated in Fig.~\ref{fig:lesser_spin_and_charge_sus_time_evolution_2D_3D_TPSC} for the dimensional ramp simulated with TPSC. In this figure, we show the spectra at momentum $\mathbf{k}_{\pi}=(\pi,\pi,\pi)$. The lesser component of the spin susceptibility (top subplot) shows that the peak at $\omega\simeq 0$ melts when going from 2D to 3D, which we attribute to the lower $T_x$ in the 3D system. Since the bandwidth increases, the energy range of the spin and charge excitations also increases. The bottom subplot shows the result for the lesser component of the charge susceptibility.  The peak of the charge excitation spectrum is shifted up in energy when going from 2D to 3D and is reduced in height. Furthermore, the peak is broadened in 3D because of the larger bandwidth.

\begin{figure}[h!]
\includegraphics[width=\linewidth]{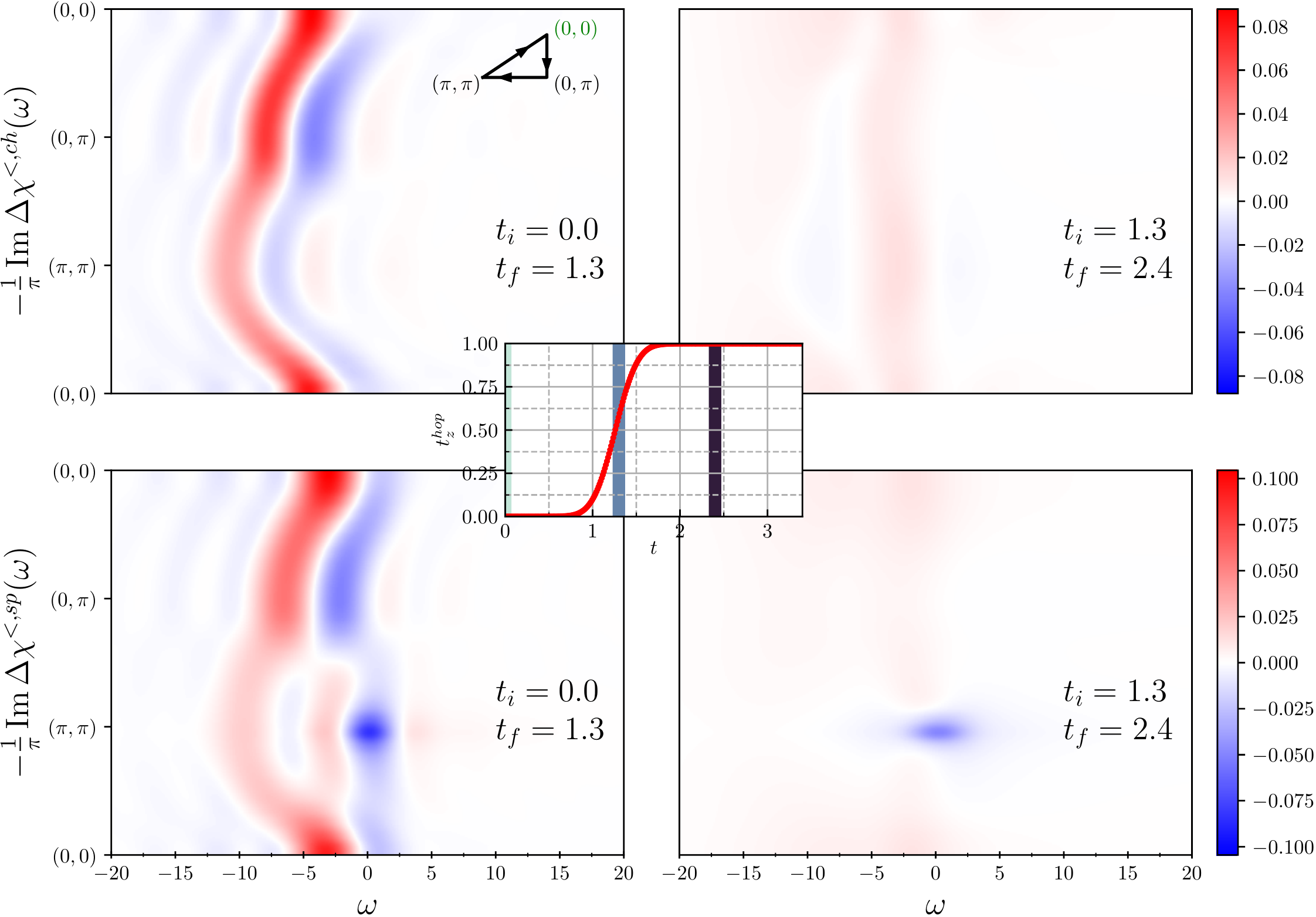}
\caption{ 
Top (Bottom) panels: Difference spectra of the lesser component of the charge (spin) susceptibility after the interaction ramp shown in the inset. The inset black triangle illustrates the path in reciprocal space -- within the $k_z=\pi$ plane -- along which the spectra are displayed. The times $t_i$ and $t_f$ used in the calculation of the difference spectra are annotated in each panel. The time window used in the Fourier transformation is $\Delta t=2.5$. Each row of panels uses the same color scale. The method used here is TPSC.
} 
\label{fig:k_resolved_retarded_charge_2_3_D_tpsc}
\end{figure}

The $\mathbf{k}$-dependent spectral evolution of the spin and charge susceptibilities obtained with TPSC is displayed in Fig.~\ref{fig:k_resolved_retarded_charge_2_3_D_tpsc}, 
along the momentum path indicated in the inset ($k_z=\pi$). We plot the change in the spectra during the ramp, defined as $\Delta Q(t_f,t_i;\omega)\equiv Q(t_f;\omega)-Q(t_i;\omega)$. The top panels show the results for the charge susceptibility ($Q=\chi^{\text{ch}}$), while the bottom panels show those for the spin susceptibility ($Q=\chi^{\text{sp}}$). On the left-hand side, the difference $\Delta\chi(t_f,t_i;\omega)$ is plotted for $t_i=0$ and $t_f=1.3$, whereas $t_i=1.3$ and $t_f=2.4$ on the right-hand side. The vertical bars in the inset indicate the time snapshots $t_i$ and $t_f$ relative to the ramp profile. One striking feature is the qualitative difference between the left and right panels; much of the change happens in the first half of the ramp, while only small changes occur in the second half of the ramp. This can be partly explained by the fact that these spectra are computed using a forward Fourier transform defined as 
\begin{align}
\label{eq:forward_FFT}
Q^{R,<}(\omega,t^{\prime})=\int_{t^{\prime}}^{t^{\prime}+\Delta t}\mathrm{d}t \ e^{i\omega(t-t^{\prime})} \ Q^{R,<}(t,t^{\prime}),
\end{align}
using a time window $\Delta t$ that is larger than the duration of the ramp; these transforms take into account the state after the ramp, even at early times $t'$. The two-time quantity $Q$ in Eq.~\eqref{eq:forward_FFT} represents the Green's function or spin/charge susceptibility. Since the relative weight of the ripples appearing at $\left|\omega\right|\gtrsim 15$ varies a lot with the time window $\Delta t$ used in the forward Fourier transform, we believe that these are artifacts of the Fourier transformation. These ripples however only appear in the TPSC simulations. In the case of the charge susceptibility, the excitations are redistributed to larger absolute energies. The same is true for the spin excitation spectra, which in addition exhibit a strong decrease at $\mathbf{k}_{\pi}$, consistent with the top panel of Fig.~\ref{fig:lesser_spin_and_charge_sus_time_evolution_2D_3D_TPSC}.

\begin{figure}[h!]
\includegraphics[width=1\linewidth]{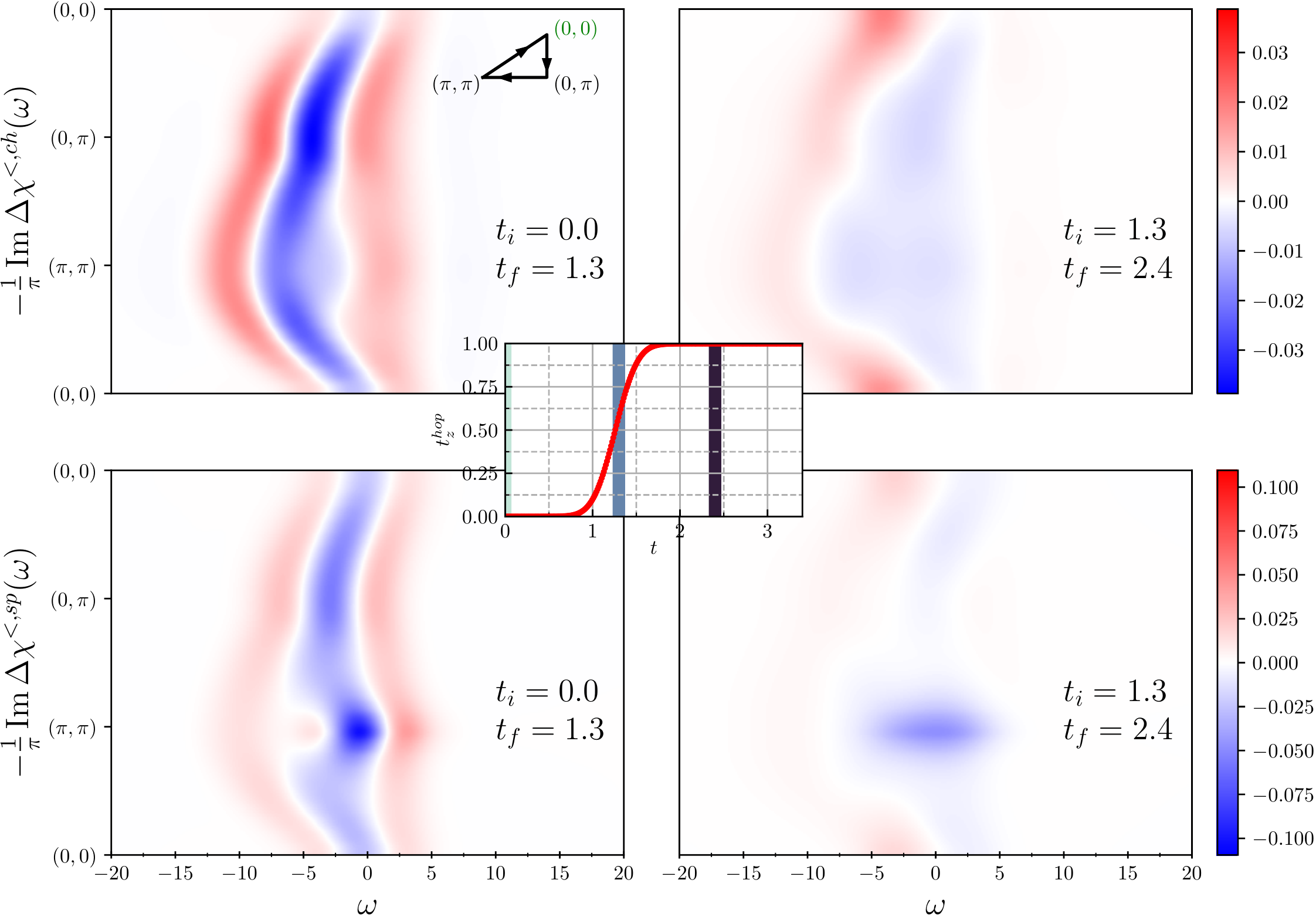} 
\caption{Top (Bottom) panels: Difference spectra of the DMFT+TPSC lesser component of the charge (spin) susceptibility after the perpendicular lattice hopping ramp from $t^{\text{hop}}_z=0$ to $t^{\text{hop}}_z=1$ shown in the inset. The time window employed in the Fourier transformation is $\Delta t=2.5$. Each row of panels uses the same color scale. The initial temperature is $T=0.2$.}
\label{fig:up_quench_time_diff_lesser_susceptibilities_dmft_tpsc_tp_0_1_U_2p5}
\end{figure}

The corresponding data obtained with DMFT+TPSC are shown in Fig.~\ref{fig:up_quench_time_diff_lesser_susceptibilities_dmft_tpsc_tp_0_1_U_2p5} (for the evolution of the local quantities, see Fig.~\ref{fig:up_quench_loc_dmft_tpsc_tp_0_1_U_2p5}). The results obtained from TPSC+GG are quantitatively almost the same (not shown). The time differences $\Delta\chi^{\text{ch/sp},<}(t_f,t_i;\mathbf{k})$ of the \textit{lesser} charge susceptibility (top panels) and spin susceptibility (bottom panels) are shown for times $t_i=0$ and $t_f=1.3$ in the left panels and for times $t_i=1.3$ and $t_f=2.4$ in the right panels. Similar to the TPSC results shown in Fig.~\ref{fig:k_resolved_retarded_charge_2_3_D_tpsc}, the dominant changes occur during the first time interval. The results from DMFT+TPSC display less oscillations in the spectra than TPSC, especially for the charge susceptibility. As in the case of TPSC (Fig.~\ref{fig:k_resolved_retarded_charge_2_3_D_tpsc}), the spin-spin correlations in the vicinity of $\mathbf{k}_{\pi}$ are substantially reduced when going from 2D to 3D, since at fixed $U$, the crossover temperature $T_x$ is reduced (\textit{cf.}~Fig.~\ref{fig:static_spin_sus_fermi_surface_dmft_tpsc_3D}) and the system heats up. In Appendix~\ref{appendice:ch:interaction_quench_comparisons} we show comparisons between TPSC, TPSC+GG and DMFT+TPSC results for a $U$-ramp going from $U=1$ to $U=3$ in the half-filled square lattice Hubbard model.

%%%%%%%%%%%%%%%%%%%%%%%%%%%%%%%%%%%%%%%%%%%%%%%% Conclusion %%%%%%%%%%%%%%%%%%%%%%%%%%%%%%%%%%%%%%%%%%%%%%%%%%%%%%%%%%%%%%%%%%
%%

\section{Conclusions}
\label{sec:conclusion}

The nonequilibrium formulation of TPSC and its variants on the Kadanoff-Baym contour has been detailed. We also introduced nonequilibrium DMFT+TPSC, which makes use of the TPSC self-energy to incorporate nonlocal electronic correlations into the DMFT framework in a self-consistent manner, or, alternatively speaking, replaces the local component of the TPSC self-energy by the DMFT counterpart. Focusing on the weak-to-intermediate correlation regime, we employed $2^{\text{nd}}$-order or $3^{\text{rd}}$-order IPT to solve the DMFT impurity problem. In equilibrium, our self-consistent version of DMFT+TPSC gives similar results to the non-self-consistent scheme recently introduced in Ref.~\onlinecite{https://doi.org/10.48550/arxiv.2211.01919}.

We have extensively tested the different TPSC variants and provided benchmarks against more sophisticated methods to check the accuracy. For the 2D Hubbard model, it was demonstrated that the momentum-dependent self-energy of TPSC+GG and DMFT+TPSC match very well the DiagMC results, especially in the case of TPSC+GG. Moreover, it was shown that the growth of the antiferromagnetic correlation length as temperature is lowered is significantly improved in TPSC+GG and DMFT+TPSC when compared to OG TPSC, which overestimates the spin correlations. 

TPSC and its variants were then tested in nonequilibrium settings, by applying interaction ramps and lattice hopping ramps designed to switch between 2D and 3D lattices. While in this case, we lack exact benchmark results, the comparison to established approximate schemes like $\Sigma^{(2)}$ or DMFT could provide some useful insights. It turns out that the transient dynamics of the double occupancy is substantially improved in both DMFT+TPSC and TPSC+GG, compared to OG TPSC, which produces seemingly unphysical features in the time evolution. DMFT+TPSC yields double occupancies very close to DMFT, which shows that for this local quantity, the feedback from the nonlocal components of the self-energy has only minor effects. More generally, we found that in the weak-to-intermediate correlation regime, TPSC+GG and DMFT+TPSC lead to very similar results both for momentum-resolved two-particle and single-particle spectral functions, and for time-dependent two-body local quantities. 

A conceptual problem of the DMFT+TPSC approach lies in the fact that the double occupancy measured from the impurity problem can deviate from the one estimated from the lattice quantities, thereby creating an ambiguity in the definition of the potential energy. Calculating all the energy contributions from the lattice Green's functions and self-energies, we found that the state after a ramp to intermediate interactions (e.g. $U=3$ in the 2D Hubbard model) is not consistent with a thermalized state, even though the post-ramp evolution of physical observables is almost constant. Since a thermalization bottleneck at intermediate couplings is not expected, this points to a breakdown of the formalism, which may be related to the aforementioned ambiguity in the calculation of the potential energy contribution, the non-conserving nature of the formalism, or the perturbative impurity solver, which becomes unreliable at intermediate $U$. The mismatch between the post-ramp observables and the expected thermalized values is much reduced within TPSC+GG, where it might (at weak coupling) originate from slow thermalization. An attempt to enforce consistency between the impurity and lattice double occupancies within a DMFT+TPSC$\alpha$ scheme resulted in an algorithm which suffers from an unstable time propagation on the real axis. 

In further studies, different avenues to overcome the issues with the effective temperature at intermediate coupling will be investigated. For instance, the spin and charge irreducible vertices could be extracted in the same fashion as discussed in Ref.~\onlinecite{kusunose_influence_2006}, \textit{i.e.} directly from the impurity self-energy, bypassing the two-particle sum-rules. More accurate impurity solvers should be employed within DMFT+TPSC to access the intermediate and strong coupling regime. Furthermore, the consequences of the approximate solution of the BSE (Eq.~\eqref{eq:bethe_Salpeter_eq_approximated}) need to be investigated. For this purpose nonequilibrium setups in which this approximation can be circumvented, such as nonequilibrium steady-state solutions, are of particular interest. 

While this study presented the current status in the development of TPSC based nonequilibrium methods, and revealed a certain number of challenges and inconsistencies, it also demonstrated the potential of TPSC and DMFT+TPSC approaches as a promising and computationally efficient new method to access nonequilibrium dynamics of correlated lattice systems. In particular, this approach enables calculations with self-consistently renormalized spin and charge vertices and full momentum resolution. 

\begin{acknowledgments}
The calculations have been performed on the Beo05 cluster at the University of Fribourg. OS and PW acknowledge support from ERC Consolidator Grant No.~724103.
\end{acknowledgments}
    
\pagebreak

%--------------------------------------------------------------------------------------------------

\appendix

%--------------------------------------------------------------------------------------------------
\section{Weak-coupling self-energy expansion}
\label{appendice:ch:weak_coupling_self_expansion}
%----------------------------------------------------------------------------------------

The weak-coupling self-energy expansion can be derived starting from the general physical ($\phi\to 0$) self-energy expression \eqref{eq:four_point_self_G}, whose Fock term vanishes in the case of the Hubbard model. Substituting the first-order term describing the susceptibility \eqref{eq:bethe_Salpeter_equation} into \eqref{eq:combination_bethe_Salpeter_two_particle_corr}, one gets

\begin{align}
\label{eq:appendice:weak_coupling_self_expansion:self_energy_sus_substituted}
&\Sigma_{\sigma\sigma^{\prime}}(z_1,z_2) = -iU(z_1)\mathcal{G}_{-\sigma}(z_1,z_1^+)\delta^{\mathcal{C}}(z_1,z_2)\delta_{\sigma,\sigma^{\prime}}\notag\\
&+ U(z_1)\mathcal{G}_{\sigma\bar{\sigma}_3}(z_1,\bar{z}_3)\Gamma_{\bar{\sigma}_3\sigma^{\prime}\bar{\sigma}_4\bar{\sigma}_5}(\bar{z}_3,z_2,\bar{z}_4,\bar{z}_5)\notag\\
&\times\left[-i\mathcal{G}_{\bar{\sigma}_4-\sigma}(\bar{z}_4,z_1^+)\mathcal{G}_{-\sigma\bar{\sigma}_5}(z_1,\bar{z}_5)\right].
\end{align}
Note that the propagators appearing in this self-energy expression are boldified, \textit{i.e.}, they are dressed with self-energy insertions according to the Dyson equation~\eqref{eq:modified_Dyson_equation}. Recalling that the vertex function appearing in Eq.~\eqref{eq:appendice:weak_coupling_self_expansion:self_energy_sus_substituted} is defined as 

\begin{align}
\label{eq:appendice:weak_coupling_self_expansion:vertex_function_def}
\Gamma_{\sigma_3\sigma^{\prime}\sigma_4\sigma_5}(z_3,z_2;z_4,z_5) = -\frac{\delta\Sigma_{\sigma_3\sigma^{\prime}}(z_3,z_2)}{\delta\mathcal{G}_{\sigma_4\sigma_5}(z_4,z_5)},
\end{align}
one obtains $\Sigma^{(2)}$, as defined in Eq.~\eqref{eq:nonequilibrium_quantum_many_body_physics:IPT:second_order_bubble}, by selecting the Hartree term in Eq.~\eqref{eq:appendice:weak_coupling_self_expansion:self_energy_sus_substituted} as the differentiated self-energy component in Eq.~\eqref{eq:appendice:weak_coupling_self_expansion:vertex_function_def}. Doing so and using the first-order term in the Dyson equation~\eqref{eq:modified_Dyson_equation}, Eq.~\eqref{eq:appendice:weak_coupling_self_expansion:self_energy_sus_substituted} becomes

\begin{align}
\label{eq:appendice:weak_coupling_self_expansion:self_energy_2nd_order_1}
&\Sigma^{(2)}_{\sigma\sigma^{\prime}}(z_1,z_2)=U(z_1)\mathcal{G}^0_{\sigma\bar{\sigma}_3}(z_1,\bar{z}_3)\notag\\
&\times\bigl[iU(\bar{z}_3)\delta^{\mathcal{C}}(\bar{z}_3,\bar{z}_4)\delta^{\mathcal{C}}(\bar{z}_3^+,\bar{z}_5)\delta^{\mathcal{C}}(\bar{z}_3,z_2)\delta_{\bar{\sigma}_3,\sigma^{\prime}}\delta_{-\bar{\sigma}_3,\bar{\sigma}_4}\notag\\
&\times\delta_{-\bar{\sigma}_3,\bar{\sigma}_5}\bigr]\left[-i\mathcal{G}^0_{\bar{\sigma}_4-\sigma}(\bar{z}_4,z_1^+)\mathcal{G}^0_{-\sigma\bar{\sigma}_5}(z_1,\bar{z}_5)\right].
\end{align}
Since $\sigma^{\prime}$ needs to be equal to $\sigma$ for a nonzero self-energy, Eq.~\eqref{eq:appendice:weak_coupling_self_expansion:self_energy_2nd_order_1} reduces to Eq.~\eqref{eq:nonequilibrium_quantum_many_body_physics:IPT:second_order_bubble}.

Next, to determine the second-order Hartree term $\Sigma^{(2)}_H$ defined in Eq.~\eqref{eq:nonequilibrium_quantum_many_body_physics:IPT:Hartree_2nd_order}, one needs to use the Dyson equation to expand the boldified propagator, whereby the Hartree term constitutes the self-energy (second term in the $\mathcal{G}$ expansion):

\begin{align}
\label{eq:appendice:weak_coupling_self_expansion:Hartree_second_order_Dyson_expansion}
&\mathcal{G}^{(2)}_{-\sigma}(z_1,z_1^+) = \mathcal{G}^0_{-\sigma\bar{\sigma}}(z_1,\bar{z}_2)\bigl[-iU(\bar{z}_2)\mathcal{G}^0_{-\bar{\sigma}}(\bar{z}_2,\bar{z}_2^{+})\notag\\
&\times\delta_{\bar{\sigma},\bar{\sigma}^{\prime}}\delta^{\mathcal{C}}(\bar{z}_2,\bar{z}_3)\bigr]\mathcal{G}^0_{\bar{\sigma}^{\prime}-\sigma}(\bar{z}_3,z_1^+).
\end{align}
Reinserting the Green's function expansion \eqref{eq:appendice:weak_coupling_self_expansion:Hartree_second_order_Dyson_expansion} into the Hartree term of Eq.~\eqref{eq:appendice:weak_coupling_self_expansion:self_energy_sus_substituted}, one finds the following $\Sigma^{(2)}_H$ term

\begin{align}
\label{eq:appendice:weak_coupling_self_expansion:Hartree_second_order}
&\Sigma_{H,\sigma,\sigma^{\prime}}^{(2)}(z_1,z_2) =\notag\\
&-iU(z_1)\biggl[\mathcal{G}^0_{-\sigma}(z_1,\bar{z}_2)\left[-iU(\bar{z}_2)\mathcal{G}^0_{\sigma}(\bar{z}_2,\bar{z}_2^{+})\delta^{\mathcal{C}}(\bar{z}_2,\bar{z}_3)\right]\notag\\
&\times\mathcal{G}^0_{-\sigma}(\bar{z}_3,z_1^+)\biggr]\delta^{\mathcal{C}}(z_1,z_2)\delta_{\sigma,\sigma^{\prime}}.
\end{align}

Moving on to the $3^{\text{rd}}$-order diagrams, the first set of diagrams, comprised of two elements, uses the $2^{\text{nd}}$-order diagram \eqref{eq:appendice:weak_coupling_self_expansion:self_energy_2nd_order_1} in the vertex calculation \eqref{eq:appendice:weak_coupling_self_expansion:vertex_function_def}. Carrying out the functional derivatives and changing all propagators to $\mathcal{G}^0$, one gets

\begin{align}
\label{eq:appendice:weak_coupling_self_expansion:vertex_functional_derivative_2nd_order_Sigma}
&\Gamma_{\sigma_3\sigma^{\prime}\sigma_4\sigma_5}(z_3,z_2;z_4,z_5) = \notag\\
&-U(z_3)U(z_2)\mathcal{G}_{-\sigma_3}^0(z_2,z_3)\mathcal{G}_{-\sigma_3}^0(z_3,z_2^+)\delta_{\sigma_3,\sigma_4}\delta_{\sigma_3,\sigma_5}\delta_{\sigma_3,\sigma^{\prime}}\notag\\
&\times\delta^{\mathcal{C}}(z_3,z_4)\delta^{\mathcal{C}}(z_2,z_5)\notag\\
&-U(z_3)U(z_2)\mathcal{G}_{\sigma_3}^0(z_3,z_2)\mathcal{G}_{-\sigma_3}^0(z_3,z_2^+)\delta_{-\sigma_3,\sigma_4}\delta_{-\sigma_3,\sigma_5}\delta_{\sigma_3,\sigma^{\prime}}\notag\\
&\times\delta^{\mathcal{C}}(z_2,z_4)\delta^{\mathcal{C}}(z_3^+,z_5)\notag\\
&-U(z_3)U(z_2)\mathcal{G}_{\sigma_3}^0(z_3,z_2)\mathcal{G}_{-\sigma_3}^0(z_2,z_3^+)\delta_{-\sigma_3,\sigma_4}\delta_{-\sigma_3,\sigma_5}\delta_{\sigma_3,\sigma^{\prime}}\notag\\
&\times\delta^{\mathcal{C}}(z_3,z_4)\delta^{\mathcal{C}}(z_2,z_5).
\end{align}
The first term of Eq.~\eqref{eq:appendice:weak_coupling_self_expansion:vertex_functional_derivative_2nd_order_Sigma} vanishes because $\sigma_4$ and $\sigma_5$ cannot have the same spin projection as $\sigma^{\prime}$. Otherwise, the first-order bubble term appearing in the susceptibility vanishes (see Eq.~\eqref{eq:appendice:weak_coupling_self_expansion:self_energy_sus_substituted}). Substituting the third term featuring in Eq.~\eqref{eq:appendice:weak_coupling_self_expansion:vertex_functional_derivative_2nd_order_Sigma} into the self-energy expression \eqref{eq:appendice:weak_coupling_self_expansion:self_energy_sus_substituted} leads to the self-energy $\Sigma^{3a}$ (Eq.~\eqref{eq:nonequilibrium_quantum_many_body_physics:IPT:3a_diagram})

\begin{align}
\label{eq:appendice:weak_coupling_self_expansion:3a_self_energy_determination}
&\Sigma_{\sigma\sigma^{\prime}}^{3a}(z_1,z_2) =\notag\\
&iU(z_1)\mathcal{G}_{\sigma\bar{\sigma}_3}(z_1,\bar{z}_3)\bigl[U(\bar{z}_3)U(z_2)\mathcal{G}_{\sigma_3}^0(\bar{z}_3,z_2)\mathcal{G}_{-\bar{\sigma}_3}^0(z_2,\bar{z}_3^+)\notag\\
&\times\delta_{-\bar{\sigma}_3,\bar{\sigma}_4}\delta_{-\bar{\sigma}_3,\bar{\sigma}_5}\delta_{\bar{\sigma}_3,\sigma^{\prime}}\delta^{\mathcal{C}}(\bar{z}_3,\bar{z}_4)\delta^{\mathcal{C}}(z_2,\bar{z}_5)\bigr]\notag\\
&\times\left[\mathcal{G}_{\bar{\sigma}_4-\sigma}(\bar{z}_4,z_1^+)\mathcal{G}_{-\sigma\bar{\sigma}_5}(z_1,\bar{z}_5)\right],
\end{align}
while the second term of Eq.~\eqref{eq:appendice:weak_coupling_self_expansion:vertex_functional_derivative_2nd_order_Sigma} inserted into Eq.~\eqref{eq:appendice:weak_coupling_self_expansion:self_energy_sus_substituted} gives the self-energy $\Sigma^{3b}$ (Eq.~\eqref{eq:nonequilibrium_quantum_many_body_physics:IPT:3b_diagram})

\begin{align}
\label{eq:appendice:weak_coupling_self_expansion:3b_self_energy_determination}
&\Sigma_{\sigma\sigma^{\prime}}^{3b}(z_1,z_2) =\notag\\
&iU(z_1)\mathcal{G}_{\sigma\bar{\sigma}_3}(z_1,\bar{z}_3)\bigl[U(\bar{z}_3)U(z_2)\mathcal{G}_{\bar{\sigma}_3}^0(\bar{z}_3,z_2)\mathcal{G}_{-\bar{\sigma}_3}^0(\bar{z}_3,z_2^+)\notag\\
&\delta_{-\bar{\sigma}_3,\bar{\sigma}_4}\delta_{-\bar{\sigma}_3,\bar{\sigma}_5}\delta_{\bar{\sigma}_3,\sigma^{\prime}}\delta^{\mathcal{C}}(z_2,\bar{z}_4)\delta^{\mathcal{C}}(\bar{z}_3^+,\bar{z}_5)\bigr]\notag\\
&\times\left[\mathcal{G}_{\bar{\sigma}_4-\sigma}(\bar{z}_4,z_1^+)\mathcal{G}_{-\sigma\bar{\sigma}_5}(z_1,\bar{z}_5)\right].
\end{align}

The second set of $3^{\text{rd}}$-order self-energy diagrams is generated by substituting the second term of the expanded boldified Green's function \eqref{eq:appendice:weak_coupling_self_expansion:Hartree_second_order_Dyson_expansion} into each interacting Green's function making up the second-order self-energy diagram. This produces 3 different diagrams, whose expressions are

\begin{align}
\label{eq:appendice:weak_coupling_self_expansion:3c_self_energy_determination}
&\Sigma^{3c}_{\sigma\sigma^{\prime}}(z_1,z_2) = U(z_1)U(z_2)\notag\\
&\times\biggl[\mathcal{G}^0_{\sigma\bar{\sigma}}(z_1,\bar{z}_1)\bigl[-iU(\bar{z}_1)\mathcal{G}^0_{-\bar{\sigma}}(\bar{z}_1,\bar{z}_1^{+})\delta^{\mathcal{C}}(\bar{z}_1,\bar{z}_2)\delta_{\bar{\sigma},\bar{\sigma}^{\prime}}\bigr]\notag\\
&\times\mathcal{G}^0_{\bar{\sigma}^{\prime}\sigma}(\bar{z}_2,z_2)\biggr]\mathcal{G}^0_{-\sigma}(z_2,z_1^+)\mathcal{G}^0_{-\sigma}(z_1,z_2^+),
\end{align}
corresponding to Eq.~\eqref{eq:nonequilibrium_quantum_many_body_physics:IPT:3c_diagram}, 

\begin{align}
\label{eq:appendice:weak_coupling_self_expansion:3d_self_energy_determination}
&\Sigma^{3d}_{\sigma\sigma^{\prime}}(z_1,z_2) = U(z_1)U(z_2)\mathcal{G}^0_{\sigma}(z_1,z_2)\notag\\
&\times\biggl[\mathcal{G}^0_{-\sigma\bar{\sigma}}(z_2,\bar{z}_1)\bigl[-iU(\bar{z}_1)\mathcal{G}^0_{\bar{\sigma}}(\bar{z}_1,\bar{z}_1^{+})\delta^{\mathcal{C}}(\bar{z}_1,\bar{z}_2)\delta_{\bar{\sigma},\bar{\sigma}^{\prime}}\bigr]\notag\\
&\times\mathcal{G}^0_{\bar{\sigma}^{\prime}-\sigma}(\bar{z}_2,z_1^+)\biggr]\mathcal{G}^0_{-\sigma}(z_1,z_2^+),
\end{align}
corresponding to Eq.~\eqref{eq:nonequilibrium_quantum_many_body_physics:IPT:3d_diagram}, and 

\begin{align}
\label{eq:appendice:weak_coupling_self_expansion:3e_self_energy_determination}
&\Sigma^{3e}_{\sigma\sigma^{\prime}}(z_1,z_2) = U(z_1)U(z_2)\mathcal{G}^0_{\sigma}(z_1,z_2)\mathcal{G}^0_{-\sigma}(z_2,z_1^+)\notag\\
&\times\biggl[\mathcal{G}^0_{-\sigma\bar{\sigma}}(z_1,\bar{z}_1)\bigl[-iU(\bar{z}_1)\mathcal{G}^0_{\bar{\sigma}}(\bar{z}_1,\bar{z}_1^{+})\delta^{\mathcal{C}}(\bar{z}_1,\bar{z}_2)\delta_{\bar{\sigma},\bar{\sigma}^{\prime}}\bigr]\notag\\
&\times\mathcal{G}^0_{\bar{\sigma}^{\prime}-\sigma}(\bar{z}_2,z_2^+)\biggr],
\end{align}
corresponding to Eq.~\eqref{eq:nonequilibrium_quantum_many_body_physics:IPT:3e_diagram}.

Next, turning to the $3^{\text{rd}}$-order Hartree self-energy diagrams, the top Green's function of $\Sigma^{(2)}_H$ (Eq.~\eqref{eq:appendice:weak_coupling_self_expansion:Hartree_second_order}) is dressed by a Hartree self-energy insertion

\begin{align}
\label{eq:appendice:weak_coupling_self_expansion:3a_Hartree}
&\Sigma_{H,\sigma\sigma^{\prime}}^{3a}(z_1,z_2) = -iU(z_1)\mathcal{G}^0_{-\sigma}(z_1,\bar{z}_2)\notag\\
&\biggl[-iU(\bar{z}_2)\mathcal{G}^0_{\sigma\bar{\sigma}}(\bar{z}_2,\bar{z}_3)\left[-iU(\bar{z}_3)\mathcal{G}^0_{-\bar{\sigma}}(\bar{z}_3,\bar{z}_3^+)\delta^{\mathcal{C}}(\bar{z}_3,\bar{z}_4)\delta_{\bar{\sigma},\bar{\sigma}^{\prime}}\right]\notag\\
&\times\mathcal{G}^0_{\bar{\sigma}^{\prime}\sigma}(\bar{z}_4,\bar{z}_2^{+})\delta^{\mathcal{C}}(\bar{z}_2,\bar{z}_3)\biggr]\mathcal{G}^0_{-\sigma}(\bar{z}_3,z_1^+)\delta^{\mathcal{C}}(z_1,z_2)\delta_{\sigma,\sigma^{\prime}}
\end{align}
and this simplifies to $\Sigma^{3a}_H$ defined in Eq.~\eqref{eq:nonequilibrium_quantum_many_body_physics:IPT:Hartree_3a}. The next $3^{\text{rd}}$-order Hartree diagram is obtained by expanding the Dyson equation up to third order. The third-order term reads

\begin{align}
\label{eq:appendice:weak_coupling_self_expansion:Hartree_third_order_Dyson_expansion}
&\mathcal{G}^{(3)}_{-\sigma}(z_1,z_1^+) = \mathcal{G}^0_{-\sigma\bar{\sigma}}(z_1,\bar{z}_2)\notag\\
&\times\left[-iU(\bar{z}_2)\mathcal{G}^0_{-\bar{\sigma}}(\bar{z}_2,\bar{z}_2^{+})\delta_{\bar{\sigma},\bar{\sigma}^{\prime}}\delta^{\mathcal{C}}(\bar{z}_2,\bar{z}_3)\right]\mathcal{G}^0_{\bar{\sigma}^{\prime}\bar{\sigma}^{\prime\prime}}(\bar{z}_3,\bar{z}_4)\notag\\
&\times\left[-iU(\bar{z}_4)\mathcal{G}^0_{-\bar{\sigma}^{\prime\prime}}(\bar{z}_4,\bar{z}_4^{+})\delta_{\bar{\sigma}^{\prime\prime},\bar{\sigma}^{\prime\prime\prime}}\delta^{\mathcal{C}}(\bar{z}_4,\bar{z}_5)\right]\mathcal{G}^0_{\bar{\sigma}^{\prime\prime\prime}-\sigma}(\bar{z}_5,z_1^+).
\end{align}
Replacing the Green's function in the Hartree diagram of Eq.~\eqref{eq:appendice:weak_coupling_self_expansion:self_energy_sus_substituted} by $\mathcal{G}^{(3)}$ (Eq.~\eqref{eq:appendice:weak_coupling_self_expansion:Hartree_third_order_Dyson_expansion}), one obtains $\Sigma^{3b}$ as described by Eq.~\eqref{eq:nonequilibrium_quantum_many_body_physics:IPT:Hartree_3b},

\begin{align}
\label{eq:appendice:weak_coupling_self_expansion:Hartree_third_order_Dyson_expansion_3b}
&\Sigma^{3b}_{H,\sigma\sigma^{\prime}}(z_1,z_2) = -iU(z_1)\mathcal{G}^0_{-\sigma\bar{\sigma}}(z_1,\bar{z}_2)\notag\\
&\times\left[-iU(\bar{z}_2)\mathcal{G}^0_{-\bar{\sigma}}(\bar{z}_2,\bar{z}_2^{+})\delta_{\bar{\sigma},\bar{\sigma}^{\prime}}\delta^{\mathcal{C}}(\bar{z}_2,\bar{z}_3)\right]\mathcal{G}^0_{\bar{\sigma}^{\prime}\bar{\sigma}^{\prime\prime}}(\bar{z}_3,\bar{z}_4)\notag\\
&\times\left[-iU(\bar{z}_4)\mathcal{G}^0_{-\bar{\sigma}^{\prime\prime}}(\bar{z}_4,\bar{z}_4^{+})\delta_{\bar{\sigma}^{\prime\prime},\bar{\sigma}^{\prime\prime\prime}}\delta^{\mathcal{C}}(\bar{z}_4,\bar{z}_5)\right]\notag\\
&\times\mathcal{G}^0_{\bar{\sigma}^{\prime\prime\prime}-\sigma}(\bar{z}_5,z_1^+)\delta^{\mathcal{C}}(z_1,z_2)\delta_{\sigma,\sigma^{\prime}}.
\end{align}
For Eq.~\eqref{eq:appendice:weak_coupling_self_expansion:Hartree_third_order_Dyson_expansion_3b} to be nonzero, since the off-diagonal spin component of the Green's function is zero within the Hubbard model, it is easy to deduce that $\bar{\sigma}=\bar{\sigma}^{\prime}=\bar{\sigma}^{\prime\prime}=\bar{\sigma}^{\prime\prime\prime}=-\sigma$.

Finally, the very last $3^{\text{rd}}$-order Hartree self-energy diagram comes from the insertion of the $2^{\text{nd}}$-order self-energy diagram \eqref{eq:nonequilibrium_quantum_many_body_physics:IPT:second_order_bubble} into the second term of the Dyson equation expansion

\begin{align}
\label{eq:appendice:weak_coupling_self_expansion:Hartree_second_order_Dyson_expansion_2}
&{\mathcal{G}^{(2)}_{-\sigma}}^{\prime}(z_1,z_1^+) = \mathcal{G}^0_{-\sigma\bar{\sigma}}(z_1,\bar{z}_2)\notag\\
&\times\left[U(\bar{z}_2)\mathcal{G}^0_{\bar{\sigma}}(\bar{z}_2,\bar{z}_3)U(\bar{z}_3)\mathcal{G}^0_{-\bar{\sigma}}(\bar{z}_3,\bar{z}_2^+)\mathcal{G}_{-\bar{\sigma}}^0(\bar{z}_2,\bar{z}_3^+)\right]\notag\\
&\times\mathcal{G}^0_{\bar{\sigma}-\sigma}(\bar{z}_3,z_1^+).
\end{align}
Inserting ${\mathcal{G}^{(2)}}^{\prime}$ from \eqref{eq:appendice:weak_coupling_self_expansion:Hartree_second_order_Dyson_expansion_2} into the Hartree term of Eq.~\eqref{eq:appendice:weak_coupling_self_expansion:self_energy_sus_substituted}, one obtains $\Sigma^{3c}_H$ (Eq.~\eqref{eq:nonequilibrium_quantum_many_body_physics:IPT:Hartree_3c})

\begin{align}
\label{eq:appendice:weak_coupling_self_expansion:Hartree_third_order_Dyson_expansion_3c}
&\Sigma^{3c}_{H,\sigma\sigma^{\prime}}(z_1,z_2) = -iU(z_1)\mathcal{G}^0_{-\sigma\bar{\sigma}}(z_1,\bar{z}_2)\notag\\
&\times\left[U(\bar{z}_2)\mathcal{G}^0_{\bar{\sigma}}(\bar{z}_2,\bar{z}_3)U(\bar{z}_3)\mathcal{G}^0_{-\bar{\sigma}}(\bar{z}_3,\bar{z}_2^+)\mathcal{G}_{-\bar{\sigma}}^0(\bar{z}_2,\bar{z}_3^+)\right]\notag\\
&\times\mathcal{G}^0_{\bar{\sigma}-\sigma}(\bar{z}_3,z_1^+)\delta^{\mathcal{C}}(z_1,z_2)\delta_{\sigma,\sigma^{\prime}}.
\end{align}

\section{Interaction quench comparisons}
\label{appendice:ch:interaction_quench_comparisons}
%----------------------------------------------------------------------------------------

\begin{figure}[h!]
\includegraphics[width=1\linewidth]{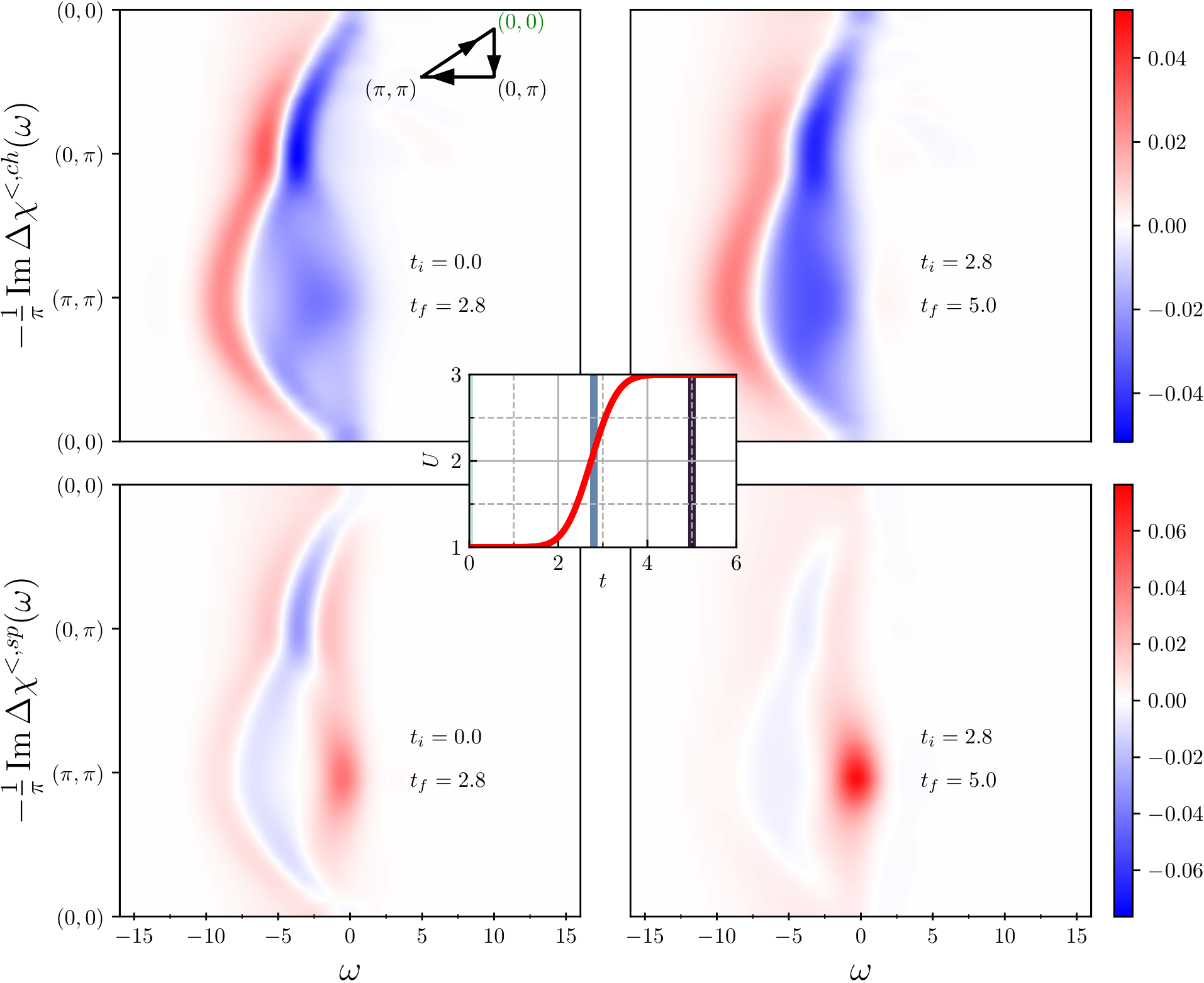} 
\caption{Top (Bottom) panels: Difference spectra of the DMFT+TPSC lesser component of the charge (spin) susceptibility after the interaction ramp from $U=1$ to $U=3$ shown in the inset. The time window employed in the Fourier transformation is $\Delta t=5$. Each row of panels uses the same color scale. The initial temperature is $T=0.33$.}
\label{fig:up_quench_time_diff_lesser_susceptibilities_dmft_tpsc_U_1_3}
\end{figure}

\begin{figure}[t]
\begin{minipage}[h]{1.0\columnwidth}
\begin{center}
\includegraphics[width=1\columnwidth]{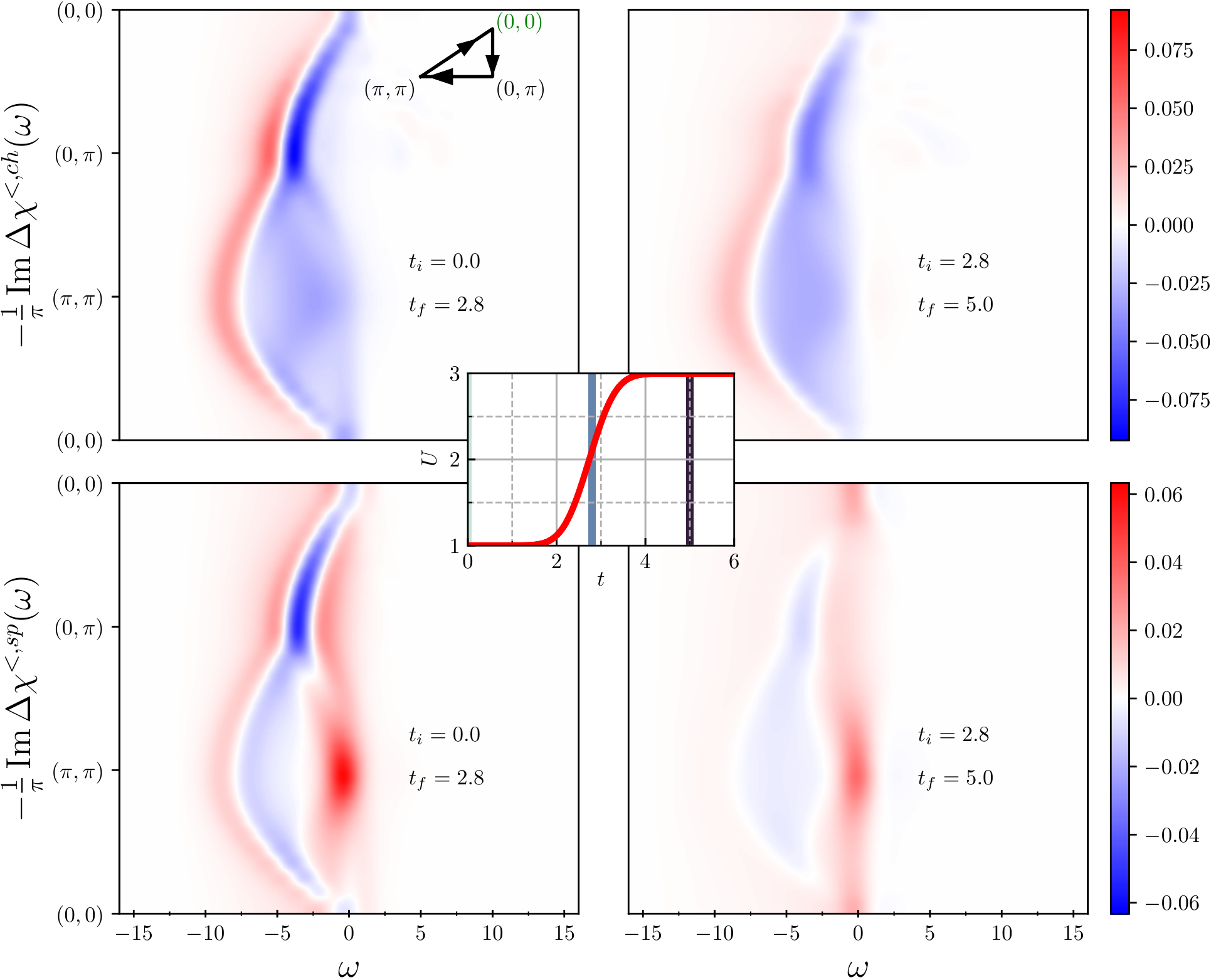} 
\end{center} 
\end{minipage}
\vfill
%\hspace{0.1 cm}
\begin{minipage}[h]{1.0\columnwidth}
\begin{center}
\includegraphics[width=1\columnwidth]{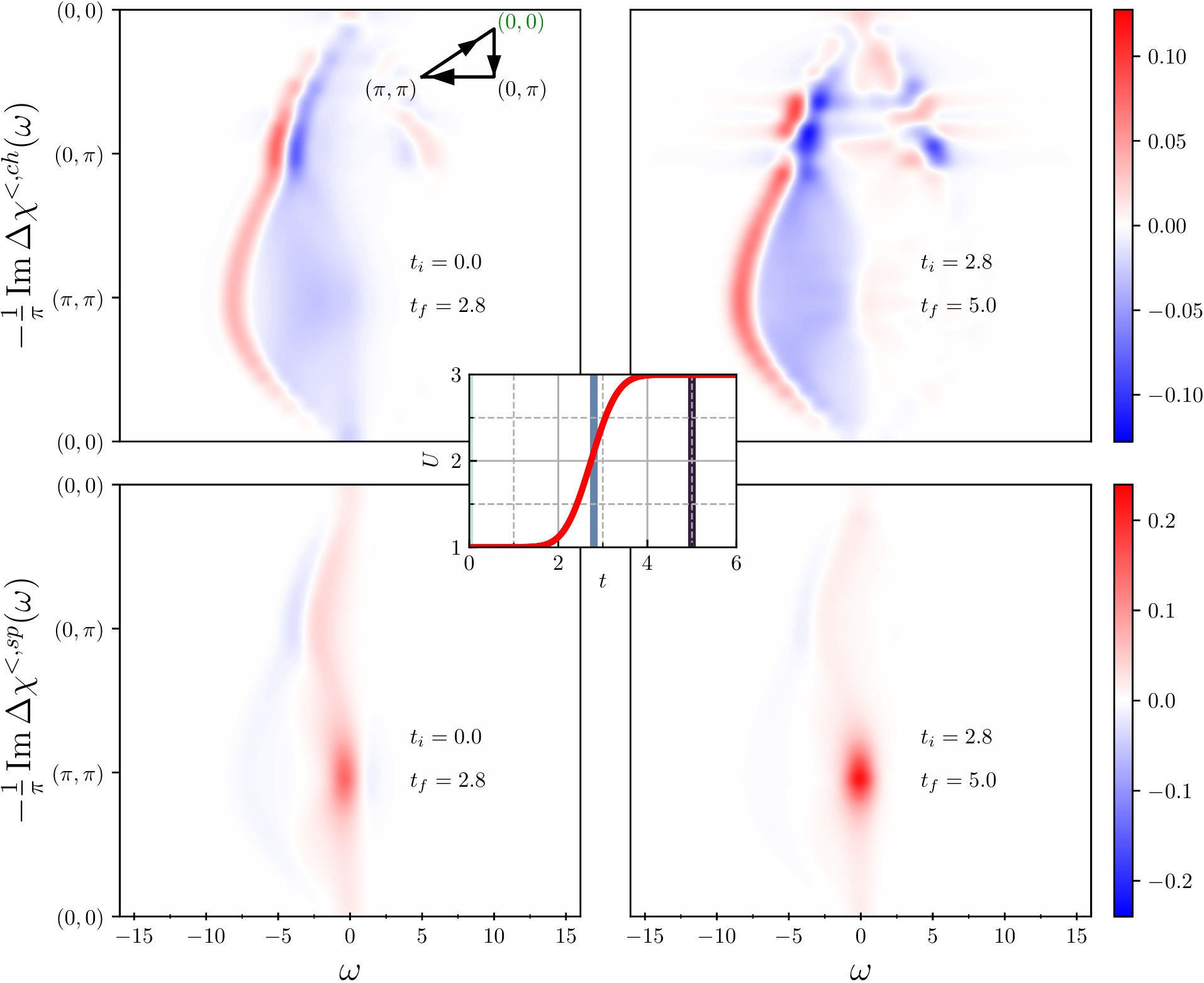} 
\end{center}
\end{minipage}
\caption{Top (Bottom) panels: Difference spectra of the lesser component of the charge (spin) susceptibility after the interaction ramp from $U=1$ to $U=3$ shown in the inset. The top (bottom) subplot shows the results obtained using TPSC+GG (TPSC). The time window employed in the Fourier transformation is $\Delta t=5$. Each row of panels uses the same color scale. The initial temperatures are $T=0.33$.
%\textcolor{blue}{[is the inset correct? should use the same ramp profile as in Fig. 26 REP: Yes, I have included a DMFT+TPSC calculation using the same ramp as in Fig.~27.]}
}
\label{fig:up_quench_time_diff_lesser_susceptibilities}
\end{figure}

In this section, we use DMFT+TPSC to calculate the time differences in the $\mathbf{k}$-resolved susceptibility spectra for the interaction ramp from $U=1$ to $U=3$ ($\gamma=1.5$ and $\delta=0.675$ in Eq.~\eqref{eq:ramp_profile_function_erf}) in the 2D system. The results are shown in Fig.~\ref{fig:up_quench_time_diff_lesser_susceptibilities_dmft_tpsc_U_1_3}. Analogous plots with TPSC+GG (top subplot) and TPSC (bottom subplot) data for the same ramp are plotted in Fig.~\ref{fig:up_quench_time_diff_lesser_susceptibilities}. The interaction ramps used are shown in the inset plots of Figs.~\ref{fig:up_quench_time_diff_lesser_susceptibilities_dmft_tpsc_U_1_3} and \ref{fig:up_quench_time_diff_lesser_susceptibilities}. Here, we consider the time differences $\Delta\chi^{\text{ch/sp},<}(t_f,t_i;\mathbf{k})$ with $t_i=0$ and $t_f=2.8$ in the left panels, and $t_i=2.8$ and $t_f=5$ in the right panels. The fact that the interaction ramp spans over a longer time window in Figs.~\ref{fig:up_quench_time_diff_lesser_susceptibilities_dmft_tpsc_U_1_3} and \ref{fig:up_quench_time_diff_lesser_susceptibilities}, compared to Fig.~\ref{fig:up_quench_time_diff_lesser_susceptibilities_dmft_tpsc_tp_0_1_U_2p5}, explains why the two panels corresponding to the first and second time window look more similar, just like is the case for the perpendicular lattice hopping ramp $t_z^{\text{hop}}$. Both the TPSC+GG and the DMFT+TPSC results are quantitatively very similar. TPSC shows a qualitatively similar time evolution, namely a growth of spin correlations with increasing interaction strength and a shift of the charge excitation spectra to higher energies due to the enhancement of the correlations. The ripples appearing in the TPSC results (bottom subplot of Fig.~\ref{fig:up_quench_time_diff_lesser_susceptibilities}) are of the same nature as those showing up in the $t_z^{\text{hop}}$ ramp (Fig.~\ref{fig:k_resolved_retarded_charge_2_3_D_tpsc}).

%----------------------------------------------------------------------------------------
\section{Nonequilibrium approximation to the TPSC irreducible vertices}
\label{appendice:ch:noneq_approx_TPSC_vertices}
%----------------------------------------------------------------------------------------

In this section, we motivate the approximation employed in the Bethe-Salpeter equations so as to satisfy the two-particle sum-rules~\eqref{eq:fluctuation_dissipation_two_particle} on the real-time axis. Since the local sum-rules~\eqref{eq:fluctuation_dissipation_two_particle} involve \textit{lesser} components,\footnote{Equivalently, because the susceptibilities are bosonic contour-time objects, the \textit{greater} component could also be used in the sum-rules~\eqref{eq:fluctuation_dissipation_two_particle} ($\chi^<(t,t)=\chi^>(t,t)$).} the Langreth rule for the \textit{lesser} component of the spin/charge susceptibility is used:
\begin{align}
\label{eq:appendice:ch:noneq_approx_TPSC_vertices:convolution_chi_lesser}
&\chi^{\text{sp/ch},<(>)}_{\mathbf{q}}(t,t^{\prime}) = \int_{0}^{t}\mathrm{d}\bar{t} \ \chi^{0,R}_{\mathbf{q}}(t,\bar{t})\Gamma^{\text{sp/ch}}(\bar{t})\chi^{\text{sp/ch},<(>)}_{\mathbf{q}}(\bar{t},t^{\prime})\notag\\
&\hspace{1.0cm} + \int_{0}^{t^{\prime}}\mathrm{d}\bar{t} \ \chi^{0,<(>)}_{\mathbf{q}}(t,\bar{t})\Gamma^{\text{sp/ch}}(\bar{t})\chi_{\mathbf{q}}^{\text{sp/ch},A}(\bar{t},t^{\prime})\notag\\
&\hspace{1.0cm} - i\int_{0}^{\beta}\mathrm{d}\bar{\tau} \ \chi^{0,\neg}_{\mathbf{q}}(t,\bar{\tau})\Gamma^{\text{sp/ch}}(0^-)\chi_{\mathbf{q}}^{\text{sp/ch},\invneg}(\bar{\tau},t^{\prime}).
\end{align}
\noindent
Note that in these expressions, the general contour-time arguments $z$ have been replaced by real-time variables $t$. Now, the local-time two-particle sum-rules apply at equal time, \textit{i.e.}, when $t=t^{\prime}$ in Eq.~\eqref{eq:appendice:ch:noneq_approx_TPSC_vertices:convolution_chi_lesser}. In the time-stepping scheme, these sum-rules have to be fulfilled at each time step by varying the local vertices $\Gamma^{\text{sp/ch}}$ at the latest time $t$. Since the susceptibilities $\chi^{\text{sp/ch}}$ and $\chi^{0}$ are bosonic quantities, their equal-time retarded/advanced components give $0$, because 
$$\chi^{R}_{\mathbf{q}}(t,t^{\prime})=\Theta(t-t^{\prime})\left[\chi^{>}_{\mathbf{q}}(t,t^{\prime})-\chi^{<}_{\mathbf{q}}(t,t^{\prime})\right] \stackrel{t^{\prime}\to t}{=} 0$$
\noindent
and
$\chi^{R}_{\mathbf{q}}(t,t^{\prime})^{\ast}=\chi^{A}_{\mathbf{q}}(t^{\prime},t).$
\noindent
This property makes it numerically difficult to fix the vertex at time $t$ from the solution of the BSE \eqref{eq:appendice:ch:noneq_approx_TPSC_vertices:convolution_chi_lesser}. 
We thus change $\chi^0_\mathbf{q}(z,\bar z)\Gamma^\text{sp/ch}(\bar z)$ to $\Gamma^\text{sp/ch}(z)\chi^0_\mathbf{q}(z,\bar z)$ in the BSE which defines $\chi_\textbf{q}$ to obtain Eq.~\eqref{eq:bethe_Salpeter_eq_approximated}, which is an ad-hoc modification of the original TPSC scheme.

Overcoming this approximation might involve resorting to modified two-particle sum-rules more suitable to nonequilibrium set-ups, or alternative schemes for extracting the vertices directly from the impurity self-energy, as done in Ref.~\onlinecite{kusunose_influence_2006} at equilibrium.

\bibliography{Bibliography}
\end{document}